\documentclass[aps,prx,superscriptaddress,twocolumn,balancelastpage]{revtex4-2}

\usepackage{times}
\usepackage{amsmath,amssymb}
\usepackage{graphicx}
\graphicspath{{figures/}{/}}
\usepackage{verbatim}
\usepackage[dvipsnames]{xcolor}
\usepackage{float}
\usepackage{mathtools}
\usepackage[export]{adjustbox}
\usepackage{hyperref}
\usepackage{placeins}  
\usepackage{flafter}     
\usepackage{physics}
\usepackage{dsfont} 
\usepackage{wasysym}
\usepackage{color}
\usepackage{dblfloatfix}
\usepackage{longtable}
\usepackage{multirow}

\newcommand{\ue}{\text{e}}
\newcommand{\ui}{\text{i}}

\newcommand{\newc}{\newcommand}

\newc{\beq}{\begin{equation}}
\newc{\eeq}{\end{equation}}
\newc{\kt}{\rangle}
\newc{\br}{\langle}
\newc{\beqa}{\begin{eqnarray}}
\newc{\eeqa}{\end{eqnarray}}
\newc{\pr}{\prime}
\newc{\longra}{\longrightarrow}
\newc{\ot}{\otimes}
\newc{\rarrow}{\rightarrow}
\newc{\h}{\hat}
\newc{\bom}{\boldmath}
\newc{\btd}{\bigtriangledown}
\newc{\al}{\alpha}
\newc{\be}{\beta}
\newc{\ld}{\lambda}
\newc{\sg}{\sigma}
\newc{\p}{\psi}
\newc{\eps}{\epsilon}
\newc{\om}{\omega}
\newc{\mb}{\mbox}
\newc{\tm}{\times}
\newc{\hu}{\hat{u}}
\newc{\hv}{\hat{v}}
\newc{\hk}{\hat{K}}
\newc{\ra}{\rightarrow}
\newc{\non}{\nonumber}
\newc{\hs}{\hspace}
\newc{\longla}{\longleftarrow}
\newc{\ts}{\textstyle}
\newc{\f}{\frac}
\newc{\df}{\dfrac}
\newc{\ovl}{\overline}
\newc{\bc}{\begin{center}}
\newc{\ec}{\end{center}}
\newc{\dg}{\dagger}
\newc{\rz}{\textcolor{red}{0}}
\newc{\ro}{\textcolor{red}{1}}
\newc{\CC}{C $\otimes$ C }
\newc{\RR}{R $\otimes$ R }
\newc{\EE}{E $\otimes$ E }

\newcommand{\HIDDEN}[1]{}

\makeatletter
\let\Hy@backout\@gobble
\makeatother

\begin{document}

\title{Quantum coherence controls the nature of equilibration in coupled chaotic
systems}

\author{Jethin J. Pulikkottil}
\affiliation{Department of Physics and Astronomy, Washington State University,
             Pullman, WA~99164-2814}
\author{Arul Lakshminarayan}
\affiliation{Department of Physics, Indian Institute of Technology Madras, 
Chennai 600036, India}
\author{Shashi C. L. Srivastava}
\affiliation{Variable Energy Cyclotron Centre, 1/AF Bidhannagar,
             Kolkata 700064, India.}
\affiliation{Homi Bhabha National Institute, Training School Complex,
		 Anushaktinagar, Mumbai - 400094, India}
\author{Maximilian F. I. Kieler}
\affiliation{Technische Universit\"at Dresden, Institut f\"ur Theoretische
             Physik and Center for Dynamics, 01062 Dresden, Germany}
\author{Arnd B\"acker}
\affiliation{Technische Universit\"at Dresden, Institut f\"ur Theoretische
             Physik and Center for Dynamics, 01062 Dresden, Germany}
\author{Steven Tomsovic}
\affiliation{Department of Physics and Astronomy, Washington State University,
             Pullman, WA~99164-2814}

\date{\today}

\begin{abstract}
A bipartite system whose subsystems are fully quantum chaotic and coupled by a 
perturbative interaction with a tunable strength is a paradigmatic
model for investigating how isolated quantum systems relax towards an 
equilibrium. It is found that quantum coherence of the initial product states in
the uncoupled eigenbasis can be viewed as a resource for equilibration and 
approach to thermalization as manifested by the entanglement. Results are given 
for four distinct perturbation strength regimes, the ultra-weak, weak, 
intermediate, and strong regimes. For each, three types of initially unentangled 
states are considered, coherent random-phase superpositions, random 
superpositions, and eigenstate products. A universal time scale is identified
involving the interaction strength parameter. Maximally coherent initial states 
thermalize for any perturbation strength in spite of the fact that in the 
ultra-weak perturbative regime the underlying eigenstates of the system have a 
tensor product structure and are not at all thermal-like; though the time taken 
to thermalize tends to infinity as the interaction vanishes. In contrast to the 
widespread linear behavior, in this regime the entanglement initially grows 
quadratically in time.
\end{abstract}

\maketitle


\section{Introduction \label{sec:Introduction} }

Thermalization in isolated quantum many-body systems has been an 
active area of research for many years~\cite{Rigol08}. The essential question is
whether the system prepared in some initial state of interest reaches a thermal 
equilibrium after a sufficiently long time.  As proposed roughly three decades 
ago~\cite{Deutsch1991, Srednicki1994}, thermalization really happens at the 
eigenstate level and is indicative of quantum chaotic nature of the system under 
consideration. Thus, the system relaxes to a thermal state irrespective of the 
initial state and without having to do any initial state ensemble averaging. 
This thermalization is seen in the subsystem states of such isolated systems 
where the reduced density matrix of the subsystem follows quantum statistical 
mechanics~\cite{Srednicki1994}. There are numerous contemporary studies on the 
process of thermalization in isolated quantum many-body systems and for 
different classes of initial states~\cite{Rigol2016, HeRig2013, TorSan2013, 
ColKorCal2014, RigSre2012, Abanin_RMP2019, Khare_2020}.  Whereas any lack of 
thermalization is often attributed to disorder induced many-body 
localization~\cite{Abanin_RMP2019}, it may originate from memory effects in the 
initial state purely from weakness of the interactions~\cite{Rau96}.

Compelling insights pertaining to the foundations of quantum statistical 
mechanics can be gained through the study of paradigmatic bipartite systems 
whose subsystems are quantum chaotic.  By adding an interaction between 
the subsystems with a tunable strength, relaxation towards an equilibrium in the 
full system of various classes of initial states can be studied over the 
complete range from vanishing interactions to the opposite limit of extremely 
strong interactions. There are classes of initially unentangled (product) 
states of such subsystems that respond quite differently to weak interaction 
strengths. Some may thermalize achieving near maximal entanglement as random
states, others may equilibrate but with smaller entanglement, whereas others 
practically develop no entanglement at all.

In this context, taking advantage of the universality of chaotic single-particle
or many-body dynamics, a random matrix 
model that highlights some of the major scenarios in the case of bipartite 
weakly coupled chaotic systems is explored analytically and numerically. 
In particular, three sets of unentangled initial pure states constructed from 
the subsystem eigenstates in the absence of interactions are contrasted: 
(i) tensor products of coherent random-phase superpositions (\CC
type), (ii) tensor products of random superpositions (\RR type), 
and (iii) tensor product of individual subsystem eigenstates (\EE 
type). In each of these cases, when interactions are turned on, the 
interest is in studying the entanglement production and its time scale, its 
long-time average, and the nature of the fluctuations. Naturally 
the interaction strength plays a crucial role and the full range is 
explored from the perturbative regime $\Lambda \ll 1$, 
to $\Lambda \gtrsim 1$ where $\Lambda$ is a scaled universal 
transition parameter identified earlier 
\cite{French88a, SriTomLakKetBae2016, LakSriKetBaeTom2016}. 
In this regime, a spectral transition is observed from the Poisson to Wigner 
level statistics, although the subsystem dynamics is always fully chaotic.

The set (i) corresponds to an ensemble of initial states of maximal coherence 
\cite{BaiDu_2015}. Coherence in a state (represented as density matrix) is 
quantified by the off-diagonal elements of its density matrix and is a basis
dependent notion. However, fixing a preferred basis, it has been found to be 
useful to develop coherence measures in quantum information theory, and states 
that are diagonal are incoherent.
In parallel to the resource theory of entanglement, a resource theory for 
quantum coherence has been developed; for a review see~\cite{Streltsov_RMP2017}.
For both sets (i) and (ii), the infinite-time averaged entanglement can be
nearly maximal for arbitrarily small interactions, although the approach to 
the long time limit can be arbitrarily slow. 
It turns out that set (i) of random-phases with maximal coherence 
(C $\otimes$ C) engenders a large amount of entanglement, and already reaches 
the typical thermalized entangled state value. This occurs in spite of the 
perturbative nature of the interactions, i.e.~$\Lambda \ll 1$ 
and Poissonian level statistics~\cite{Tkocz12}. 
For the set (ii) of random superpositions (R $\otimes$ R) a
smaller amount of entanglement is obtained in comparison to set (i) states. 
The set (iii) of subsystem eigenstate products (E $\otimes$ E) are 
incoherent from this point of view as they have diagonal density matrices. 
Under perturbative interactions, their entanglement remains essentially 
perturbative~\cite{Jethin_PRE2020}. Thus, the results suggest investigating 
quantum coherence in the uncoupled eigenbasis as a ``resource" for 
thermalization. In addition, the universal rescaled time that 
was identified in~\cite{Jethin_PRE2020} holds true in the current study for all 
initial states considered. 

Previous studies have shown a 
linear-in-time entanglement growth in systems with signatures of classical chaos
\cite{Zurek_1994,MillerSarkar1999,Monteoliva_2000,Tanaka_2002,Fujisaki_2003,
Bandyopadhyay_2004}, and in many-body systems \cite{Calabrese_2005,KimHus2013,
Kaufman_2016}. This study reveals that the initial entanglement growth is 
controlled by both the transition parameter $\Lambda$ and the quantum coherence 
in the initial state. The former leads to a linear growth and the latter to a 
quadratic one where a competition between these two is observed and a time scale 
is derived depicting the transition between linear and quadratic growths. In the 
ultra-weak regime, the linear growth is suppressed and a dominant quadratic 
growth is seen.

The structure of this paper is as follows: the next section briefly presents 
essential background regarding thermalization, quantum coherence, 
quantum chaos, bipartite systems, the transition parameter, and the 
universal rescaled time.  In Sect.~\ref{sec:generalities}, a relation is 
given for the infinite time averaged purity, and hence linear entropy as 
well, along with a summary of the four perturbation regimes. This is 
followed with a section on analytical and numerical results for the 
limiting extremes of ultra-weak and ultra-strong interaction strengths.  
The remaining perturbation regimes, weak and intermediate strength, are 
covered in Sect.~\ref{sec:wipr}.  The final section summarizes the 
results of this work and gives a brief outlook.


\section{Background \label{sec:Background} }

It is helpful to review key background information regarding
thermalization and equilibration in isolated quantum systems, quantum coherence
as a resource, bipartite systems and linear entropy, and set up notation
to be used throughout the rest of the paper.  Also included are the relevant 
random matrix transition ensembles, and the concepts of symmetry 
breaking, the transition parameter, and universal rescaled time.


\subsection{\label{subsec:thermalization} 
Thermalization in isolated quantum many-body systems}

An isolated many-body system prepared in an initial pure state thermalizes 
when evolved for a sufficiently long time if the eigenstates of the system 
are quantum chaotic in nature, and behave according to the
\emph{eigenstate thermalization hypothesis} (ETH)~\cite{Deutsch1991, 
Srednicki1994}. For such systems, any generic initial state
will approach thermal equilibrium in the strong sense, meaning almost all the
initial states relax to equilibrium beyond some time, thus exhibiting 
thermal distributions such as Maxwell, or Bose-Einstein, or Fermi-Dirac 
depending on the exchange symmetry and stationary expectation values. In 
contrast to strong thermalization, \emph{weak} thermalization has been found to 
exist in some types of initial product states~\cite{Banuls2011, Lin2017}.  Weak 
thermalization occurs when the observable of interest fluctuates about the 
thermal average and only with long-time averaging gives the thermal result, 
in contrast to the stationarity of strong thermalization. Numerical simulations
show that a weakly interacting bipartite system may achieve an equilibrium 
~\cite{Linden2009, Deutsch2018} -- in either the weak or strong sense -- that
is different from a thermal one, and may be characterized similarly to that of
thermal fluctuations in quantum chaotic systems~\cite{Srednicki1996}.

Given a generic initial product state $\ket{\alpha}$ and an observable or
a measure of interest, the system (of size sufficiently large 
\cite{Linden2009}) may reach an equilibrium after a long time and can be 
identified by looking at the infinite time average of the quantum expectation 
value of the observable or measure in the time evolved state $\ket{\alpha(t)}$. 
In this paper, the linear entropy (introduced ahead) serves as a suitable
(entanglement) measure for the time evolved state $\ket{\alpha(t)}$ and is 
denoted by $S_2(t)$. The infinite time average of $S_2$ is computed as
\begin{align}
    \overline{S}_2 = \lim_{\tau \rightarrow \infty} \frac{1}{\tau} \int_0^\tau
    S_2(t) \dd{t},
\end{align}
and the equilibrium value for an ensemble of initial states can be taken as the 
initial state ensemble average of
$\overline{S}_2$ denoted as $\langle \overline{S}_2 \rangle$, where the angular
brackets represent the initial state ensemble averaging.

This prompts an immediate question as to how $\overline{S}_2$ is distributed
across various initial states from an ensemble.  For example, does the 
probability density of $\overline{S}_2$ behave as a power-law (indicating 
heavy-tails) or more localized exponential type?  If the density contains a 
power-law, where the fluctuations can be quite conspicuous compared with 
$\langle \overline{S}_2 \rangle$, then the notion of equilibrium becomes 
suspect.  To the extent that the various initial states of an ensemble generate 
an $\overline{S}_2$ closer and closer to $\langle \overline{S}_2 \rangle$, the 
sharper the notion of equilibrium becomes. To study fluctuations, a 
dimensionless normalized variance is defined as
\begin{align}
    \sigma^2(X) = \frac{\langle (X-\langle X \rangle)^2 \rangle}
    {\langle X \rangle^2} \label{eq:sigma2_def}
\end{align}
for a quantity $X$ distributed as $P_X(x)$. This fluctuation measure is
also used in the studies of optical and acoustic scintillation, or irradiance 
fluctuations caused by small temperature variations in a random medium; 
for example see \cite{Andrews1999}. 

In the context of quantifying equilibrium, the fluctuation measure 
$\sigma^2(\overline{S}_2)$ is employed, which measures the scaled variance of 
the probability density across initial states of the linear entropy's infinite 
time average and is referred to as the \emph{equilibrium measure}. 
If $\sigma^2(\overline{S}_2) \sim 1$, the equilibrium is quite weak. On the 
other hand, if $\sigma^2(\overline{S}_2) \ll 1$, an equilibrium is 
possible since a majority of the initial states in the ensemble of interest 
generate an $\overline{S}_2$ close to $\langle \overline{S}_2 \rangle$. 
Similar to the weak thermalization mentioned earlier, $S_2(t)$ may exhibit 
oscillations from the equilibrium, even after a long time. Performing infinite 
time averaging removes any temporal fluctuations about the equilibrium, and 
thus, examining the characteristics of the $\overline{S}_2$ probability density 
alone is insufficient to reveal whether the system relaxes to an equilibrium.

To explore the relaxation to an equilibrium value, the infinite time average of 
the $S_2(t)$ temporal fluctuation about $\langle S_2(t) \rangle$ is useful, and 
is given by
\begin{align}
    \overline{\sigma^2\big(S_2\big)} = \lim_{\tau \rightarrow \infty}
    \frac{1}{\tau} \int_0^\tau \dd{t} \sigma^2\big(S_2(t)\big),
    \label{eq:sigma2InfAvg}
\end{align}
which is referred to as \emph{relaxation measure}. 
If $\overline{\sigma^2(S_2)} \sim 1$ the system relaxes 
to the equilibrium in the weak sense characterized by 
glaring temporal fluctuations about the equilibrium and is referred to as
\emph{weak equilibration} in the spirit of weak thermalization discussed 
earlier. If the more stringent 
condition, $\overline{\sigma^2\big(S_2\big)} \ll 1$, is 
satisfied, then the system relaxes to 
equilibrium in the strong sense for almost all the initial states in the 
ensemble of interest, and is referred to as \emph{strong equilibration}.
Moreover, if the equilibrium value coincides with the thermal value, which is
based on random pure state Haar measure average of $S_2$ \cite{Lubkin1978}, then
the system thermalizes in the strong sense. This is referred to as \emph{strong
thermalization}.

\subsection{Quantum coherence as a resource}

A formal resource theory of quantum coherence, and its quantification was 
developed recently \cite{Aberg2006, Baumgratz2014, Streltsov_RMP2017}, which is 
not to be confused with the concept of coherent states for bosonic many-body 
systems \cite{Sudarshan1963, Glauber1963}.
Fundamentally, quantum coherence is a basis dependent quantity, and depending on
the problem at hand a \emph{preferred} basis (denote as $\mathbb{B}$) is 
identified.  For example, an energy eigenbasis may be preferable for studying 
coherence in thermodynamics.  Coherence resource theory stems from identifying 
a set of incoherent states (denote as $\mathcal{I}_\mathbb{B}$ in the preferred 
basis), maximally coherent states, and incoherent operations, which will be 
briefly summarized here. For more details, see the review on this topic 
\cite{Streltsov_RMP2017}.

Consider a preferred basis $\mathbb{B} = \{\ket{m}\}_{m=1, \ldots, N}$ for a 
given $N$-dimensional Hilbert space $\mathcal{H}$. An incoherent state
represented by the density matrix $\varrho \in \mathcal{I}_\mathbb{B}$ is 
diagonal in the basis $\mathbb{B}$, i.e. 
$\varrho = \sum_{m=1}^N p_m \ket{m}\bra{m}$, with $\{p_m\}$ being probabilities
such that $\sum p_m = 1$.
For the states of the form (C-type)
\begin{align}
    \ket{\alpha_K}_\text{C} =\frac{1}{\sqrt{K}} \sum_{m=1}^K \ue^{\ui \varphi_m} 
    \ket{m} \label{eq:K-alphaState}
\end{align}
involving an equal superposition, with random phases 
$0 \leqslant \varphi_m < 2\pi$, of $K$ basis kets, the amount of coherence 
increases with $K$ 
and becomes maximal for $K = N$. A quantum operation, $\Phi$, on a state 
that does not generate any 
coherence, but may consume it, is regarded as an incoherent operation. More 
precisely, for a quantum operation that admits a set of Kraus operators 
$\{ \mathcal{K}_i \}$ (also known as operation elements) \cite{Nielsen2010}, 
i.e., $\varrho' = \Phi[\varrho] = \sum_i \mathcal{K}_i \varrho 
\mathcal{K}_i^\dagger$ such that $\sum_i \mathcal{K}_i^\dagger \mathcal{K}_i = 
\mathds{1}$ (trace-preserving), is an incoherent operation if $\mathcal{K}_i 
\varrho \mathcal{K}_i^\dagger/ \tr(\mathcal{K}_i \varrho \mathcal{K}_i^\dagger) 
\in \mathcal{I}_\mathbb{B}$ for all $i$.

Although there are several quantifiers of coherence, the following measure, 
although lacking some desired properties, is sufficient for our purposes and 
based on the Hilbert-Schmidt norm. It is a valid coherence monotone for 
trace-preserving operations such as unitary evolutions~\cite{Baumgratz2014}.
Furthermore, this coherence measure was used in recent 
studies~\cite{Styliaris_2019, Anand2021} (termed as 2-coherence) to establish 
a connection between quantum coherence and either localization or quantum chaos,
depending upon the circumstances. Using the notation introduced 
in~\cite{Styliaris_2019}, this coherence measure of a 
quantum state $\rho$ is given by
\begin{align}
    c^{(2)}_\mathbb{B}(\rho) = \sum_{\substack{m,m'\\m\neq m'}} 
    |\rho_{m m'}|^2, \label{eq:Generic2Coherence}
\end{align}
which for incoherent states is zero, and for a state of the form
Eq.~\eqref{eq:K-alphaState} equals $1-1/K$.

\subsection{Bipartite systems, linear entropy}

Consider pure states $\ket{\alpha}$ of a bipartite system whose Hilbert space
is a tensor product space, $\mathcal{H}^A \otimes \mathcal{H}^B$ with subsystem
dimensionalities $N_A$ and $N_B$, respectively.  Without loss of generality,
let $N_A \leq N_B$.  The dynamics of such a generic conservative system could 
be governed by a Hamiltonian or by a unitary Floquet operator in the case of 
periodically driven systems whose time evolution produces a quantum map. 
Specifically, a bipartite Hamiltonian system is of the form,
\begin{equation}
H_\epsilon = H_A \otimes \mathds{1}_B + \mathds{1}_A \otimes H_B + 
    \epsilon V_{AB} \ , \label{eq:GenericHamiltonian}
\end{equation}
where the non-interacting limit is $\epsilon = 0$. For a quantum map, the 
dynamics can be described by a unitary Floquet 
operator~\cite{SriTomLakKetBae2016},
\begin{equation}
\mathcal{U}_\epsilon = (U_A \otimes U_B) \,U_{AB}\ ,
\label{eq:GenericFloquet}
\end{equation}
for which the non-interacting limit is $U_{AB} \rightarrow \mathds{1}$. Assume 
that both $\epsilon V_{AB}$ and $U_{AB}$ are entangling interaction operators 
for $\epsilon > 0$~\cite{LakSriKetBaeTom2016}.

The Schmidt decomposition of a pure state \cite{Nielsen2010} is given by
\begin{equation}
 \ket{\alpha} = \sum_{l=1}^{N_A} \sqrt{\lambda_l} \, \ket{l^A}
    \ket{l^B},
  \label{eq:GenericSchmidtDecomposition}
\end{equation}
with Schmidt eigenvalues $\lambda_l$ ordered such that 
$\lambda_1 \ge \lambda_2 \ge \ldots \ge \lambda_{N_A}$ and 
$\sum_l \lambda_l = 1$. To simplify the notation of direct product
states, $\ket{l^A}\ket{l^B} \equiv \ket{l^A \l^B}$ notation with first entry for
subsystem $A$ and second for $B$ will be adopted in the paper, and superscripts
$A$ and $B$ are dropped whenever it is understood. The state
Eq.~\eqref{eq:GenericSchmidtDecomposition} is unentangled iff the largest 
eigenvalue $\lambda_1 = 1$ (all others vanishing), 
and maximally entangled if $\lambda_l = 1/N_A$ for all $l$. By partial 
traces, it follows that the reduced density matrices
\begin{equation}
\rho_A = \tr_B(\ket{\alpha}\bra{\alpha}), \qquad 
    \rho_B = \tr_A(\ket{\alpha}\bra{\alpha}) \,
\end{equation}
have the property
\begin{equation}
\rho_A \ket{l^A} = \lambda_l \ket{l^A}, \quad 
    \text{and} \quad \rho_B \ket{l^B} = \lambda_l \ket{l^B},
\end{equation}
respectively.  They are positive semi-definite, share the same non-vanishing 
(Schmidt) eigenvalues $\lambda_l$ and $\{ \ket{l^A} \}$, and 
$\{ \ket{l^B} \}$ form orthonormal basis sets in the respective Hilbert 
spaces.  For subsystem $B$ there are $N_B-N_A$ additional vanishing 
eigenvalues and associated eigenvectors. 

As previously mentioned, in this study the linear entropy $S_2$ of a state
$\ket{\alpha}$ is a suitable observable and given by
\begin{align}
    S_2 = 1-\mu_2, \label{eq:Generic_S2}
\end{align}
where $\mu_2$ is the purity defined by
\begin{align}
    \mu_2 = \tr_A(\rho_A^2) = \tr_B(\rho_B^2) = \sum_l \lambda_l^2.
    \label{eq:Generic_mu2}
\end{align}

\subsection{Quantum chaos, RMT and universality}

Random matrix theory (RMT) can be employed to model complex systems that 
exhibit quantum chaos, and in general, the statistical properties of a quantum
chaotic system does not depend on the system details except for the presence of 
fundamental symmetries that the system may respect~\cite{PorterBook, Brody81, 
Altland97}. In particular, in this study the focus is on Floquet systems of the 
form Eq.~\eqref{eq:GenericFloquet}.  Since the subsystems are assumed to be 
quantum chaotic in nature, the subsystem unitary operators $U_A$ and $U_B$ can
be regarded as members of the circular RMT ensembles. Furthermore, consider 
systems that are time-reversal non-invariant, hence circular 
unitary ensembles (CUE). Thus the dynamics of bipartite systems (generic in
the above context) is well captured by the random matrix transition ensemble
given by \cite{SriTomLakKetBae2016, LakSriKetBaeTom2016}
\begin{equation}
\mathcal{U}^{\text{RMT}}_\epsilon = (U_A^\text{CUE} \otimes U_B^\text{CUE}) 
\, U_{AB}.
\label{eq:GenericFloquetRMT}
\end{equation}
The interaction operator is assumed to be of the form 
$U_{AB} = \exp(\ui \epsilon \mathcal{V})$, where $\mathcal{V}$ is a hermitian
operator. In the direct product basis of the two subsystem ensembles
$\{\ket{ab}\}$ where $U_A$ and $U_B$ are represented in 
Eq.~\eqref{eq:GenericFloquetRMT}, $\mathcal{V}$ is assumed to be diagonal. That 
is, 
\begin{equation}
    \mathcal{V}_{ab,a'b'} = 2\pi \xi_{ab} \delta_{ab,a'b'}, \label{eq:V_def_ab}
\end{equation}
where $\xi_{ab}$ is a random independent number uniformly distributed in 
$(-1/2,\,1/2]$ for subsystem ensemble basis indexes $(a,\,b)$ such that 
$1\le a \le N_A$ and similarly for $b$ index.

Let the eigenvalues and corresponding eigenstates of the unitary operators
$U_A$ and $U_B$ for the subsystems and of $\mathcal{U}_\epsilon$ of the full 
bipartite system Eq.~\eqref{eq:GenericFloquetRMT} be
\begin{eqnarray}
 U_A \ket{j^A} &=& \exp(\ui \,\theta_j^A)  \ket{j^A},\quad j=1,2,3,\ldots,N_A
    \nonumber \\
 U_B \ket{k^B} &=& \exp(\ui \,\theta_k^B) \ket{k^B}, \quad k=1,2,3,\ldots,N_B
    \nonumber \\
 \mathcal{U}_\epsilon \ket{jk(\epsilon)} &=& \exp[\ui \, \theta_{jk}(\epsilon)]
    \ket{jk(\epsilon)}.
\end{eqnarray}
To simplify the notation, the superscripts $A$ and $B$ are dropped for both 
eigenkets, and the eigenvalues $\theta_j^A \equiv \theta_j$ ($\theta_k^B \equiv 
\theta_k$). From here on, it is understood that the labels $j$ and $k$ are 
reserved for the subsystems $A$ and $B$, respectively. Furthermore, the 
eigenbasis $\{\ket{j}\}$ is denoted as $\mathbb{B}_A$, and similarly for
subsystem $B$. Given the form Eq.~\eqref{eq:GenericFloquetRMT} of the unitary 
operator $\mathcal{U}_\epsilon$, in the limit $\epsilon \rightarrow 0$ one has
$\ket{jk(\epsilon)} \rightarrow \ket{jk}$ which is a product eigenstate of the 
unperturbed system and forms a complete basis (denoted as $\mathbb{B}_{AB}$) 
with spectrum $\theta_{jk}(0) = \theta_j+\theta_k \,
\text{mod} \,2\pi$.

\subsection{\label{subsec:SymmetryTP} Symmetry breaking and the transition 
parameter}

Earlier studies on spectral statistics \cite{SriTomLakKetBae2016} of weakly
interacting bipartite systems, of the type considered in this study, revealed 
that when the interaction is turned off between the subsystems, the system 
enjoys a dynamical symmetry. This symmetry (for $\epsilon = 0$) can be viewed 
as having two subsystem total energies that are separately conserved in 
Eq.~\eqref{eq:GenericHamiltonian}, or having product structure of subsystem
Floquet operators for the system Floquet operator in 
Eq.~\eqref{eq:GenericFloquet}. Introducing a weak interaction between the 
subsystems weakly breaks this symmetry, and in the limit of strong interaction,
the symmetry is completely broken. Since the subsystems are assumed quantum 
chaotic, a universal scaling parameter, the so-called transition parameter -- 
a concept originally appearing in statistical nuclear physics
~\cite{Pandey83, French88a, Tomsovicthesis} -- governs the influence of the 
symmetry breaking on the system's statistical properties. The transition 
parameter is defined as
\begin{align}
    \Lambda = \frac{v^2(\epsilon)}{D^2}, \label{eq:LambdaDef}
\end{align}
where $D$ is the local mean level spacing and $v^2(\epsilon)$ is the (local) 
average of off-diagonal (but close to the diagonal) intensities of the 
symmetry-breaking operator represented in the symmetry-preserving eigenenergy 
basis. For unitary systems that are of the type considered here, the mean 
(quasi-energy) level spacing is uniform and equals $D = 2\pi/(N_A N_B)$ 
\cite{Tkocz12}.

For the random matrix transition ensemble in Eq.~\eqref{eq:GenericFloquetRMT} 
with $U_A$ and $U_B$ members of CUE~\cite{SriTomLakKetBae2016},
\begin{align}
    \Lambda = \frac{N_A^2 N_B^2}{4\pi^2(N_A+1)(N_B+1)} 
    \Big[1- \frac{\sin^2(\pi \epsilon)}{\pi^2
    \epsilon^2} \Big] \approx \frac{\epsilon^2  N_A N_B}{12},
    \label{eq:LambdaExpCUE}
\end{align}
where the last result is in the limit of large $N_A,\,N_B$, $\epsilon \ll
1$, and $\Lambda$ ranges over $0 \leqslant \Lambda \leqslant N_A N_B /4\pi^2$,
where limiting cases are the fully symmetry preserving, and the fully broken
symmetry, respectively.

The transition parameter facilitates the comparison of quantum chaotic 
systems' statistical properties regardless of size and kind, i.e.~regardless 
of whether it is single particle or many-body, fermionic or bosonic.  The 
system dependent details can be mapped onto a value for the universal transition 
parameter in such a way that all quantum chaotic systems possessing the same 
value of $\Lambda$ possess identical statistical properties.  It has been 
calculated for both weakly broken  fundamental symmetries~\cite{French88b, 
Tomsovicthesis, Bohigas95, Tomsovic00} and dynamical symmetries
~\cite{Bohigas93, Tomsovic94}. Calculations of $\Lambda$ for weakly 
interacting coupled kicked rotors~\cite{SriTomLakKetBae2016} and coupled 
kicked tops~\cite{Herrman2020} have been given.  

For the RMT transition ensemble of Eq.~\eqref{eq:GenericFloquetRMT}, the
off-diagonal matrix elements of the symmetry-breaking operator $\mathcal{V}$
in the unperturbed product subsystem eigenbasis behave as complex Gaussian 
random variables and the diagonal ones as zero-centered Gaussian random 
variables. The transformed $\mathcal{V}_{jk,j'k'}$ is given by
\begin{align}
    \mathcal{V}_{jk,j'k'} = \sum_{a,b} u^{A*}_{ja} \, u^{B*}_{kb} \, u^A_{j'a}
    \, u^B_{k'b} \, (2 \pi \xi_{ab}) \label{eq:Vmatjk_def}
\end{align}
where $u^A$ and $u^B$ are the unitary transformation matrices for subsystem $A$ 
and $B$, respectively. Since the subsystems are quantum chaotic in nature, both 
the transformation matrices are Haar measure distributed on respective subsystem 
unitary groups. For large $N_A$ and $N_B$, the real and/or 
imaginary parts (depending on diagonal element or not) are Gaussian distributed
with certain variance, where the variance is computed with respect to 
uniformly distributed $\xi_{ab}$ and the Haar measure on the unitary groups 
for both $u^A$ and $u^B$ \cite{Puchala11}. It can be shown explicitly that (see
App.~\ref{App:HaarUnitary}),
\begin{align}
    \langle |\mathcal{V}_{jk,j'k'}|^2  \rangle = \frac{\pi^2
    (1+\delta_{jj'})(1+\delta_{kk'})}{3(N_A+1)(N_B+1)}
    ,  \label{eq:Vmatjk_var}
\end{align}
for any given pairs of indexes $jk$ and $j'k'$.
Furthermore, the off-diagonal elements can be rewritten as
\begin{equation}
|\mathcal{V}_{jk,j'k'} |^2 = \langle |\mathcal{V}_{jk,j'k'}|^2 \rangle \,
    w_{jk,j'k'},
\end{equation}
where $w_{jk,j'k'}$ is distributed as an exponential $P_w(x) = \exp(-x)$. Note
that the off-diagonal elements that are close to the diagonal that enter into the 
$\Lambda$ definition given in Eq.~\eqref{eq:LambdaDef} 
has index pair such that $j \neq j'$ and $k \neq k'$, whereas
off-diagonal elements with either $j=j'$ or $k=k'$ are much further 
away from the diagonal but can be related to $\Lambda$ via
Eq.~\eqref{eq:Vmatjk_var}. Thus the off-diagonal absolute squared matrix
elements can be rescaled as
\begin{equation}
    \epsilon^2 |\mathcal{V}_{jk,j'k'}|^2 = \Lambda D^2 (1+\delta_{jj'})
    (1+\delta_{kk'}) \, w_{jk,j'k'}. \label{eq:V_offscaling}
\end{equation}
Notice that, from Eq.~\eqref{eq:Vmatjk_var}, the variance of diagonal matrix
elements is four times that of the off-diagonal ones with $j \neq j'$ and $k
\neq k'$. Scaling the diagonal matrix elements with its standard deviation 
gives
\begin{equation}
    \mathcal{V}_{jk,jk} = x_{jk} \sqrt{\langle |\mathcal{V}_{jk,jk}|^2\rangle},
    \label{eq:V_diagscaling}
\end{equation}
in which $x_{jk}$ follows a zero-centered Gaussian distribution with 
unit variance. It is worth mentioning that the various diagonal elements are 
correlated to each other and the covariance between any two diagonal elements 
can be calculated similar to Eq.~\eqref{eq:Vmatjk_var} mentioned earlier (see
App.~\ref{App:HaarUnitary}), which in terms of rescaled variables is given by
\begin{equation}
    \langle x_{jk}\, x_{j'k'} \rangle =
    \frac{1}{4}(1+\delta_{jj'})(1+\delta_{kk'}). \label{eq:x_jkDiagCov}
\end{equation}
Moreover, the unperturbed spectrum $\{\theta_{jk}(0)\}$ is an uncorrelated
spectrum and behaves as Poissonian (for large $N_A$, and $N_B$), so adding 
in the first order perturbation corrections (i.e. 
$\epsilon \mathcal{V}_{jk,jk}$) that 
are random will not change the statistical nature of the spectrum.

With these in mind, it is useful to cast the theory in terms of 
universal parameters, namely, the transition parameter and rescaled 
time (introduced ahead in Subsec.~\ref{subsec:univtime}). Let
\begin{equation}
    s_{jk,j'k'}(\epsilon) = \frac{\theta_{jk}(\epsilon) -
    \theta_{j'k'}(\epsilon)}{D} \label{eq:UnfoldedSpectrum}
\end{equation}
be the unfolded level spacing of the perturbed spectrum whose (local)  mean 
level spacing is unity \cite{Mehta04}. Define, $s_{jk,j'k'} =
s_{jk,j'k'}(0)$. Then, in terms of $\Lambda$ and other rescaled quantities, 
the standard perturbation expression for Eq.~\eqref{eq:UnfoldedSpectrum}
is given by
\begin{align}
    s_{jk,j'k'}(\Lambda) & \approx s_{jk,j'k'} + 2\sqrt{\Lambda}  \,
    (x_{jk}-x_{j'k'} ) \nonumber \\
    & \quad + \frac{2 \Lambda w_{jk,j'k'}}{s_{jk,j'k'}},
    \label{eq:unfoldedSpectrumApprox}
\end{align}
where the approximation in Eq.~\eqref{eq:unfoldedSpectrumApprox} is obtained 
by considering up to $\mathcal{O}(\epsilon^2)$ corrections. Furthermore, the 
second order corrections due to levels other than $jk$ and $j'k'$ are ignored,
since the main effect of ignored levels is just shifting levels $jk$ and $j'k'$
back and forth, and will mostly cancel out. However, the
terms involving just the levels $jk$ and $j'k'$ push them away from each 
other and contribute to opening the gap, and is more pronounced when they are 
nearest neighbors due to the small energy denominator.

The approximation in Eq.~\eqref{eq:unfoldedSpectrumApprox} fails when the
energy levels become too close resulting in divergences, and for a Poissonian 
spectrum such close lying levels occur far more often than in the case of a CUE 
spectrum. This, however, can be regularized using degenerate perturbation 
theory giving~\cite{Tomsovicthesis, SriTomLakKetBae2016}
\begin{align}
  &  \left| s_{jk,j'k'}(\Lambda)\right| \approx  \nonumber \\
    & \quad \sqrt{\left[s_{jk,j'k'} + 2\sqrt{\Lambda}\left(x_{jk}-x_{j'k'} 
    \right)\right]^2 + 4 \Lambda w_{jk,j'k'}} 
    \label{eq:unfoldedSpectrumApproxReg} 
\end{align}
and the sign is given by 
$\text{sgn}[s_{jk,j'k'}+2\sqrt{\Lambda}\left(x_{jk}-x_{j'k'}\right)]$.

\subsection{Universal rescaled time \label{subsec:univtime}}

In a recent study of the average entanglement production of initially 
unperturbed eigenstates $\{\ket{jk}\}$ for  bipartite systems 
(same arrangement as described here -- weakly interacting chaotic subsystems)
~\cite{Jethin_PRE2020}, a universal rescaled time, $t$, was identified as
\begin{equation}
\label{eq:RescaledTime}
    t = n \, D\, \sqrt{\Lambda}, 
\end{equation}
where $n$ is the number of iterations of a unitary operator generating the 
dynamics.  Independent of system details, any two systems possessing the same 
value of $\Lambda$ have the same entropy production curve in terms of this time 
scale.  Thus, the mean level spacing times the square root of the transition 
parameter identifies the time scale of relaxation towards equilibration.  
Naturally, as the interaction strength gets weaker, this time scale gets 
longer, tending to infinity as $\epsilon \rightarrow 0$.  Furthermore, if 
normalized by the infinite time saturation value, in the 
perturbative regime, i.e.~$\Lambda \lesssim 10^{-2}$, all entropy production 
curves collapse onto the same curve as a function of time.  Even beyond the 
perturbative regime, this universal curve is only slightly altered as $\Lambda$ 
grows.

It turns out that this same rescaled time extends to the time evolution of a 
generic pure state as follows.  Consider an arbitrary initial state 
$\ket{\alpha(0)}$ whose density operator evolves after $n$ iterations as
\begin{align}
    \ket{\alpha(n;\epsilon)}\bra{\alpha(n;\epsilon)} & = \mathcal{U}_\epsilon^n
    \ket{\alpha(0)}\bra{\alpha(0)} \big(\mathcal{U}_\epsilon^\dagger\big)^n
    \nonumber \\
 & = \sum_{jk,j'k'} \exp(\ui\,n [\theta_{j'k'}(\epsilon)-\theta_{jk}(\epsilon)])
   \nonumber \\
    & \quad \times \bra{j'k'(\epsilon)}\ket{\alpha(0)} \bra{\alpha(0)}
    \ket{jk(\epsilon)} \nonumber \\
    & \quad \times \ket{j'k'(\epsilon)}\bra{jk(\epsilon)}.
\end{align}
Applying the rescalings introduced in the previous subsection, and relabeling 
the eigenstates by $\Lambda$ instead of $\epsilon$ gives
\begin{align}
    \ket{\alpha(t;\Lambda)}\bra{\alpha(t;\Lambda)} & = \sum_{jk,j'k'} \exp(\ui
    \frac{t}{\sqrt{\Lambda}}\,s_{j'k',jk}(\Lambda) ) \nonumber \\
    & \quad \times \bra{j'k'(\Lambda)}\ket{\alpha(0)} \bra{\alpha(0)}
    \ket{jk(\Lambda)} \nonumber \\
    & \quad \times \ket{j'k'(\Lambda)}\bra{jk(\Lambda)},
    \label{eq:alphaTimeEvolUniversal}
\end{align}
where it is understood that $\ket{\alpha(t;\Lambda)} \equiv
\ket{\alpha(n;\epsilon)}$ with appropriate variable changes.
For the rest of the paper, the rescaled parameters $\Lambda$ and $t$ are used 
instead of $\epsilon$ and $n$.  As shown ahead in Subsect.~\ref{subsec:ultraweak}, 
the universal rescaled time emerges naturally for ultra-weak perturbation
strengths for any kind of pure state, in the case where only the lowest order 
correction to the eigenphase
$\theta_{jk}(\epsilon)$ is relevant and no rotation to the eigenstate
$\ket{jk(\epsilon)}$ is considered. This generalizes the \emph{universal}
nature of the rescaled time beyond its relevance to the time evolution of
initial unperturbed product eigenstates $\{\ket{jk}\}$ presented in
\cite{Jethin_PRE2020}.

\section{Equilibration and Thermalization - Generalities}
\label{sec:generalities}

The central question of interest is to what extent does quantum coherence in the 
initial state play a role in the entanglement generated at long times, and thus,
the thermalization of the system with an eye on whether it happens in the 
weak or strong sense.  Various ensembles of initially unentangled states, 
based on the amount of coherence present are considered.  To begin though, 
an exact expression for the infinite time average of $S_2$ is calculated valid 
for a generic initial state and for a given interaction strength characterized 
by the transition parameter $\Lambda$.

Consider a generic initial pure state $\ket{\alpha(0)}$ whose 
density operator evolution (for Floquet systems) is given by 
Eq.~\eqref{eq:alphaTimeEvolUniversal}. The time-dependent reduced density 
matrix $\rho_A(t;\Lambda)$ of subsystem $A$ can be expressed as
\begin{align}
    \rho_A(t;\Lambda) & = \sum_{j,j',k} \ket{j}\bra{j'}
    \bra{jk}\ket{\alpha(t;\Lambda)}\bra{\alpha(t;\Lambda)}\ket{j'k} \nonumber\\
    & = \overline{\rho_A}(\Lambda) + \delta \rho_A(t;\Lambda),
\end{align}
where, $\overline{\rho_A}(\Lambda)$ is the infinite time average of
$\rho_A(t;\Lambda)$ given by
\begin{align}
    \overline{\rho_A}(\Lambda) = \sum_{jk} \rho_{A,jk} |\bra{jk(\Lambda)} 
    \ket{\alpha(0)}|^2,
\end{align}
in which $\rho_{A,jk}$ is the reduced density matrix of subsystem $A$
for the state $\ket{jk(\Lambda)}$, and let $\mu_{2,jk}(\Lambda)$ be the 
corresponding purity of the eigenstate. The matrix elements of 
$\delta \rho_A(t;\Lambda)$ are
\begin{align}
    (\delta \rho_A)_{jj'} & = 
    \sum_{\substack{j''k''\neq j'''k''' \\k}}
    \bra{jk}\ket{j''k''(\Lambda)} \bra{j'''k'''(\Lambda)}\ket{j'k} \nonumber \\
    & \qquad \quad \times \bra{j''k''(\Lambda)}\ket{\alpha(0)} 
    \bra{\alpha(0)}\ket{j'''k'''(\Lambda)} \nonumber \\
    & \qquad \qquad \times
    \exp(\ui\frac{t}{\sqrt{\Lambda}}s_{j''k'',j'''k'''}(\Lambda)).
\end{align}
The infinite time average of $\big(\delta \rho_A\big)_{jj'}$ vanishes 
assuming no degeneracy in the spectrum $\{\theta_{jk}(\Lambda)\}$. Furthermore,
assume that all possible level spacings, $s_{jk,j'k'}(\Lambda)$, 
are unique. These are 
reasonable assumptions to make  because the spectrum $\{\theta_{jk}(\Lambda)\}$ 
is a result of superposition of two uncorrelated spectra that are quantum 
chaotic in nature. This gives the infinite time average of the purity 
$\mu_2(t;\Lambda) = \tr_A[\rho_A^2(t;\Lambda)]$ as
\begin{align}
    \overline{\mu_2(t;\Lambda)} & = 
    \sum_{jk} \mu_{2,jk} |\bra{\alpha(0)}\ket{jk(\Lambda)}|^4
    \nonumber \\
    &\,\,\,+ \sum_{jk \neq j'k'}
    [\tr_A(\rho_{A,jk}\rho_{A,j'k'})+\tr_B(\rho_{B,jk}\rho_{B,j'k'})]
    \nonumber \\
    & \quad \qquad \times |\bra{\alpha(0)}\ket{jk(\Lambda)}|^2
    |\bra{\alpha(0)}\ket{j'k'(\Lambda)}|^2, \label{eq:ITA_purity}
\end{align}
where the infinite time average of $\tr_A[\delta\rho^2_A]$,
\begin{align}
    \overline{\tr_A[\delta \rho^2_A(t;\Lambda)]} & = 
    \sum_{jk \neq j'k'} \tr_B(\rho_{B,jk}\rho_{B,j'k'}) \nonumber \\
    &\,\,\, \times |\bra{\alpha(0)}\ket{jk(\Lambda)}|^2 
    |\bra{\alpha(0)}\ket{j'k'(\Lambda)}|^2,
\end{align}
is used to derive Eq.~(\ref{eq:ITA_purity}), and $\rho_{B,jk}$
is the equivalent of $\rho_{A,jk}$ but for subsystem $B$. Thus, the infinite
time average of $S_2(t;\Lambda)$ can be obtained using
Eq.~\eqref{eq:ITA_purity} in Eq.~\eqref{eq:Generic_S2}.

To gain greater insight into the range of possible behaviors, there are 
various strength of interaction regimes to consider. First, there are two 
limiting regimes, an \emph{ultra-weak perturbation strength regime} denoted as 
$\Lambda \rightarrow 0^+$, and a \emph{strong interaction regime} denoted as 
$\Lambda \gg 1$.  In the former, no rotation of the unperturbed 
eigenbasis describing the initial state needs to be taken into account, i.e.,
$\ket{jk(\Lambda)} \approx \ket{jk}$. In the study of the irreversibility in 
quantum theory by Peres \cite{Peres1984}, precisely
such a regime was analysed. There the quantity of interest was the squared 
overlap of two time evolved states via an unperturbed and its perturbed 
Hamiltonian, or the so-called fidelity. The decay law of ensemble averaged
fidelity for a chaotic Hamiltonian was found to follow a Gaussian behavior in 
time. In \cite{Cerruti02,Cerruti2003}, the fidelity of a
chaotic system was also studied where similar Gaussian behavior was derived,
and moreover associated the width of the Gaussian (also related to a 
transition parameter) to the phase space volume of
the system and the classical action diffusion coefficient. For the scenario
presented in this paper, there is a phase mixing between 
the unperturbed eigenstate components of the 
initial state whereas their respective intensities remain nearly the same.
This causes the time evolved state to depart from the product structure and 
generate entanglement, which saturates at a common rescaled time 
$t_{\text{sat}}$. Note however, since $\Lambda \rightarrow 0^+$, the actual 
(nonrescaled) saturation time $n_{\text{sat}} \rightarrow \infty$ by virtue 
of Eq.~\eqref{eq:RescaledTime}.  As a consequence, performing the infinite 
time average of $S_2$ first and then taking the limit $\Lambda \rightarrow 0$ 
gives very different results to the reverse order of the limits, which gives 
$\overline{S}_2 = 0$.  In the latter limiting regime, the full system 
eigenstate components follow a Haar measure behavior along with 
orthonormalization constraints.  This leads to the expected known results given 
ahead.

There are two further regimes, first a \emph{weak perturbation regime} 
($0^+ < \Lambda < 10^{-2}$), which can be characterized as the
regime in which an eigenstate $\ket{jk(\Lambda)}$ remains Schmidt decomposed in 
the unperturbed eigenbasis $\{\ket{jk}\}$~\cite{LakSriKetBaeTom2016, 
TomLakSriBae2018}.  Consequently, the time evolution of an 
unperturbed eigenstate, which is incoherent, will also remain Schmidt decomposed 
in the unperturbed eigenbasis~\cite{Jethin_PRE2020}. Furthermore, it was shown 
that for this regime, the majority of the contribution 
($\sim \mathcal{O}(\sqrt{\Lambda})$ on average) to 
$\ket{jk(\Lambda)}$ and its $S_2$ is due to the first 
two largest Schmidt eigenvalues.  The rest of the Schmidt eigenvalues 
contribute at a higher order ($\sim \mathcal{O}(\Lambda \ln \Lambda)$ 
on average) that can be neglected.

Finally, an \emph{intermediate perturbation regime} 
($10^{-2} \lesssim \Lambda \simeq 1$) 
occurs for interaction strengths in which an eigenstate $\ket{jk(\Lambda)}$ 
is not Schmidt decomposed in the unperturbed eigenbasis.  This regime controls 
the transition in behaviors between weak perturbation regime and the strong 
interaction limit, but is the most difficult to treat analytically as the 
eigenstates possess neither a Schmidt decomposed nor Haar measure form.

To summarize the various regimes are:
\begin{itemize}
    \item ultra-weak perturbation regime: $\Lambda \rightarrow 0^+$,
    \item strong interaction regime: $\Lambda \gg 1$,
    \item weak perturbation regime: $0^+ < \Lambda < 10^{-2}$,
    \item intermediate regime: $10^{-2} \lesssim \Lambda \simeq 1$.
\end{itemize}

\section{\label{sec:LimitingRegimes} Limiting regimes} 

In~\cite{Jethin_PRE2020}, a theory for the entropy production of direct
products of subsystem eigenstates was given, which has a vanishing 
coherence measure.  Here, much more general classes of initially 
unentangled states are considered with non-vanishing products of 
coherence measures, such as product states having the form of 
Eq.~\eqref{eq:K-alphaState}, and product states which are randomized within some
subspace of the subsystem eigenstates.

\subsection{\label{subsec:ultraweak} Ultra-weak perturbation limit}

\subsubsection{Entanglement production}

Let $\mathbb{B}'_A \subseteq \mathbb{B}_A$ be a subset of eigenstates of
subsystem $A$ containing $K_A$ elements, and similarly for $B$.  Now consider 
an arbitrary initial product state whose components are formed in these 
subspaces
\begin{equation}
\ket{\alpha(0)} = \Bigg( \sum_{\ket{j} \in \mathbb{B}'_A} z_{A,j} \ket{j} \Bigg) 
    \otimes \Bigg(\sum_{\ket{k} \in \mathbb{B}'_B} z_{B,k} \ket{k} \Bigg),
    \label{eq:alpha0Generic}
\end{equation}
where the $\{z_{A,j}\}$ are a particular set of complex numbers with no 
constraints other than satisfying the normalization condition 
$\sum_j |z_{A,j}|^2 = 1$, and likewise for $B$.  Using the approximation 
mentioned earlier that defines this regime, i.e.~a perturbed eigenstate 
remains close enough to the corresponding unperturbed eigenstate so that it is 
sufficient to consider $\ket{jk(\Lambda)} \approx \ket{jk}$, gives the reduced 
density matrix, $\rho_{A,jk} \approx \ket{j}\bra{j}$ and similarly for 
subsystem $B$.  In this case, the infinite time average of the purity in 
Eq.~\eqref{eq:ITA_purity} for an initial state of the form 
Eq.~\eqref{eq:alpha0Generic} becomes
\begin{align}
    \overline{\mu_2(t;\Lambda)} & \approx \sum_{jk} |z_{A,j}|^4 |z_{B,k}|^4
    \nonumber \\
    & +
    \sum_{j'k' \neq jk} \Big[ \delta_{jj'}(1-\delta_{kk'}) +
    \delta_{kk'}(1-\delta_{jj'})\Big] \nonumber \\
    & \quad \times |z_{A,j}|^2 |z_{A,j'}|^2 |z_{B,k}|^2
    |z_{B,k'}|^2 \nonumber \\
    & = 1 - c_A^{(2)} c_B^{(2)} \ .
    \label{eq:ITA_purity_zeroplus}
\end{align}
The initial coherence measure of subsystem $A$ (and similarly for 
subsystem $B$) in the $\mathbb{B}_A$ basis is given by
\begin{equation}
    c_A^{(2)} = \sum_{j \neq j'} |z_{A,j}|^2 |z_{A,j'}|^2 = 
    1 - \sum_j |z_{A,j}|^4
        \label{eq:ca2}
\end{equation}
and is used in the last step in Eq.~\eqref{eq:ITA_purity_zeroplus}.  Thus, the 
infinite time average of $S_2$ for an initial product state
$\ket{\alpha(0)}$ in Eq.~\eqref{eq:alpha0Generic} is given by
\begin{equation}
    \overline{S_2} \approx c^{(2)}_A c^{(2)}_B,
    \label{eq:S2inf_TP_0p}
\end{equation}
which is just the product of coherence measures of subsystem $A$ and $B$ in 
their preferred eigenbasis. Note that for initial pure states, either of whose
coherence measure of the subsystems vanishes, higher order corrections must be 
incorporated in order to find a non-vanishing infinite time average of $S_2$ 
leading to some function of the transition parameter $\Lambda$.   Thus, such 
systems saturate at values that depend on $\Lambda$, unlike systems for which 
the right hand side of Eq.~\eqref{eq:S2inf_TP_0p} vanishes.
\begin{figure}[h!]
\includegraphics[width=\columnwidth]
    {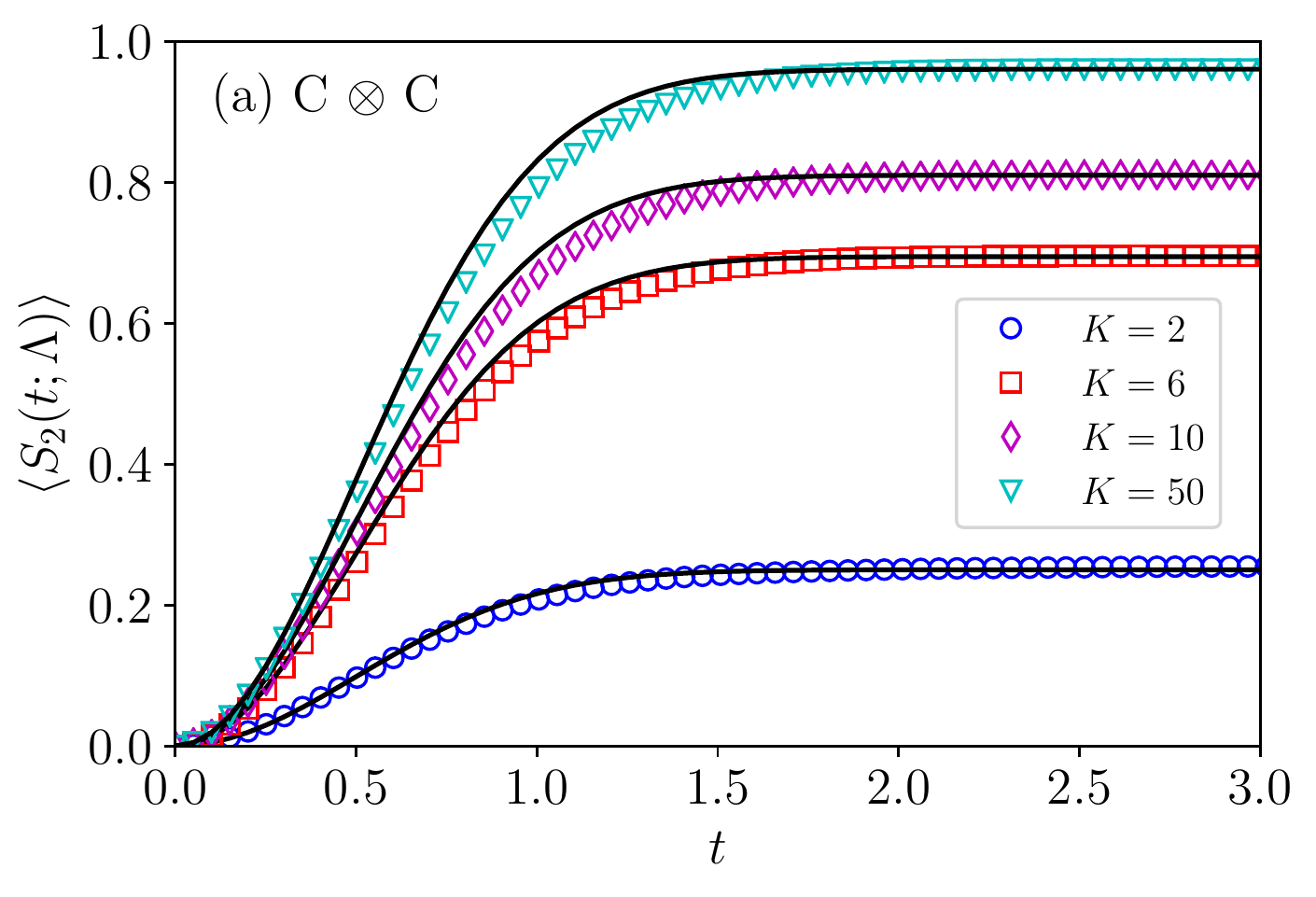}
\includegraphics[width=\columnwidth]
    {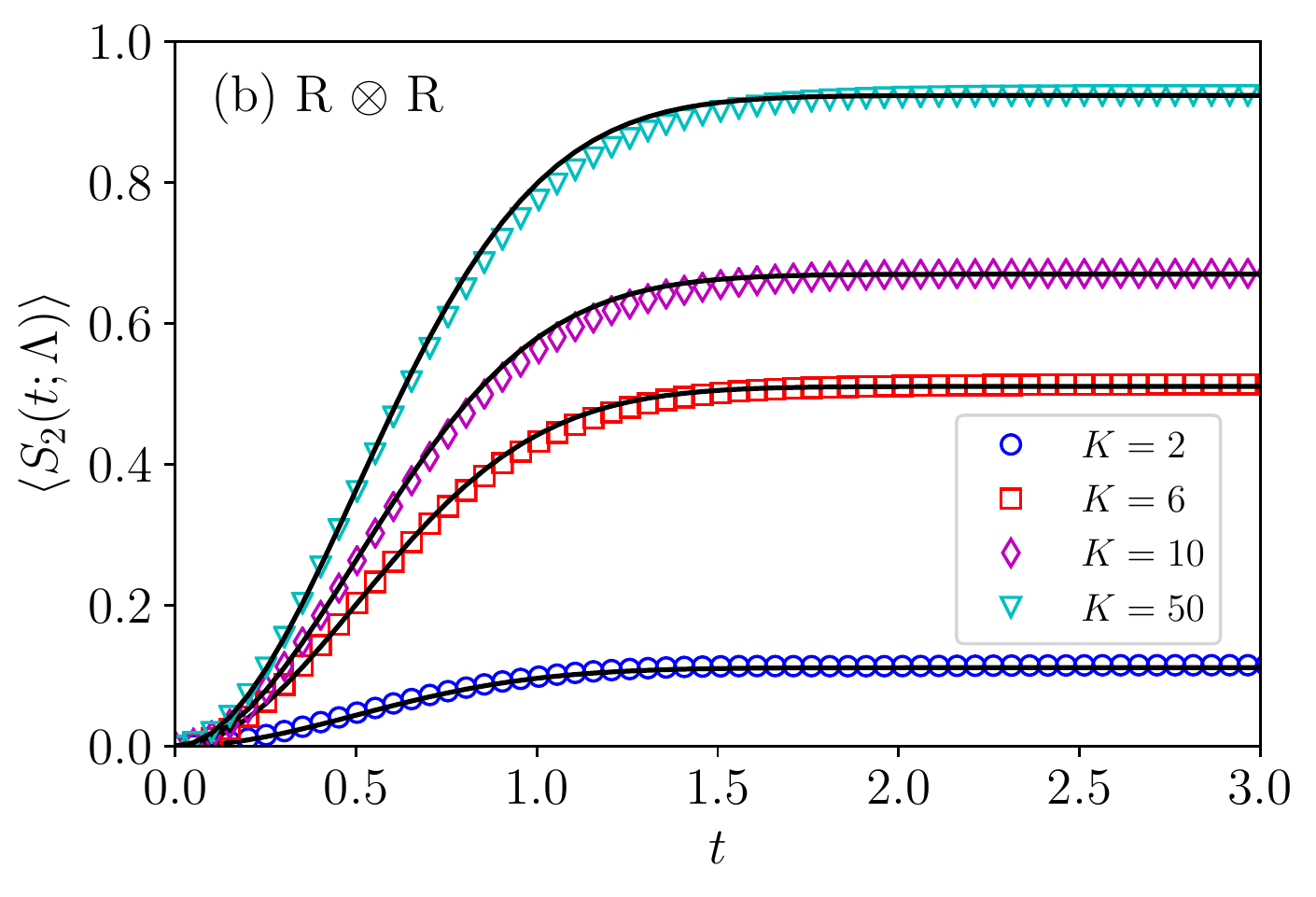}
\caption{\label{fig:avg_S2_te_TP0p} Ensemble averaged linear entropy 
    $\langle S_2(t,\Lambda)\rangle$ versus rescaled time for: (a) initial states 
   of  \CC type for the RMT transition ensemble defined in 
    Eq.~\eqref{eq:GenericFloquetRMT}, and (b) \RR type. Both use $N_A = N_B = 50$,
    $\Lambda =10^{-6}$, and various $K$ values. The black solid 
    lines show the corresponding theory of Eq.~\eqref{eq:S2_t_TP_0p_ens}, using 
    Eqs.~\eqref{eq:S2inf_TP_0p_CC} and \eqref{eq:S2inf_TP_0p_RR},
    respectively.}
\end{figure}

Ensembles of either the \CC type (coherent random phase) or the \RR type 
(random superpositions) can be created by defining the appropriate probability 
densities for the values of the $\{z_{A,j}\}$ and $\{z_{B,k}\}$ in 
Eq.~\eqref{eq:alpha0Generic}, respectively.  For either ensemble, an 
approximate expression for the ensemble averaged time evolution curve of 
$S_2$ can be derived beginning from Eq.~\eqref{eq:alphaTimeEvolUniversal}. 
This gives
\begin{align}
    S_2(t;\Lambda) & \approx c_A^{(2)} c_B^{(2)}-\sum_{\substack{j \neq j'\\
    k \neq k'}} |z_{A,j}|^2 |z_{A,j'}|^2 |z_{B,k}|^2 |z_{B,k'}|^2\nonumber\\
    & \quad \times \exp(\frac{\ui\,
    t}{\sqrt{\Lambda}} [s_{jk,j'k}(\Lambda) + s_{j'k',jk'}(\Lambda)]),
    \label{eq:S2_t_TP_0p}
\end{align}
where only the first order (in $\epsilon$) correction to the quasi-eigenenergies
is to be included.  The second-order correction to the quasi-eigenenergies is 
due to the rotation of eigenstates and are omitted in this regime.  After 
ensemble averaging, $S_2$ is given by
\begin{align}
    \langle S_2(t;\Lambda) \rangle &\approx \langle c_A^{(2)} \rangle \langle 
c_B^{(2)}\rangle  \big[1-\big\langle \exp(2\,\ui \, t\,x)\big\rangle \big],
    \label{eq:S2_t_TP_0p_preens} 
\end{align}
where $x = x_{jk} - x_{j'k} + x_{j'k'} - x_{jk'}$ is a sum of four (rescaled)
diagonal matrix elements, which behaves like 
a zero-centered Gaussian random variable of unit variance 
(shown using Eq.~\eqref{eq:x_jkDiagCov}) giving 
$\langle \exp(2 \, \ui \, t \, x) \rangle = \exp(-2 \, t^2)$. Thus, the 
\CC type or \RR type ensemble-averaged $S_2$ follows the very simple 
behavior given by
\begin{equation}
\label{eq:S2_t_TP_0p_ens}
    \langle S_2(t;\Lambda)\rangle \approx \langle c_A^{(2)}\rangle \langle
    c_B^{(2)}\rangle \big[1-\exp(-2\,t^2) \big]\ . 
\end{equation}
Surprisingly, the initial entanglement generation is quadratic in time as 
opposed to the generally expected linear
increase~\cite{Zurek_1994,MillerSarkar1999,Monteoliva_2000,Tanaka_2002,
Fujisaki_2003,Bandyopadhyay_2004}, and this is linked to the 
full system eigenstates retaining their product nature in this regime.  
Note that if either $K_A=1$ or $K_B=1$, the coherence measure vanishes and 
it is necessary to calculate the $\Lambda$ dependent functional form 
following~\cite{Jethin_PRE2020}.  

Consider \CC type initial states of the form of Eq.~\eqref{eq:alpha0Generic} 
where all $|z_{A,j}| = 1/\sqrt{K_A}$ with random, independently chosen phases, 
and likewise for subsystem $B$, i.e.
\begin{equation}
    \ket{\alpha(0)} = \ket{\alpha_{K_A}}_\text{C} \otimes
    \ket{\alpha_{K_B}}_\text{C}
\end{equation}
where $\ket{\alpha_K}_\text{C}$ is of the form Eq.~\eqref{eq:K-alphaState}.  
The long time limiting evolution for a \CC initial state is 
statistically equivalent to an entangled random phase state with equal
intensities. For $K = K_A = K_B$, the singular values and various entropies
of entangled random phase states are studied in 
\cite{LakPuchalaKZ_2014}.  For these initial states,
the ensemble average of the coherence measure is
same as the individual coherence measures (no fluctuations in coherence measures 
within the ensemble), and is given by
\begin{equation}
    \langle \overline{S}_2 \rangle = \langle c^{(2)}_A \rangle \langle c^{(2)}_B
    \rangle = \left(1-\frac{1}{K_A}\right) \left(1-\frac{1}{K_B}\right).
    \label{eq:S2inf_TP_0p_CC}
\end{equation}

The ensemble averaged time evolution of $S_2(t;\Lambda)$ for \CC type initial
states is shown in Fig.~\ref{fig:avg_S2_te_TP0p} (a) for $K = 2,\,,6,\,10,\,50$ 
compared with the combined results of Eq.~\eqref{eq:S2_t_TP_0p_ens} and 
Eq.~\eqref{eq:S2inf_TP_0p_CC}. The agreement is quite good considering 
that there should be finite $N_A,\,N_B$ and $\Lambda > 0^+$ corrections. All the
necessary details about the numerical calculations shown in
Fig.~\ref{fig:avg_S2_te_TP0p} and the other figures are provided in
App.~\ref{App:numerics}.
For $K_A, K_B$ comparable to the respective subsystem dimensionality, i.e. 
initial states that are tensor product of (nearly) maximally coherent 
states of the subsystems, the system time evolution generates entanglement 
$\langle \overline{S}_2 \rangle \approx 1 - 1/N_A -1/N_B$.  This is close to
the well-known result derived in~\cite{Lubkin1978}, shown in
Eq.~\eqref{eq:Lubkin_S2} ahead. Given that the eigenstates 
are not thermal and have the product structure, this 
is a remarkable result in the sense that for an arbitrarily small interaction 
between the subsystems, a near-maximal entanglement is achieved after a 
long time by the virtue of maximal coherence in the initial product state.

Next consider \RR type initial product states of the form of 
Eq.~\eqref{eq:alpha0Generic}, where the $\{ z_{A,j} \}$ and $\{ z_{B,k} \}$ 
are random Haar measure complex 
coefficients where the only constraint is the unit normalization, i.e.
\begin{equation}
    \ket{\alpha(0)} = \ket{\alpha_{K_A}}_\text{R} \otimes
    \ket{\alpha_{K_B}}_\text{R} \ .
\end{equation}
The ket $\ket{\alpha_{K_A}}_\text{R}$ is a random pure state in a given subspace 
of subsystem $A$ Hilbert space spanned by $\mathbb{B}'_A$ (R type) and similarly 
for $B$. Performing initial state ensemble averaging of the coherence measure 
of subsystems, the equilibrium $S_2$ value can be computed as
\begin{equation}
    \langle \overline{S}_2 \rangle = \langle c_A^{(2)}\rangle 
    \langle  c_B^{(2)} \rangle = \Big(\frac{K_A - 1}{K_A + 1}\Big) \Big(
    \frac{K_B - 1 }{K_B + 1}\Big), \label{eq:S2inf_TP_0p_RR}
\end{equation}
where the Haar measure average $\langle |z_{A,j}|^4 \rangle = 2/K_A(K_A+1)$ is
used to calculate the above equilibrium value. Figure~\ref{fig:avg_S2_te_TP0p}
(b) shows the ensemble averaged $S_2(t;\Lambda)$ for \RR type initial states, 
and an excellent agreement between the theory and numerics is found.  For 
large $K_A,\, K_B$ close to 
subsystem dimensionality, the equilibrium value is given by 
$\langle \overline{S}_2 \rangle \approx 1 - 2/N_A -2/N_B$
which is slightly less than that of the \CC type initial states. This 
can be attributed to the fluctuations in the intensities in the initial states,
and is discussed ahead.
\begin{figure}
\includegraphics[width=\columnwidth]
    {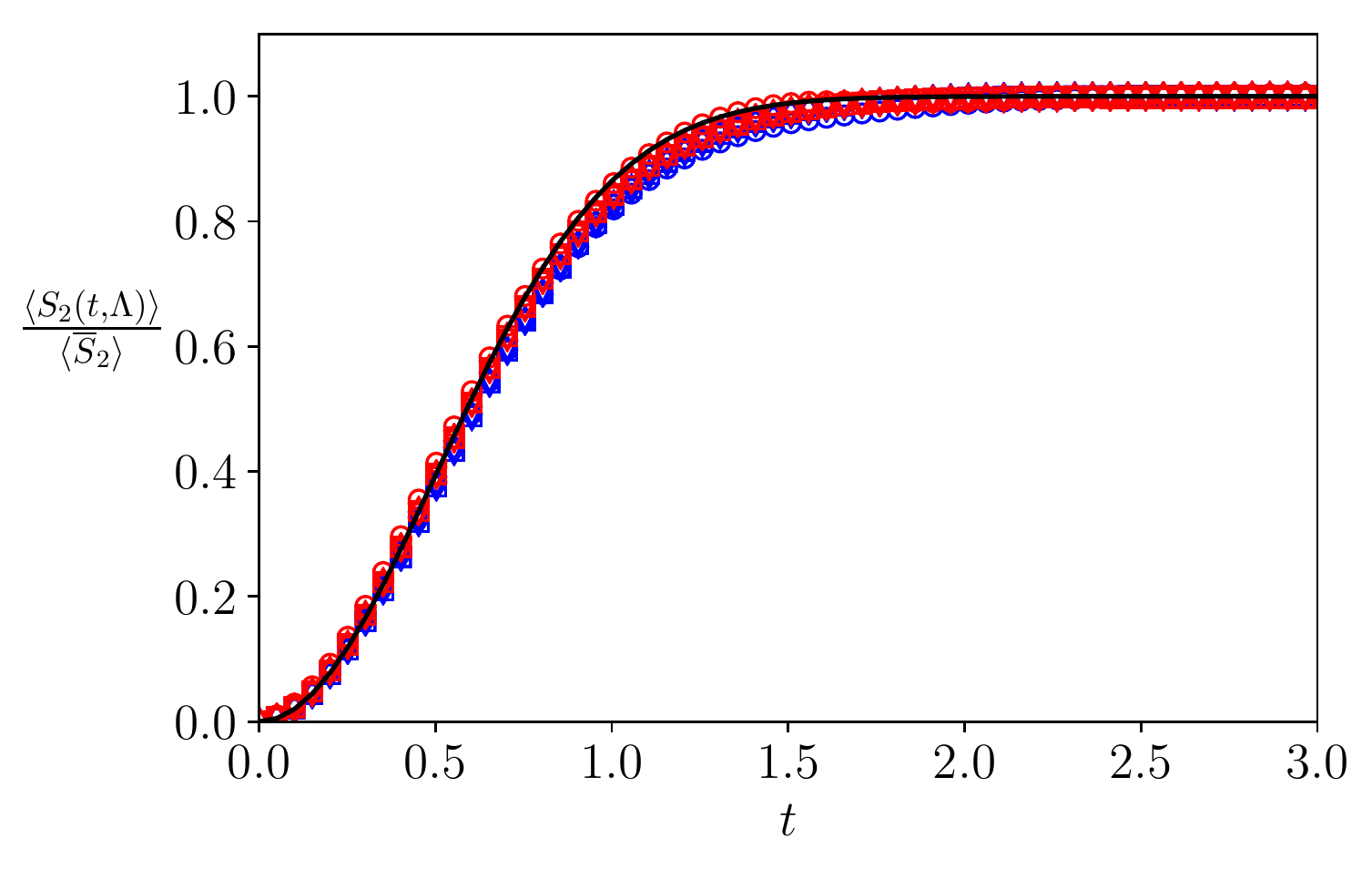}
    \caption{\label{fig:avg_S2_te_univ} Ensemble averaged linear entropy  
    $\langle S_2(t,\Lambda)\rangle$ divided by $\langle \overline{S}_2 \rangle$.
    Curves are shown for the RMT transition ensemble of 
    Eq.~\eqref{eq:GenericFloquetRMT} with $N_A = N_B = 50$, 
    $\Lambda = 10^{-6}$, and for various $K$ values.  Both \CC type 
    (blue markers) and \RR type (red markers) initial states are indicated.
    The black solid line shows corresponding theory curve based on 
    Eq.~\eqref{eq:S2_t_TP_0p_ens}.} 
\end{figure}

Remarkably, dividing Eq.~\eqref{eq:S2_t_TP_0p_ens} by the equilibrium value
$\langle \overline{S}_2 \rangle$ makes all (ensemble-averaged) entanglement 
production curves for both \CC and \RR types and various $K_A$ and $K_B$ fall 
on to one universal curve as
illustrated in Fig.~\ref{fig:avg_S2_te_univ}.  It must be emphasized that for
obtaining this universal behavior it is crucial to use the rescaled time given
by Eq.~\eqref{eq:RescaledTime}.

\subsubsection{Equilibrium and relaxation}

The nature of the equilibrium onto which the system eventually settles can be
investigated by examining the variations of $S_2$ in time and across the 
ensemble.  As discussed in Sect.~\ref{subsec:thermalization}, two 
different useful fluctuation measures are given by $\sigma^2(\overline{S}_2)$ 
(equilibrium measure) and $\overline{\sigma^2(S_2)}$ (relaxation measure).  
An equilibrium can be inferred 
from the former measure if a majority of the initial states evolve to an 
equilibrium value $\langle \overline{S}_2 \rangle$ to within small or 
negligible residual fluctuations.
In the $\Lambda \rightarrow 0^+$ regime, the equilibrium measure
$\sigma^2(\overline{S}_2)$ defined in Eq.~\eqref{eq:sigma2_def}, can be 
calculated via Eqs.~\eqref{eq:S2inf_TP_0p} 
and \eqref{eq:ca2} in a straightforward way for various initial state 
ensembles.  The relaxation measure $\overline{\sigma^2(S_2)}$ defined in 
Eq.~\eqref{eq:sigma2InfAvg} can be calculated using Eq.~\eqref{eq:S2_t_TP_0p} 
in $S^2_2(t;\Lambda)$ followed by finding its infinite-time average. This gives
\begin{equation}
    \overline{\langle S^2_2(t;\Lambda) \rangle} =
    \langle (c_A^{(2)})^2 \rangle \langle (c_B^{(2)})^2 \rangle + 
    2 \langle c_A^{(4)} \rangle \langle c_B^{(4)} \rangle,
    \end{equation}
where a higher order coherence measure $c_A^{(4)}$ is introduced based on the 
$l_4$-norm~\cite{Streltsov_RMP2017} and is defined as
\begin{equation}
    c_A^{(4)} = \sum_{j \neq j'} |z_{A,j}|^4 |z_{A,j'}|^4,
\end{equation}
and similarly for subsystem $B$. The relaxation measure can then be written as
\begin{equation}
    \overline{\sigma^2(S_2)} = \sigma^2(\overline{S}_2) + 2 \frac{\langle
    c_A^{(4)} \rangle \langle c_B^{(4)} \rangle}
    {\langle c_A^{(2)}\rangle^2  \langle c_B^{(2)} \rangle^2}.
\label{eq:sigma2InfAvg_TP_0p}
\end{equation}
It must be emphasized that fluctuation measures derived here for the $\Lambda
\rightarrow 0^+$ regime are not valid for initial states either of whose
subsystem coherence measures are vanishing. Such cases need special treatment 
due to the rotation of eigenstates, regardless of how infinitesimal $\Lambda$ is, 
and are discussed in the next section. It is worth mentioning that for initial 
unperturbed eigenstates, the $\overline{S}_2$ probability density behaves 
similarly to heavy-tailed type densities spanning $\overline{S}_2$ values from 
$\mathcal{O}(\Lambda)$ to $\mathcal{O}(1)$~\cite{Jethin_PRE2020}.
\begin{figure}[h!]
\centering
\includegraphics[width=\columnwidth]
    {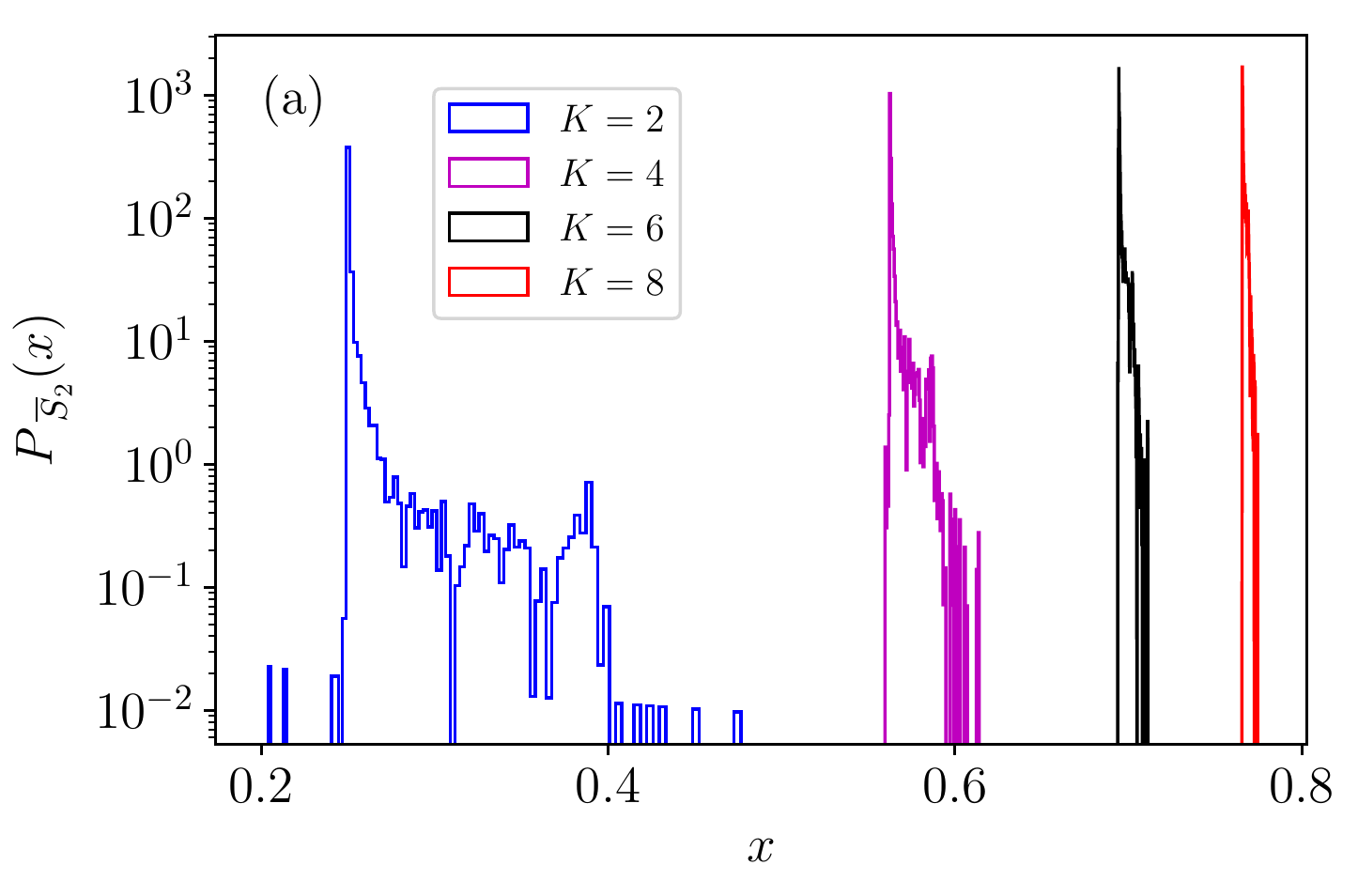}
\includegraphics[width=\columnwidth]
   {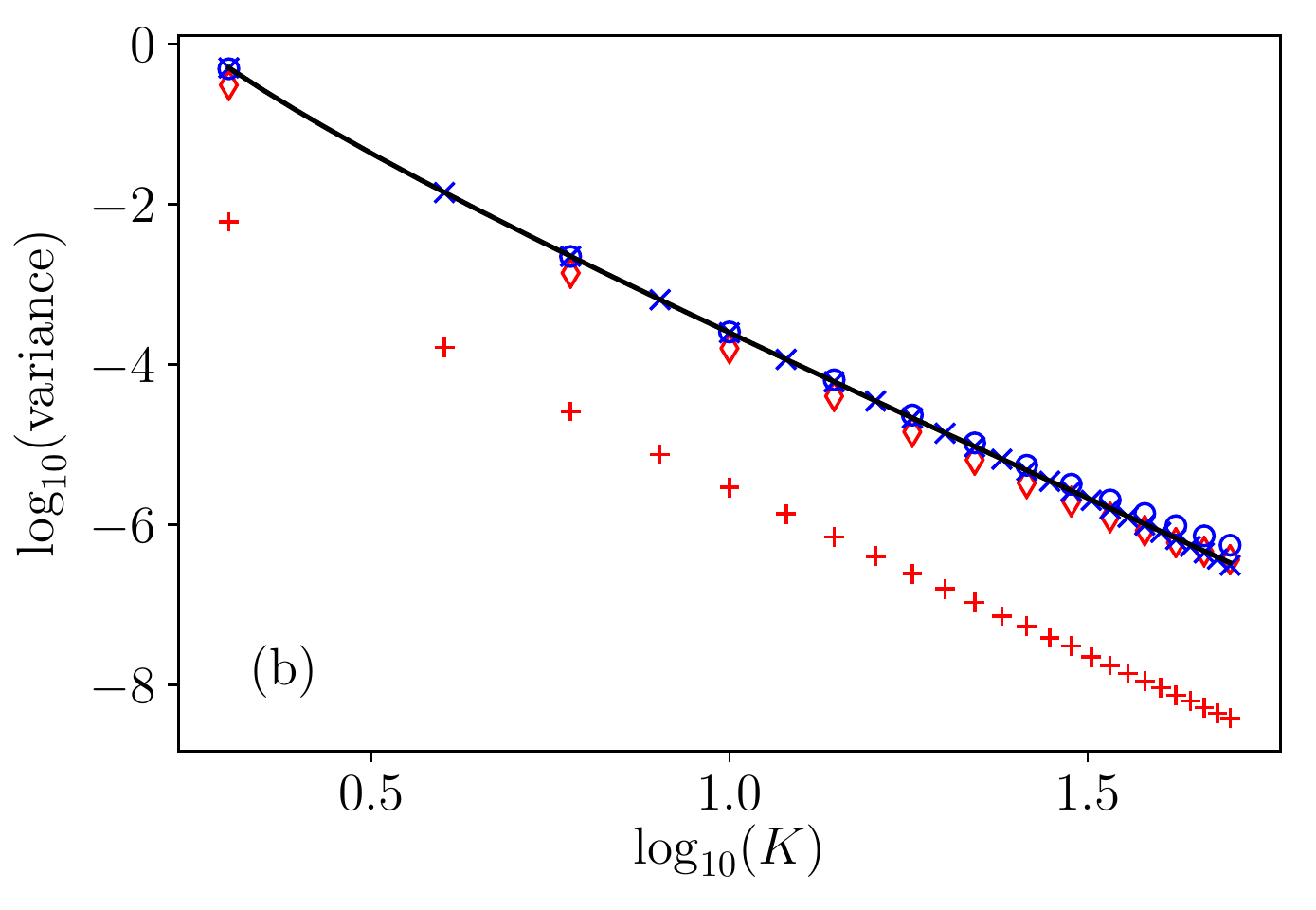}
    \caption{\label{fig:S2inf_dist_TP0p} (a) Probability density of
    $\overline{S}_2$ values, and (b) the two fluctuation measures, 
    $\overline{\sigma^2\left(S_2 \right)}$ and 
    $\sigma^2\left( \overline{S}_2 \right)$ for \CC type initial 
    states.  Various $K$ values are shown for $N_A = N_B = 50$. In (a)
    $\Lambda = 10^{-6}$.  In (b) the black solid line is from
    Eq.~\eqref{eq:sigma2InfAvg_TP_0p_CC} for $\overline{\sigma^2\left(S_2
    \right)}$. Compare this prediction to calculated
    $\overline{\sigma^2\left(S_2 \right)}$ values for $\Lambda=10^{-8}$
    ($\color{blue}{\times}$) and $\Lambda=10^{-6}$ ($\color{blue}{\Circle}$). In
    addition, calculated values for $\sigma^2(\overline{S}_2)$ (which should lie
    below $\overline{\sigma^2\left(S_2 \right)}$ ) are plotted for
    $\Lambda=10^{-8}$ ($\color{red}{+}$) and $\Lambda=10^{-6}$
    ($\color{red}{\diamondsuit}$) to illustrate its $\Lambda$-dependence.
   }
\end{figure}

The $\overline{S}_2$ probability density for \CC type initial states is 
shown in Fig.~\ref{fig:S2inf_dist_TP0p} (a) for various $K$ 
values. It is evident from 
Fig.~\ref{fig:S2inf_dist_TP0p} that for $K = 2$ 
the density is broad and reminiscent of the heavy-tailed nature of the 
$\overline{S}_2$ probability density of \EE type initial state ensemble 
discussed in the section ahead. As $K$ increases, more unperturbed eigenstates 
participate resulting in an increasingly sharper $\overline{S}_2$ density.

Now consider the two fluctuation measures, 
$\overline{\sigma^2\left(S_2 \right)}$ and $\sigma^2(\overline{S}_2)$ for \CC
type initial states. The relaxation measure can be calculated via 
Eq.~\eqref{eq:sigma2InfAvg_TP_0p} giving
\begin{equation}
    \overline{\sigma^2(S_2)} = \frac{2}{K_A (K_A-1) K_B (K_B-1)} \sim
    \frac{2}{K_A^2 K_B^2},
    \label{eq:sigma2InfAvg_TP_0p_CC}
\end{equation}
where the last expression above is valid for large $K_A$ and $K_B$.  Excellent
agreement is found between the theory and numerical calculations as illustrated
in Fig.~\ref{fig:S2inf_dist_TP0p} (b).  This shows that \CC type 
initial states evolve to an equilibrium state in the strong sense as $K$
increases. Also in Fig.~\ref{fig:S2inf_dist_TP0p} (b) is the comparison of 
$\sigma^2(\overline{S}_2)$ for $\Lambda = 10^{-8}$ and $10^{-6}$, which 
illustrates its $\Lambda$-dependent nature and that they lie below
$\overline{\sigma^2(S_2)}$. It turns out that the leading term of 
$\sigma^2(\overline{S}_2)$ vanishes for this ensemble, and the effect of 
eigenstate rotation cannot be neglected for small $K$, leading to this 
$\Lambda$-dependent behavior.
\begin{figure}[h!]
\centering
\includegraphics[width=\columnwidth]
    {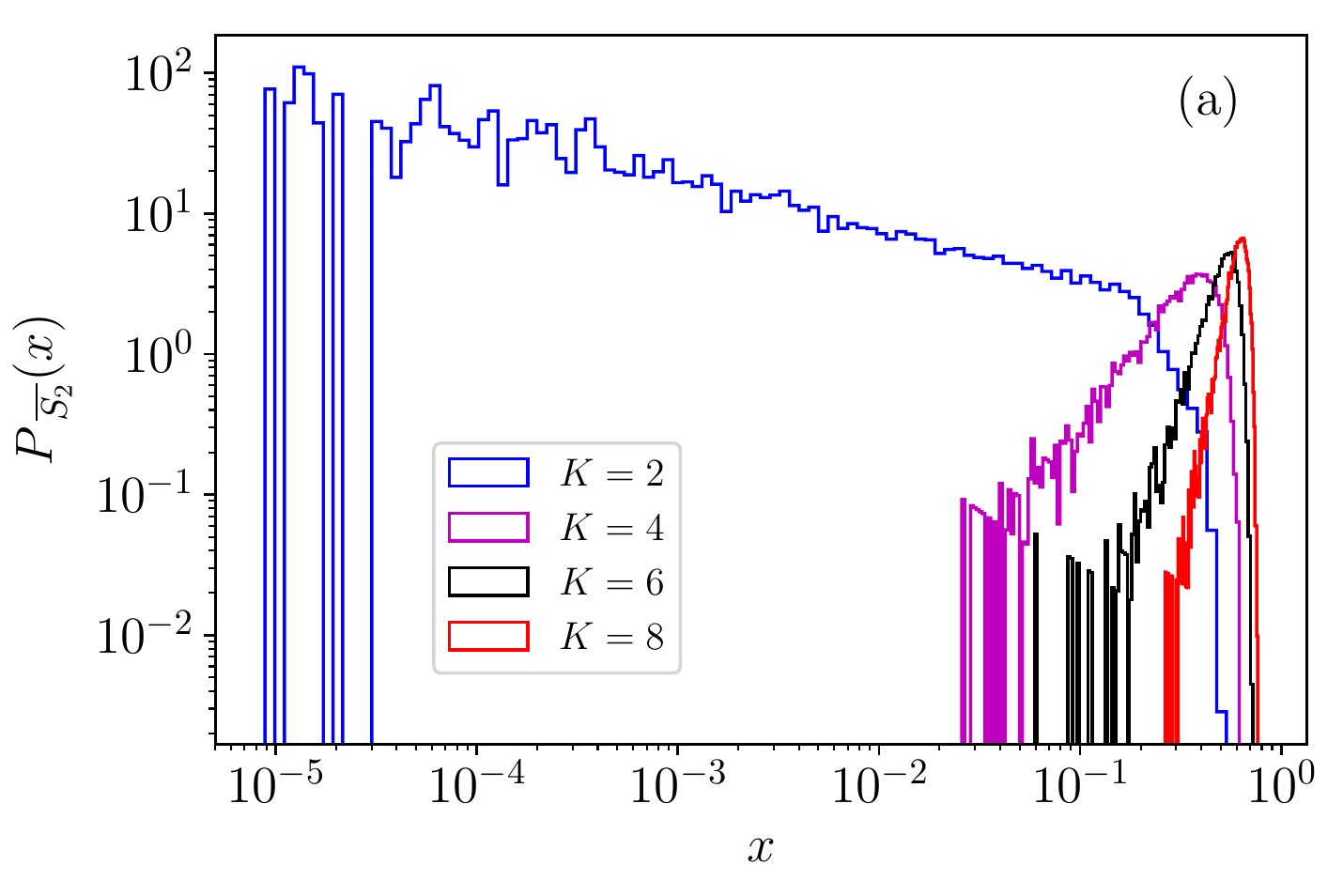}
\includegraphics[width=\columnwidth]
    {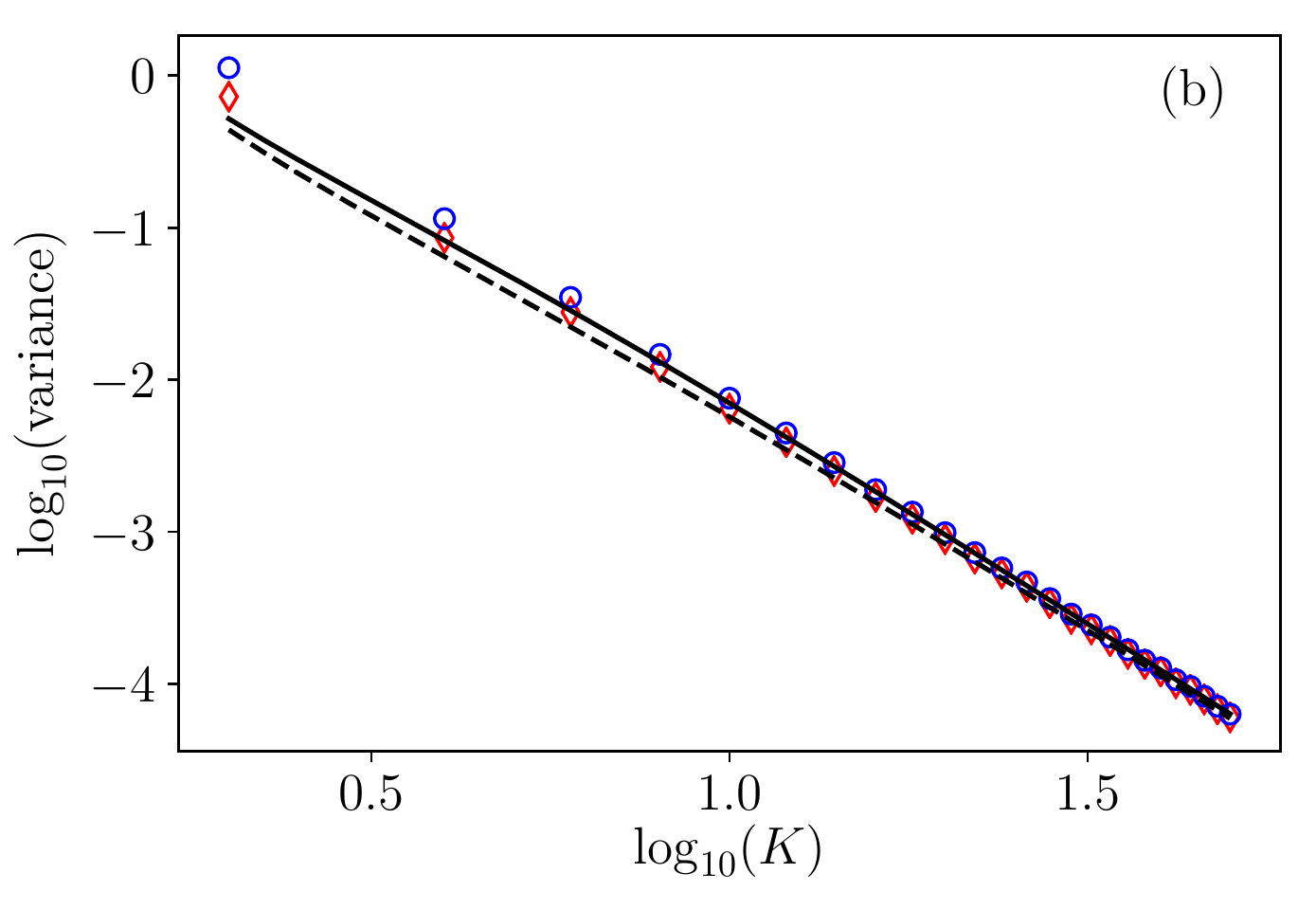}
    \caption{\label{fig:S2_TP_0p_fluct_CC_RR}  (a) Probability density of
    $\overline{S}_2$ values, and (b) the two fluctuation measures, 
    $\sigma^2\left( \overline{S}_2 \right)$ and 
    $\overline{\sigma^2\left(S_2 \right)}$ for \RR type initial 
    states.  Various $K$ values are shown for $N_A = N_B = 50$. In (a) and (b)
    $\Lambda = 10^{-6}$ for the calculations.  In (b) the solid black line 
    is the theoretical prediction for $\overline{\sigma^2\left(S_2 \right)}$ 
    and the black dashed line is for $\sigma^2\left( \overline{S}_2 \right)$.
    The $\color{blue}{\Circle}$ are the calculated values for 
    $\overline{\sigma^2\left(S_2 \right)}$ and the
    $\color{red}{\diamondsuit}$ are the values for 
    $\sigma^2\left( \overline{S}_2 \right)$.}
\end{figure}

For \RR type initial states, the probability density of $\overline{S}_2$ is 
shown for various $K$-values in Fig.~\ref{fig:S2_TP_0p_fluct_CC_RR} (a).  
Compared to the \CC type densities, the shapes are quite different and the 
width of the \RR type densities are wider for any given $K$. For $K=2$, the 
heavy-tail behavior (straight line in a log-log plot) due to the eigenstate 
rotation is quite prominent.  This is not very surprising since the 
probability density for random states has inverse square root singularities 
in the region in which one of the two states is the dominant contribution 
and the other is very small. This leads the $K=2$ case to being much closer to 
the unperturbed eigenstate case, which does have the heavy tail (discussed in
the next section). As the $K$-value increases, the probability densities 
become more narrow and the fluctuations about the mean are mainly due to the 
Haar measure probability density of the components $\{z_{A,j}\}$ and 
$\{z_{B,k}\}$. Using Eq.~\eqref{eq:S2inf_TP_0p} and various Haar measure 
moment averages (see App.~\ref{App:EigenvectorStatistics}), the fluctuation 
measure can be computed as
\begin{align}
    \sigma^2(\overline{S}_2) & = \frac{4\,[K_A^3 + K_B^3 + 4(K_A^2 +K_B^2) +
    K_A + K_B-8]}{(K_A^3+4 K_A^2 +K_A -6)(K_B^3 + 4 K_B^2 +K_B-6)} \nonumber \\
    & \sim \frac{4}{K_A^3}+ \frac{4}{K_B^3},
    \label{eq:sigma2_TP_0p_RR}
\end{align}
where the last line is good for large enough $K_A, \, K_B$. 
Similarly, it can be shown that 
\begin{align}
    \langle c_A^{(4)} \rangle &=  \frac{4 (K_A -
    1)}{(K_A+1)(K_A+2)(K_A+3)} \sim \frac{4}{K_A^2}, \label{eq:4-coherence_RR}
\end{align}
and along with Eq.~\eqref{eq:sigma2InfAvg_TP_0p} an expression for 
$\overline{\sigma^2(S_2)}$ can be found in a straightforward way. 
In contrast to the \CC type case, where the two fluctuation measures are 
significantly different, for \RR type initial states they are identical to 
leading order in large-$K_A,\, K_B$ and
\begin{equation}
    \overline{\sigma^2(S_2)} \approx \sigma^2(\overline{S}_2)\ .
    \label{eq:sigma2InfAvg_TP_0p_RR}
\end{equation}
This is shown in Fig.~\ref{fig:S2_TP_0p_fluct_CC_RR} (b) where 
a good agreement is observed between the theory and numerical values. Thus, 
both measures are dominated by their variations about the infinite time average,
and the temporal fluctuations are lower order in $K$.

For this regime, the equilibrium measure for various initial state ensembles 
with non-vanishing coherence implies equilibrium, which becomes sharper as the 
coherence increases. This is consistent with a transition from weak to strong 
equilibration as the initial state coherence increases. For initial product 
states with maximal coherence, the entanglement saturates to that of 
thermalized states, and they exhibit a strong relaxation (strong 
thermalization), although the non-scaled relaxation time lengthens to infinity 
as $\Lambda \rightarrow 0$.

\subsection{Strong interaction regime \label{subsec:strong}}

\begin{figure}[h!]
\centering
\includegraphics[width=\columnwidth]
    {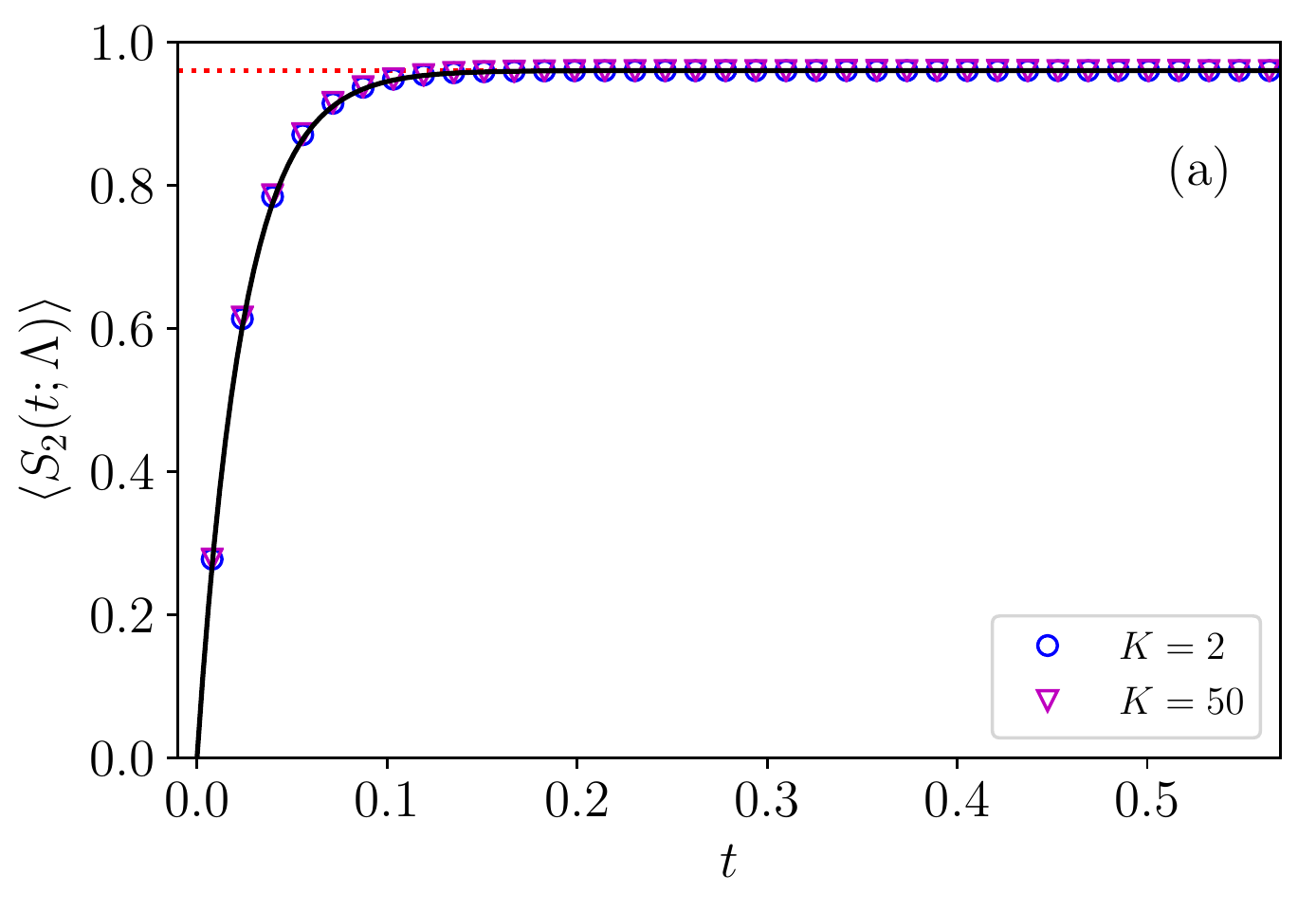}
\includegraphics[width=\columnwidth]
    {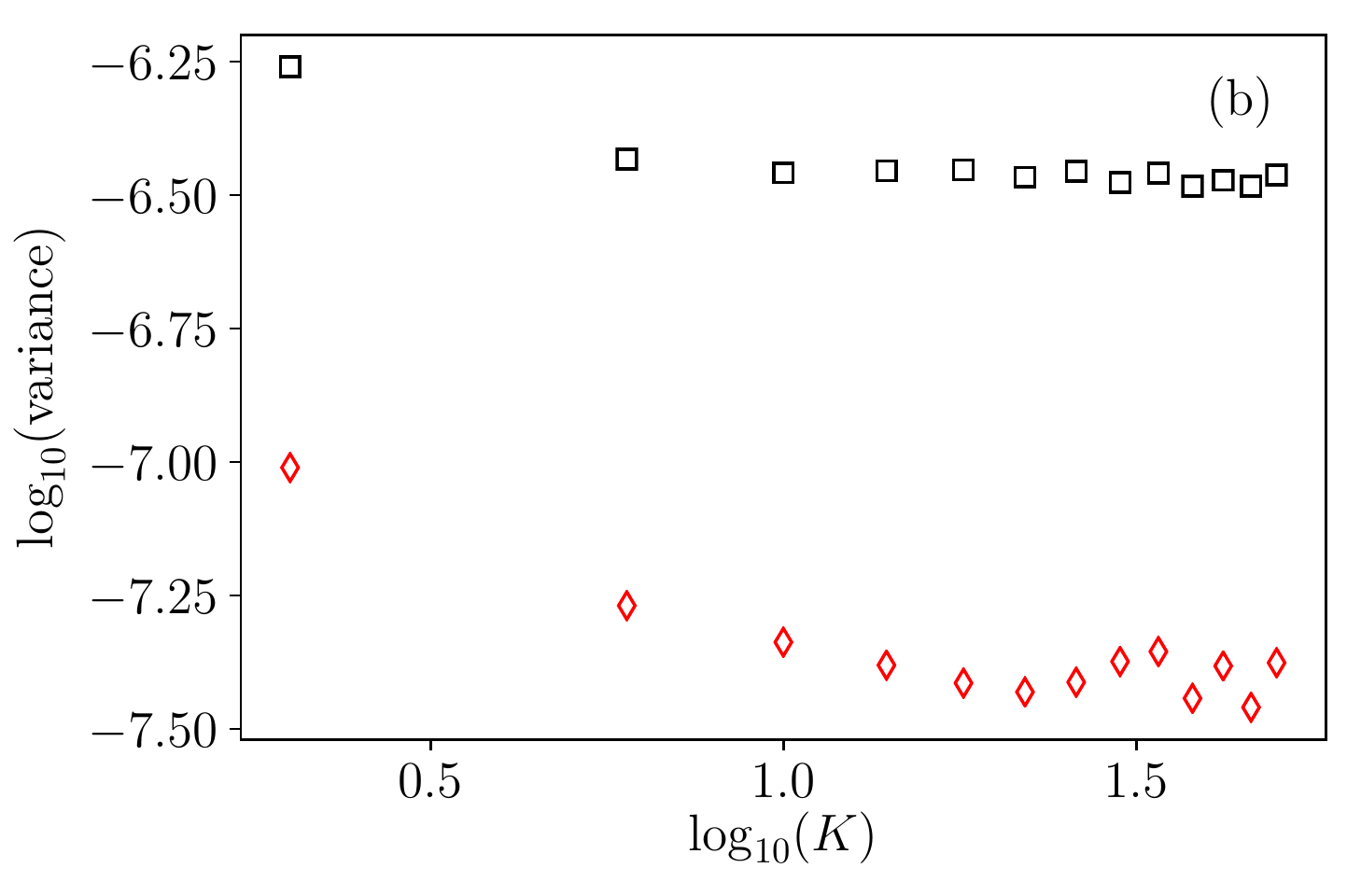}
    \caption{\label{fig:avg_S2_te_RR_strong} Ensemble averaged linear entropy
    $\langle S_2(t;\Lambda)\rangle$ in (a) and the two variance measures,
    $\sigma^2(\overline{S}_2)$ and $\overline{\sigma^2(S_2)}$ in (b) for
    \RR type initial states and $\Lambda = 10$, $N_A = N_B = 50$. 
    In (a), $K = 2 , 50$ are shown along with theory curve 
    (of product eigenstates) in Eq.~\eqref{eq:avg_S2_te_intermediate}. The red
    dotted line shows the saturation value $\langle \overline{S}_2^\infty
    \rangle$. In (b), $\sigma^2(\overline{S}_2)$
    ($\color{red}{\diamondsuit}$) and $\overline{\sigma^2(S_2)}$ 
    ($\color{black}{\square}$) are shown as a function of $K$.}
\end{figure}

In this interaction regime, the eigenstates $\{ \ket{jk(\Lambda)} \}$
essentially behave just like that of $N_A N_B$ - dimensional CUE matrices, and
the time evolution of the system shows vanishingly small initial state 
dependence. Using eigenvector statistics of unitary 
ensembles for the full system space, the limiting behavior of 
$\langle \overline{S}_2 \rangle$ can be derived.  The complex coefficients 
$\bra{\alpha(0)}\ket{jk(\Lambda)}$ in this limit behave the same as the 
eigenvector components of an $N_A N_B$-dimensional CUE.  
For $\ket{jk(\Lambda)} = \sum_{j'k'} z_{jk} \ket{j'k'}$, the average purity of 
the eigenstate is given by~\cite{Lubkin1978}
\begin{align}
    \langle \mu_{2,jk} \rangle = \frac{N_A + N_B}{N_A N_B + 1} \approx 
    \frac{1}{N_A} + \frac{1}{N_B}. \label{eq:Lubkin_S2}
\end{align}
The average cross-term trace of the reduced density matrices in
Eq.~\eqref{eq:ITA_purity} can be calculated as
\begin{align}
    \tr_A(\rho_{A,jk} \rho_{A,j'k'}) & = \big\langle \sum_{jk,j'k'} z_{jk} 
    z_{j'k}^* z'_{j'k'} z_{jk'}'^{*} \big\rangle \nonumber \\
    & = \big \langle \sum_{j,k,k'} |z_{jk}|^2 |z'_{jk'}|^2 \big\rangle 
    \nonumber \\
    & \approx \sum_{j,k,k'} \Big(\frac{1}{N_A N_B}\Big)^2 = \frac{1}{N_A}
\end{align}
for large $N_A,\,N_B$, and similarly for the subsystem $B$ trace term.
Putting all these together, it can be shown that
\begin{equation}
    \langle \overline{S}_2^\infty \rangle = \lim_{\Lambda \rightarrow \infty} 
    \langle \overline{S_2(t;\Lambda)}\rangle \sim 1 - \frac{1}{N_A} -
    \frac{1}{N_B}, \label{eq:HaarAverageS2}
\end{equation}
as expected. Furthermore, the expression for average linear entropy derived in
\cite{Jethin_PRE2020} in the non-perturbative regime for an initial state 
ensemble of product eigenstates (shown ahead in
Eq.~\eqref{eq:avg_S2_te_intermediate}) can be used to describe the situation
here with an approximation to the function $C(2;t)$ (defined in
Eq.~\eqref{eq:S2_unpert_eig_evol} and given by Eq.~\eqref{eq:C2_func}) that 
appears in the expression.  Since the saturation 
happens quickly in this regime, the small $t$ approximation of $C(2;t)$ 
shown in Eq.~\eqref{eq:S2_jk_te_short} will suffice giving
\begin{equation}
    \langle S_2(t;\Lambda) \rangle \approx \langle \overline{S}_2^\infty \rangle
 \Bigg[1 - 
 \exp(-\frac{4\,\pi \,t\, \sqrt{\Lambda}}{\langle \overline{S}_2^\infty\rangle})
    \Bigg]. \label{eq:S2_strong}
\end{equation}
This is in good agreement with numerical simulations whose initial states are of 
\RR type with $K = 2,\,50$ as illustrated in Fig.~\ref{fig:avg_S2_te_RR_strong}
(a). No initial state dependence is seen and it 
saturates as expected to $\langle \overline{S}_2^\infty \rangle$. 
As $\Lambda \rightarrow \infty$, where
the transition becomes complete, the time evolution curve approaches a 
Heavyside step function scaled by $\langle \overline{S}_2^\infty \rangle$,
saturating almost instantly. The expression in Eq.~\eqref{eq:S2_strong}, 
originally derived with a regularized perturbation theory, extends to the 
non-perturbative regime using an `embedding technique' developed in 
\cite{LakSriKetBaeTom2016}. 

Quite remarkably, the expression agrees with a
recent study based on the entangling power of sequentially applied random
diagonal nonlocal operators interlaced with random local unitary 
operations~\cite{Bhargavi2019}. Note that entangling power $e_p(U)$ is 
defined as the average entanglement (here linear entropy) produced 
by the action of a nonlocal gate $U$ on product states 
sampled from the Haar measure on the subspaces. 
For $U$ sampled from the diagonal ensemble $U_{AB}$ defined in this paper around 
Eq.~\eqref{eq:GenericFloquetRMT} with $\epsilon \ll 1$, the entangling 
power is quite small  $e_p(U) \ll 1$. For large enough subsystem dimensions
(and hence satisfying $\Lambda \gg 1$), Eq. (55) in \cite{Bhargavi2019} 
representing $e_p(U)$ averaged over local unitaries is same as
Eq.~\eqref{eq:S2_strong} here with the identification 
$4 \pi t \sqrt{\Lambda} = 2 \pi^2 \epsilon^2 n/3$ where $n$ is the actual 
time. Note that the average of $e_p(U)$ over a
Haar distribution of $U$ denoted as $\overline{e_p}$ in~\cite{Bhargavi2019}
is approximately the same as $\langle \overline{S}_2^\infty \rangle$. It may be noted
that the entangling power approach involves the local unitaries $U_A$ and $U_B$
to be different at each time step and this leads to a decorrelation that gives
rise to the initial linear growth, also evident in Eq.~\eqref{eq:S2_strong}
here. Though $U_A$ and $U_B$ are the same at each time step here in this paper,
in the strong interaction regime the memory from previous time steps is essentially
washed out justifying this connection to the entangling power approach.

The equilibrium and relaxation measures are negligibly small compared to the 
mean value $\langle \overline{S}_2 \rangle$, i.e. $\sigma^2(\overline{S}_2),
\,\overline{\sigma^2(S_2)} \ll 1$, as evident from
Fig.~\ref{fig:avg_S2_te_RR_strong} (b), and they 
show a lack of initial state dependence. Thus the initial state coherence does 
not play any significant role in the entanglement production in this regime and
saturates to the thermal value rapidly.
Furthermore, as revealed by the fluctuation measures, the system possesses a 
very sharp equilibrium and thermalizes in the strong sense where almost all the
initial states regardless of their coherence relaxes rapidly to the thermal
value.

\section{Weak and intermediate perturbation regimes \label{sec:wipr}}
\subsection{Weak perturbation regime}
\subsubsection{\EE type initial states} 
\label{subsec:1by1}

Consider the initial state ensemble of unperturbed eigenstates $\{ \ket{jk}\}$.
The coherence measures of both the subsystems are vanishing in the respective 
preferred subsystem eigenbasis. As mentioned earlier, the entanglement produced
for this type of initial states is entirely due to the rotation of eigenstates
instigated by the interaction. For sufficiently weak $\Lambda$, the leading 
order effects arise from the rotation of the initial state $\ket{jk}$ with its 
energetically nearest neighbor.  Since the unperturbed spectrum 
is a direct product of two independent spectra, the level statistics are 
Poissonian~\cite{Tkocz12}. There is an absence of level repulsion and two-level 
near degeneracies are quite frequent. Depending on the interaction strength 
between the initial state $\ket{jk}$ and its nearest neighbor, the rotation of 
the pair to perturbed eigenstates can range from little to complete 
($\pi/4$ rotation angle).

For an ensemble of eigenstates $\{\ket{jk(\Lambda)}\}$, it was shown 
in~\cite{TomLakSriBae2018} that the linear entropy ranges over the full 
possible interval $0\le S_2 \le 1/2$.  Its probability density displays a 
heavy-tailed behavior towards large values.  In addition,  the near degeneracies 
increased the order of the average eigenstate entanglement from 
$\mathcal{O}(\Lambda)$ to $\mathcal{O}(\sqrt{\Lambda})$. 
In~\cite{Jethin_PRE2020}, it was shown that the mean entanglement production 
rate increased as
\begin{equation}
    \langle S_2(t;\Lambda) \rangle \approx C(2;t) \,\sqrt{\Lambda} \,\,,
    \label{eq:S2_unpert_eig_evol}
\end{equation}
where $C(2;t)$ is a function of rescaled time and is given in 
Eq.~\eqref{eq:C2_func}. The short-time behavior of $\langle S_2(t;\Lambda)
\rangle$ is linear-in-time,
\begin{equation}
    \langle S_2(t;\Lambda) \rangle \approx 4\, \pi\, t\, \sqrt{\Lambda}\,\,,
    \label{eq:S2_jk_te_short}
\end{equation}
and after long time it saturates to
\begin{equation}
    \langle \overline{S}_2 \rangle \approx \frac{5}{8}\,\pi^{3/2}\,
    \sqrt{\Lambda} \,\,.\label{eq:S2_jk_inf}
\end{equation}
In this $\Lambda$ regime, the time evolution of such initial states will 
remain Schmidt decomposed in the unperturbed 
eigenstate basis $\mathbb{B}_{AB}$ to $\mathcal{O}(N_A^{-1})$ 
(up to some phase) and is given by
\begin{align}
    \ket{jk(t;\Lambda)} \approx \sum_{l=1}^{N_A}
    \sqrt{\lambda_{l,jk}(t;\Lambda)}\,\ket{(jk)_l}, \label{eq:TimeEvoljk}
\end{align}
where $\{\ket{(jk)_l}\}$ for $l > 1$ are unperturbed eigenstates that are 
energetically close to the initial state $\ket{(jk)_1} = \ket{jk}$. The 
expression for Schmidt eigenvalues $\lambda_{l,jk}(t;\Lambda)$ derived in 
\cite{Jethin_PRE2020}, is given by
\begin{align}
    \lambda_{l,jk}(t;\Lambda) & \approx \frac{4 \Lambda w_{jk,(jk)_l}}
    {s_{jk,(jk)_l}^2+4\Lambda w_{jk,(jk)_l}} \nonumber \\
    & \quad \times \sin^2\Big(\frac{t}
    {2\sqrt{\Lambda}}\sqrt{s^2_{jk,(jk)_l}+4\Lambda w_{jk,(jk)_l}}\,\Big)
    \label{eq:SchmidtEig_EE}
\end{align}
for $l > 1$ and $\lambda_{1,jk} = 1 - \sum_{l>1}\lambda_{l,jk}$. This expression
is derived using a degenerate perturbation theory where the
divergences due to near two-level degeneracies are regularized in a self-consistent
manner. The expression for Schmidt eigenvalues in Eq.~\eqref{eq:SchmidtEig_EE} 
shows non-self-averaging and oscillatory behavior (quite similar to Rabi 
oscillations). The linear entropy can be computed using
\begin{align}
    S_2(t;\Lambda) \approx 2 \sum_{l>1} \Big( \lambda_{l,jk}(t;\Lambda) -
    \lambda_{l,jk}^2(t;\Lambda)\Big) \label{eq:S2_jk_te} \,.
\end{align}
The cross terms
$\lambda_{l,jk}(t;\Lambda) \lambda_{l',jk}(t;\Lambda)$ 
are neglected in the above since on average they contribute to higher order,  
$\mathcal{O}(\Lambda)$~\cite{Jethin_PRE2020}.
\begin{figure}[h!]
\centering
\includegraphics[width=\columnwidth]
    {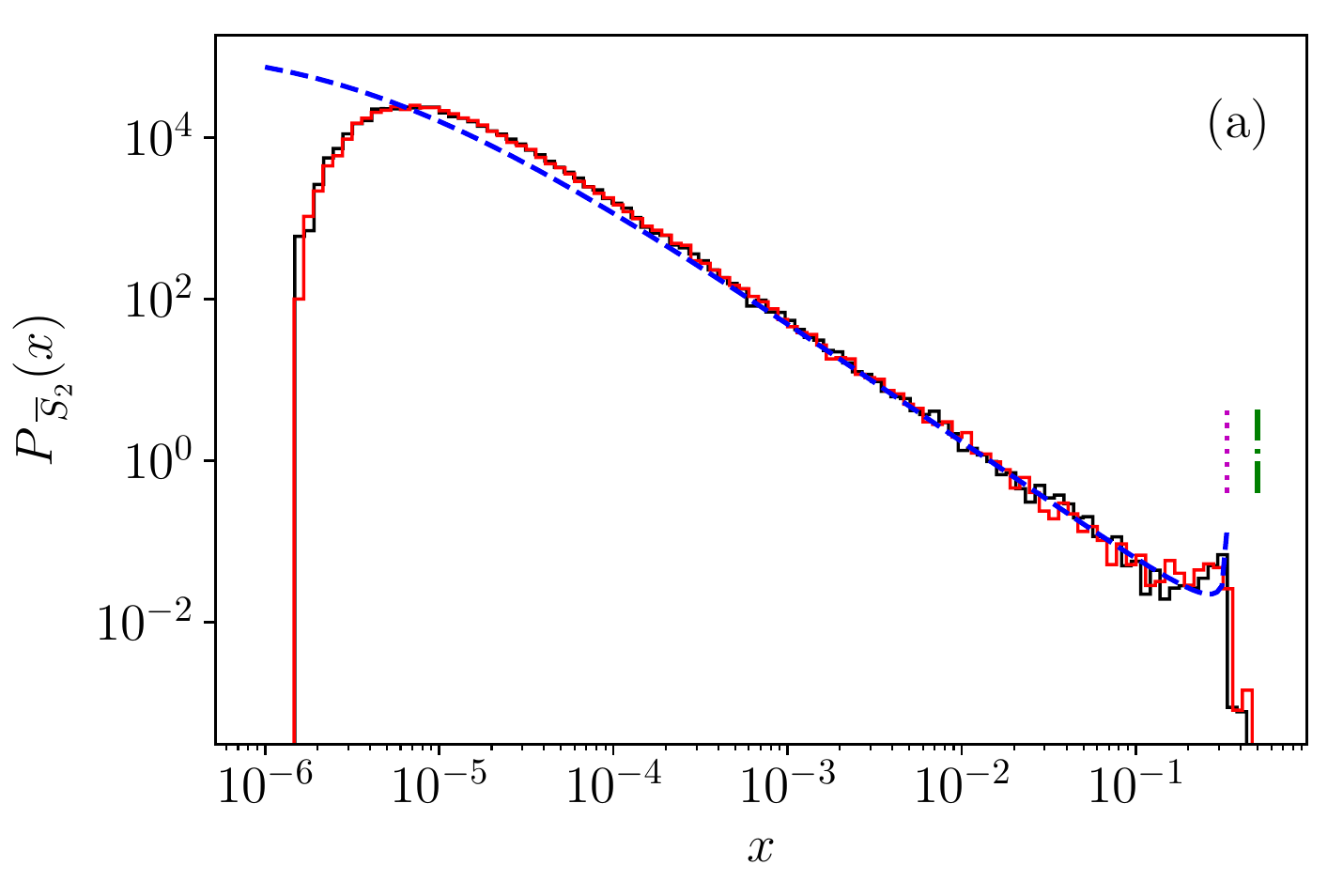}
\includegraphics[width=\columnwidth]
    {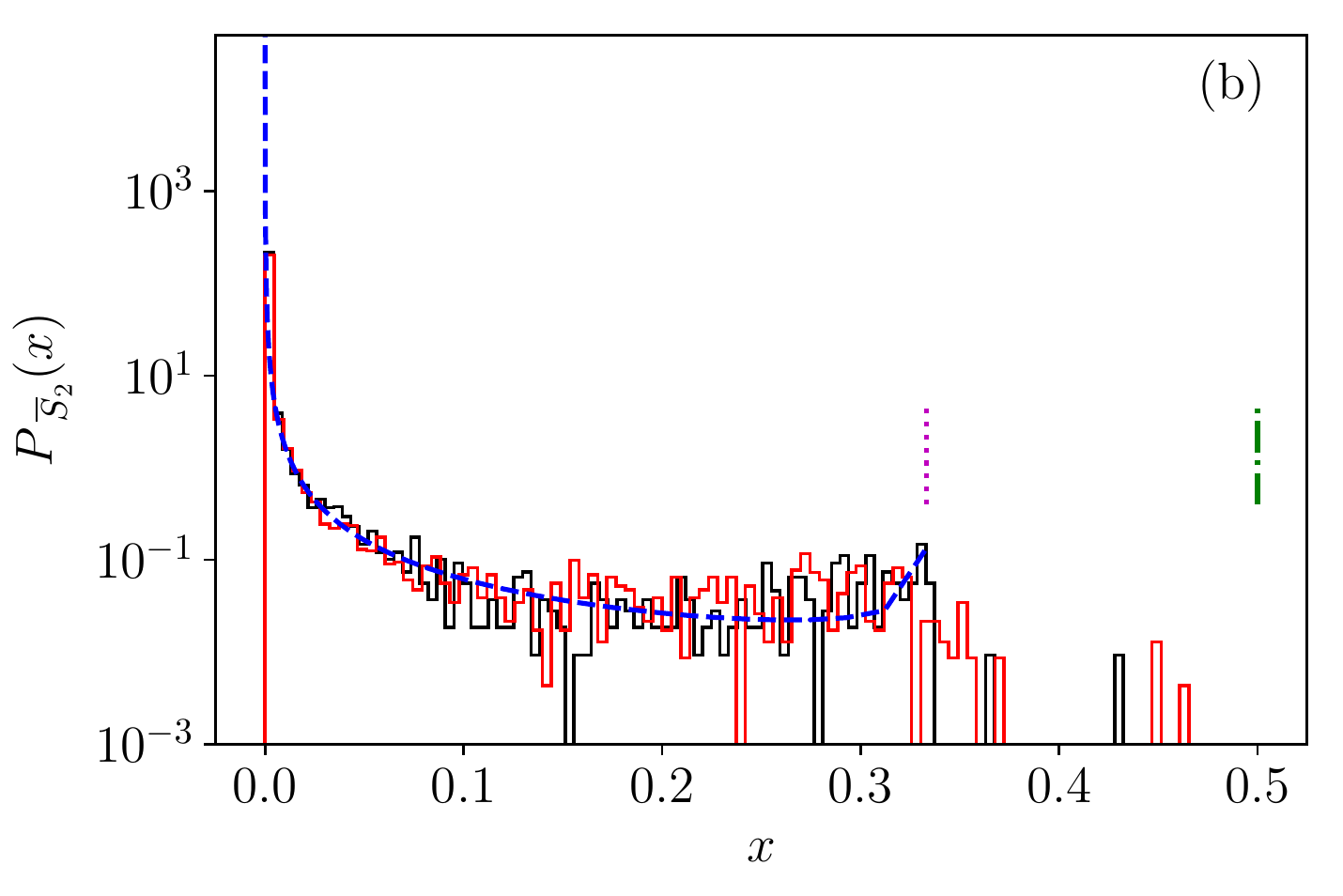}
    \caption{\label{fig:S2inf_EE_dist} Two views of the $\overline{S}_2$ 
    probability density of the \EE type initial state ensemble. In (a)
    log-log plot illustrating the heavy tail whereas in (b) log-linear which 
    is better for illustrating the endpoint behaviors at $0$ and $1/3$. The 
    blue dashed line shows the theory prediction, 
    Eq.~\eqref{eq:Prob_S2inf_dist_EE}. $\Lambda = 10^{-6}$ and 
    $N_A = N_B = 50$. The red solid line is the histogram obtained via numerical 
    averaging of individual time evolved curve of initial states after 
    saturation ($t > 2$), and the black solid line via 
    Eq.~\eqref{eq:ITA_purity}. Both sets of data reveal a cutoff around 
    $\mathcal{O}(\Lambda)$. The vertical magenta dotted line and green 
    dot-dashed line shows $x = 1/3$ and $1/2$, respectively.}
\end{figure}

Following the approach in \cite{TomLakSriBae2018}, an approximate expression 
for the probability density of $\overline{S}_2$ can 
be derived in this regime where only the closest unperturbed 
eigenstate to the initial state $\ket{jk}$ is relevant. In this limit, the 
infinite time average of Eq.~\eqref{eq:S2_jk_te} can be calculated as
\begin{equation}
    \overline{S}_2 \approx 4 u (1 - 3 u), \label{eq:S2inf_EE_approx}
\end{equation}
where the variable $u$ is given by
\begin{equation}
    u = \Big(4 + \frac{s^2}{\Lambda w} \Big)^{-1}\,.
\end{equation}
The closest neighbor spacing probability density is 
$P_{\text{CN}}(s) = 2 \exp(-2 s)$~\cite{LakSriKetBaeTom2016} for a 
Poissonian sequence. The rescaled off-diagonal $\mathcal{V}$-matrix element 
$w$ has a probability density of an exponential as discussed in 
Sect.~\ref{subsec:SymmetryTP}. The probability density of $u$, $P_u(x)$, can 
be calculated from the relation
\begin{equation}
    P_u(x) = \int_0^\infty \dd{s} \int_0^\infty \dd{w} 2 \ue^{-2 s} \ue^{-w}
    \delta\Big(x-\Big[4 + \frac{s^2}{\Lambda w} \Big]^{-1}\Big).
\end{equation}
As defined in \cite{TomLakSriBae2018}, let
\begin{equation}
    f_\Lambda(x) = \frac{x}{\Lambda(1-4x)},
\end{equation}
then upon carrying out the integral over $w$
\begin{equation}
    P_u(x) = \int_0^\infty \dd{s}\, \exp(-2 s -s^2 f_\Lambda(x)) \Big[
        \frac{2 s^2}{\Lambda (1-4x)^2}\Big],
\end{equation}
which is similar to Eq.~(84) in \cite{TomLakSriBae2018} with appropriate
variable transformations.
From Eq.~\eqref{eq:S2inf_EE_approx}, $u$ can be written in terms of
$x = \overline{S}_2$ as
\begin{equation}
    u = \frac{1}{6}\Big(1-\sqrt{1-3 x} \Big)
\end{equation}
and the solution inconsistent with the non-degenerate perturbation theory is
discarded. This gives the $\overline{S}_2$ probability density as
\begin{equation}
    P_{\overline{S}_2}(x) = P_u\Big(\frac{1}{6}\big(1-\sqrt{1-3 x} \big)\Big) 
    \Big| \dv{u}{x}\Big|. \label{eq:Prob_S2inf_dist_EE}
\end{equation}
Figure~\ref{fig:S2inf_EE_dist} illustrates $P_{\,\overline{S}_2}(x)$ for 
$\Lambda = 10^{-6}$, where a heavy-tail like distribution can be seen
with $\overline{S}_2$ covering a range of $\mathcal{O}(\Lambda)$ to
$\mathcal{O}(1)$. Notice that around $\overline{S}_2 = 1/3$ the local maximum 
predicted by Eq.~\eqref{eq:Prob_S2inf_dist_EE} appears. However, some slight 
deviations indicate that the assumption of just two unperturbed eigenstates 
participating shows some small corrections due to triple degeneracies that 
can occur with low probability.

For this ensemble of initial states, a broad distribution of $\overline{S}_2$
suggests that equilibrium is not really achieved due to the
heavy-tail-like behavior.  In addition, due to a dearth of unperturbed 
eigenstates participating in the time evolved state, a relaxation is not 
possible. As a result, a large ensemble of initial states are necessary for 
convergence to the average entanglement production curve.

\subsubsection{E $\otimes$ C and E $\otimes$ R type initial states 
\label{subsec:1byK}}

Consider an initial product state whose one of the subsystem coherence measures 
is zero in the preferred basis and the other is non-zero
\begin{equation}
 \ket{\alpha(0)} = \ket{j} \otimes \Big( \sum_{\ket{k} \in \mathbb{B}'_B }
    z_{B,k} \ket{k} \Big), \label{eq:InitialState_1K}
\end{equation}
where without loss of generality $c_A^{(2)}$ is taken to be vanishing. 
For $K_B \ll N_B$ an expression for the linear entropy can be derived as 
follows. The time evolution of the above state using the Schmidt decomposition 
in Eq.~\eqref{eq:TimeEvoljk} is given by
\begin{align}
    \ket{\alpha(t;\Lambda)} = \sum_{k} \sum_l \sqrt{\lambda_{l,jk}(t;\Lambda)}\,
    \ket{(jk)_l},
\end{align}
where any phase factor that may appear is absorbed into $\ket{(jk)_l}$. The
reduced density matrix constructed out of this time evolved state reads as
\begin{align}
    \rho_A(t;\Lambda) &= \sum_{k,k'} \sum_{l,l'} z_{B,k} z_{B,k'}^*
    \sqrt{\lambda_{l,jk}(t;\Lambda) \lambda_{l',jk'}(t;\Lambda)}
    \nonumber \\
    & \quad \times \tr_B(\ket{(jk)_l}\bra{(jk')_{l'}}). \label{eq:RDM_1byK_B}
\end{align}
Note that the set of states $\{\ket{(jk)_l}\}$ must be energetically close to 
the level corresponding to the index pair $(j,k)$ as 
mentioned earlier, which amounts to having small energy differences between them. 
Let $(j_l, k_l)$ be the index pair of the ket $\ket{(jk)_l}$, then for chaotic 
bipartite Floquet systems considered here the index pair satisfies
\begin{equation}
    j_l +k_l \approx j + k \label{eq:index_jk_l}
\end{equation}
for $N_A \approx N_B$ by the virtue of having approximately the same uniform 
mean level spacing for both the subsystems. This translates to the index pair
having a structure $(j_l,k_l) = (j \pm l,k \mp l)$ for $l = 1,2,3, \ldots$
where $l$ runs up to $N_A/2$ typically and energy differences 
$\theta_{j_l k_l} - \theta_{jk}$ are approximately $2\pi/N_A N_B$ on average 
\cite{TomLakSriBae2018}. Furthermore, for very small perturbations it is 
enough to keep first and second largest Schmidt eigenvalues in 
Eq.~\eqref{eq:TimeEvoljk} and neglect the rest \cite{Jethin_PRE2020}. This
corresponds to only including the closest neighbor, $\ket{(jk)_2}$, of
$\ket{jk}$, and the rest are insignificant. However, this argument breaks down
when triple degeneracies occur in the unperturbed spectrum.  Statistically
speaking, this is of very low probability and contributes to higher order than 
$\mathcal{O}(\sqrt{\Lambda})$.  Based on these arguments, the partial trace that 
appears in Eq.~\eqref{eq:RDM_1byK_B} can be computed as follows. 
For $k=k'$ and $l=l'$ the partial trace term is $\ket{j_l}\bra{j_l}$, and that 
for $k=k'$ and $l \neq l'$ it vanishes since $k_l = k_{l'}$ is forbidden
according to Eq.~\eqref{eq:index_jk_l}.  In the $k\neq k'$ case, the energy
difference between levels $(j,k)$ and $(j,k')$ is roughly $2\pi/N_B$, and so
the probability that the set $\{\ket{(jk)_l}\}$ and $\{ \ket{(jk')_l}\}$ share a 
common unperturbed eigenstate or more whose product $\lambda_{l,jk}
\lambda_{l',jk'}$ is also significant is quite small for $K_B \ll
N_B$. Hence, these contributions can be neglected.  As $K_B$ increases,
eventually this argument breaks down.
\begin{figure}[h!]
    \centering
  \includegraphics[width=\columnwidth]
    {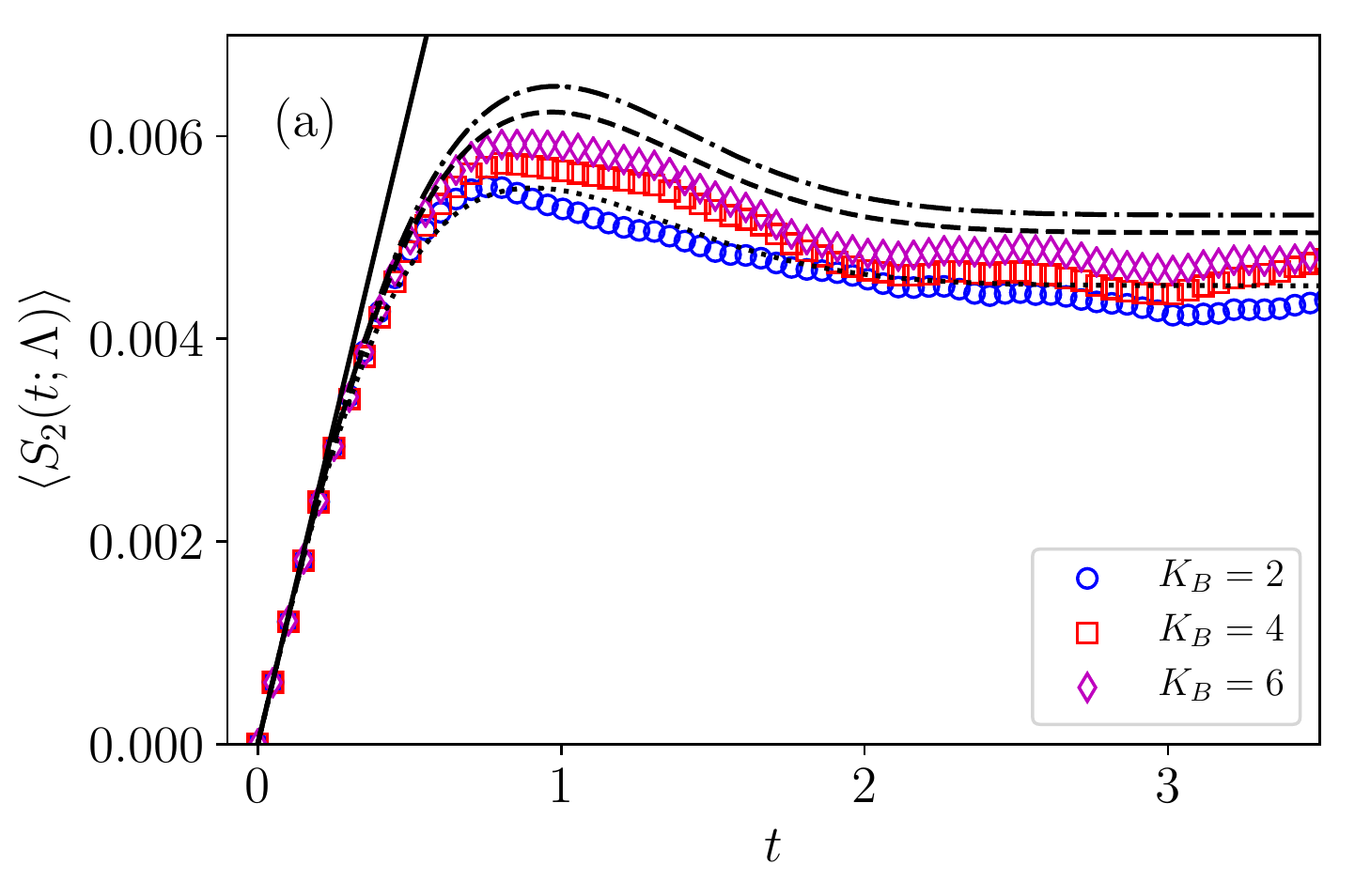}
  \includegraphics[width=\columnwidth]
    {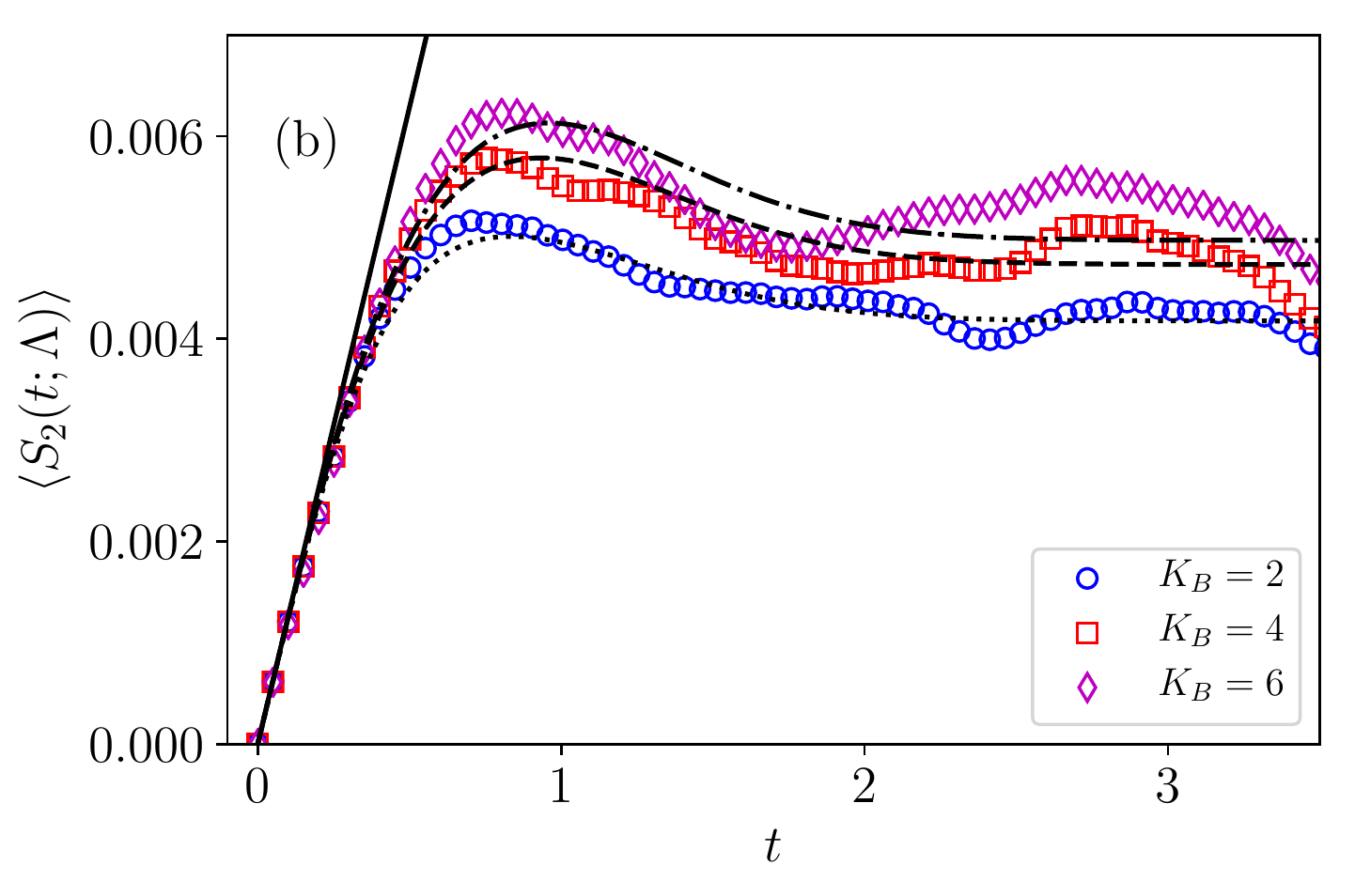}
    \caption{\label{fig:avg_S2_te_1K} Ensemble averaged linear entropy $\langle
    S_2(t;\Lambda)\rangle$ for Eq.~\eqref{eq:InitialState_1K}-type initial 
    states versus $t$ for the RMT transition ensemble of 
    Eq.~\eqref{eq:GenericFloquetRMT} (a) E $\otimes$ C and (b) E $\otimes$ 
    R-type. Both use $N_A = N_B = 50$ and $\Lambda = 10^{-6}$. The 
    theory prediction is given by Eq.~\eqref{eq:S2_1byK_B_average} in black 
    lines, $K = 2$ (dotted), $4$ (dashed), $6$ (dot-dashed), and short time 
    behavior Eq.~\eqref{eq:S2_short_time_1byK} in a black solid line.}
\end{figure}
\begin{figure}[h!]
    \centering
\includegraphics[width=\columnwidth]
    {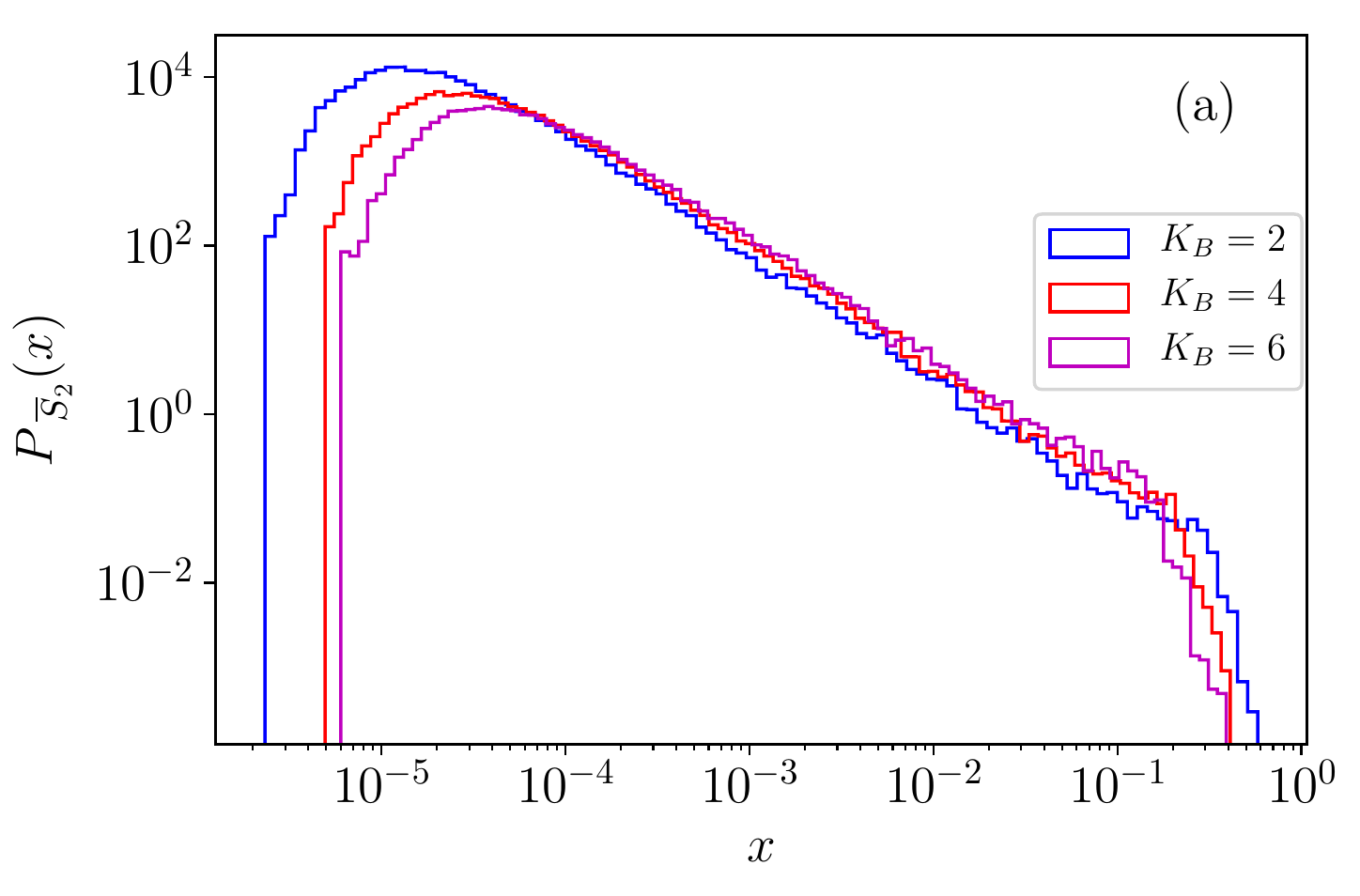}
\includegraphics[width=\columnwidth]
    {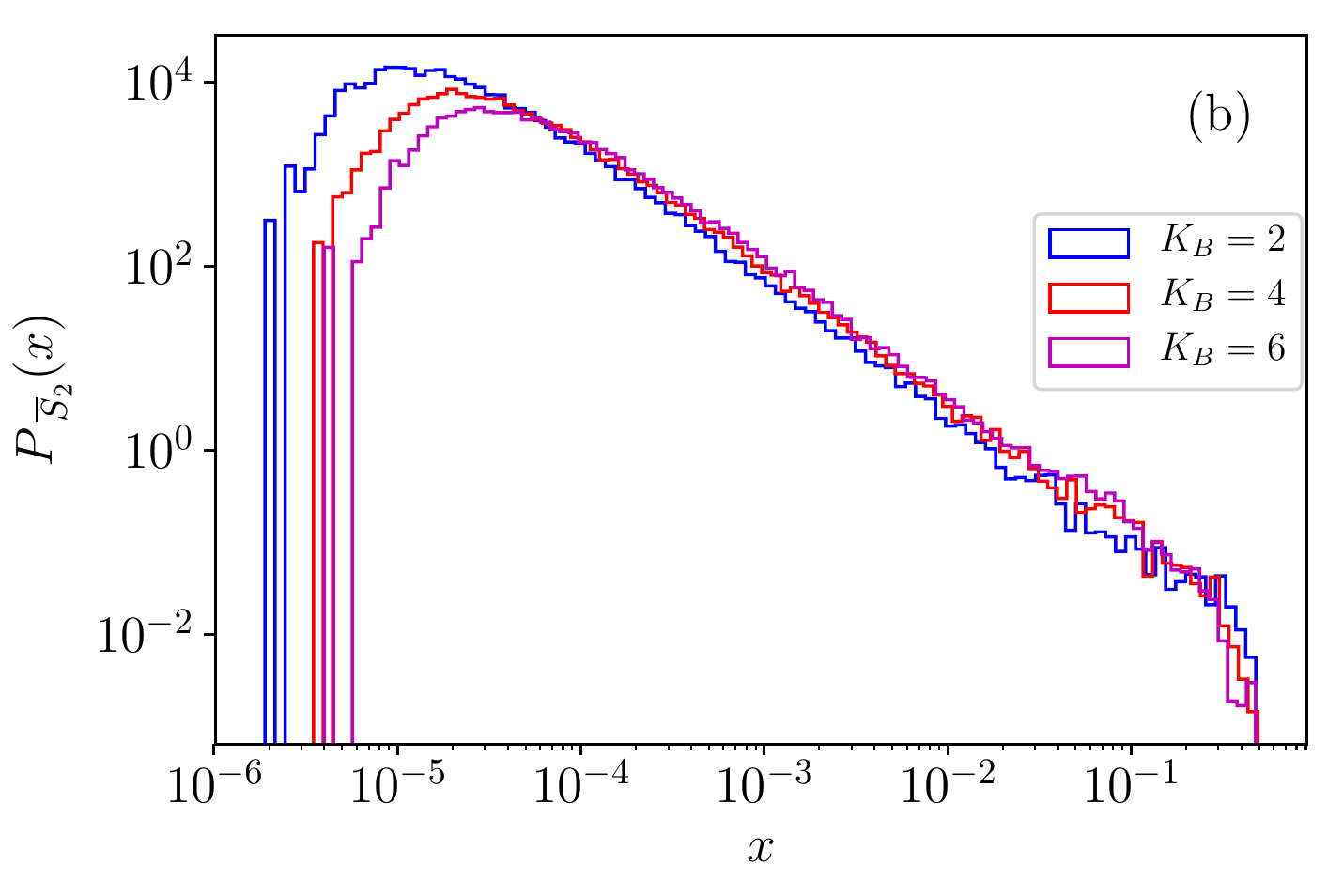}
    \caption{\label{fig:dist_S2inf_1K} Probability density of $\overline{S}_2$
    values for initial state type Eq.~\eqref{eq:InitialState_1K} (a) E $\otimes$
    C and (b) E $\otimes$ R type. Various $K_B$ values are shown for 
    $N_A = N_B = 50$ and $\Lambda = 10^{-6}$.}
\end{figure}

Putting all this together, it can be shown that the reduced density matrix in
Eq.~\eqref{eq:RDM_1byK_B} is approximately diagonal in the $\mathbb{B}_A$ basis,
and the largest Schmidt eigenvalue of $\rho_A(t;\Lambda)$ in
Eq.~\eqref{eq:RDM_1byK_B} is
\begin{align}
    \lambda_1(t;\Lambda) \approx \sum_k |z_{B,k}|^2 \lambda_{1,jk}(t;\Lambda),
\end{align}
and rest of the (relevant) Schmidt eigenvalues are
\begin{align}
    \lambda_l(t;\Lambda) = |z_{B,k}|^2 \lambda_{2,jk}(t;\Lambda),
\end{align}
where for each $l=2, \ldots, K_B+1$ there is a corresponding  pair $(j,k)$ in 
$\{ \lambda_{2,jk}(t;\Lambda)\}_{k=1,\ldots,K_B}$, however the ordering is
\emph{a priori} unknown. 
This gives the linear entropy as
\begin{align}
    S_2(t;\Lambda) & \approx 2 \sum_k |z_{B,k}|^2 \lambda_{2,jk}(t;\Lambda) -
   2 \sum_k |z_{B,k}|^4 \lambda_{2,jk}^2(t;\Lambda),
\end{align}
where $\lambda_{1,jk}(t;\Lambda) \approx 1 - \lambda_{2,jk}(t;\Lambda)$ is 
used, and the cross-terms $\lambda_{2,jk}(t;\Lambda)\lambda_{2,jk'}(t;\Lambda)$ 
for $k \neq k'$ are neglected since they contribute to $\mathcal{O}(\Lambda)$ 
on average. Upon ensemble averaging and rearranging
\begin{align}
    \langle S_2(t;\Lambda)\rangle \approx \big[C(2;t)+2 \langle c_B^{(2)}\rangle 
    C_2(2;t) \big] \sqrt{\Lambda}, \label{eq:S2_1byK_B_average}
\end{align}
where the result 
$\langle \lambda_{2,jk}^2 \rangle \approx C_2(2;t) \sqrt{\Lambda}$ from 
\cite{Jethin_PRE2020} is used; $C_2(2;t)$ is defined in
App.~\ref{App:Cfunctions}. Expanding the averaged linear entropy 
for short-time gives
\begin{equation}
    \langle S_2(t;\Lambda) \rangle = 4 \pi t \sqrt{\Lambda} + \mathcal{O}(t^3),
    \label{eq:S2_short_time_1byK}
\end{equation}
which is same as that of \EE type initial states. The average linear entropy 
after long time saturates to
\begin{equation}
    \langle \overline{S}_2 \rangle \approx \Big( \frac{5+3 \langle c_B^{(2)}
    \rangle}{8} \Big) \sqrt{\Lambda}\,\,,
\end{equation}
which is obtained by taking $t \rightarrow \infty$ limit of
Eq.~\eqref{eq:S2_1byK_B_average}.

Relatively good agreement between the theory prediction and numerical data is 
found as illustrated in Fig.~\ref{fig:avg_S2_te_1K} where C-type 
($\ket{\alpha_{K_B}}_\text{C}$, in (a)) and R-type
($\ket{\alpha_{K_B}}_\text{R}$, in (b)) states are considered for subsystem 
$B$ initial states. The deviations from the theory curve that are seen in the 
plot can be understood by examining the probability density of 
$\overline{S}_2$ as shown in Fig.~\ref{fig:dist_S2inf_1K}. A broad 
heavy-tail-like behavior similar to that of \EE initial states is found.  Thus,
for this type of initial states, equilibrium is a questionable notion. This 
shows that the convergence to the averaged linear entropy curve 
of Fig.~\ref{fig:avg_S2_te_1K} is quite slow and requires a large sample size.
Furthermore, relaxation is not occurring and no self-averaging is apparent.  
Hence no equilibration occurs even though one of the subsystems has non-vanishing
coherence.

\subsubsection{C $\otimes$ C and R $\otimes$ R type initial states}

Consider the initial states of the form Eq.~\eqref{eq:alpha0Generic} whose 
participant unperturbed eigenstates are from the subset $\mathbb{B}_{AB}' =
\mathbb{B}_A' \otimes \mathbb{B}_B'$. As discussed earlier, for the $\Lambda
\rightarrow 0^+$ regime the time evolved state after a long time remains largely
within the subspace described by $\mathbb{B}_{AB}'$ of the full Hilbert space. 
This scenario gets modified as the interaction strength increases 
where the rotation of eigenstates becomes increasingly relevant. However, for
sufficiently small interaction strengths, a perturbed eigenstate 
$\ket{jk(\Lambda)}$ consists of just $\ket{jk}$ and its energetically closest
neighbor $\ket{(jk)_2}$, similar to the earlier assumption used for time 
evolved state $\ket{jk(t;\Lambda)}$. Thus for each $\ket{jk}$ in the subset 
$\mathbb{B}_{AB}'$, there is a corresponding set of such closest neighbors
$\{\ket{(jk)_2}\}$. The set $\{ \ket{(jk)_2}\}$ can be divided into two
subsets, where one of the subsets has eigenkets outside of $\mathbb{B}_{AB}'$ 
and the other contains the eigenkets that are also in $\mathbb{B}_{AB}'$.
The elements in these subsets are \emph{a priori} unknown. This
situation brings some non-trivial effects on the entanglement produced and its
fluctuation about the equilibrium value $\langle \overline{S}_2 \rangle$
compared to $\Lambda \rightarrow 0^+$ regime and is discussed ahead.
\begin{figure}[h!]
    \centering
\includegraphics[width=\columnwidth]
    {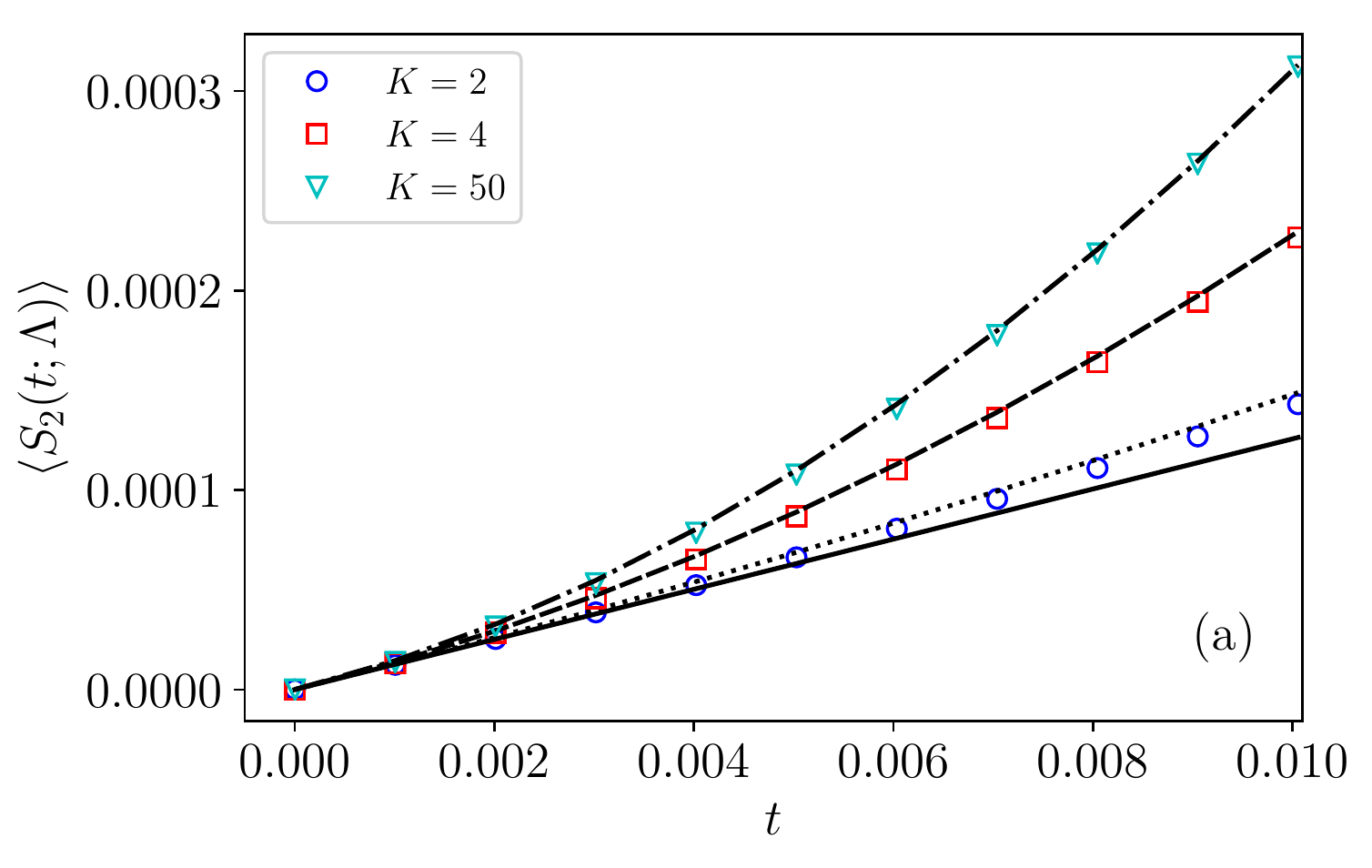}
\includegraphics[width=\columnwidth]
    {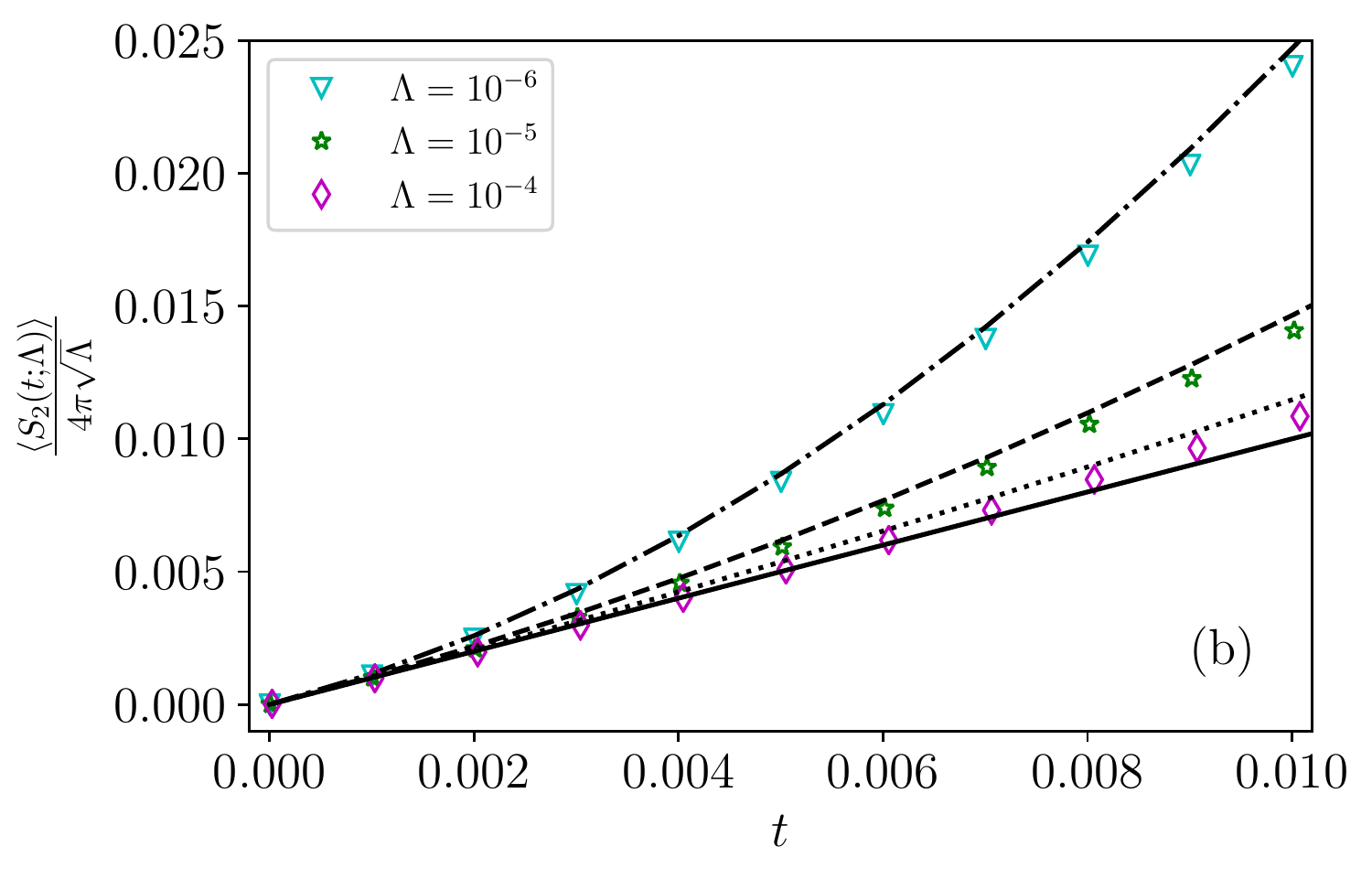}
    \caption{\label{fig:avg_S2_TP_06_short_t} 
    Short time behavior of ensemble averaged 
    $\langle S_2(t;\Lambda)\rangle$ for \RR type initial
    states for $N_A = N_B = 50$.  
    In (a) $\Lambda = 10^{-6}$ with various $K$ values. The black solid line 
    represents Eq.~\eqref{eq:S2_jk_te_short}, and other black lines indicate 
    the combined theory in Eq.~\eqref{eq:S2_weak_combined} corresponding to
    various $K$ values. In (b) $K = 50$ with the 
    $\Lambda$ values indicated in the legend. The black solid line represents
    Eq.~\eqref{eq:S2_jk_te_short} and the other black lines correspond to various
    $\Lambda$ values of the combined theory given in 
    Eq.~\eqref{eq:S2_weak_combined}.  All the curves are scaled by
    $4\pi\sqrt{\Lambda}$.
    }
\end{figure}
\begin{figure}[h!]
    \centering
\includegraphics[width=\columnwidth]
    {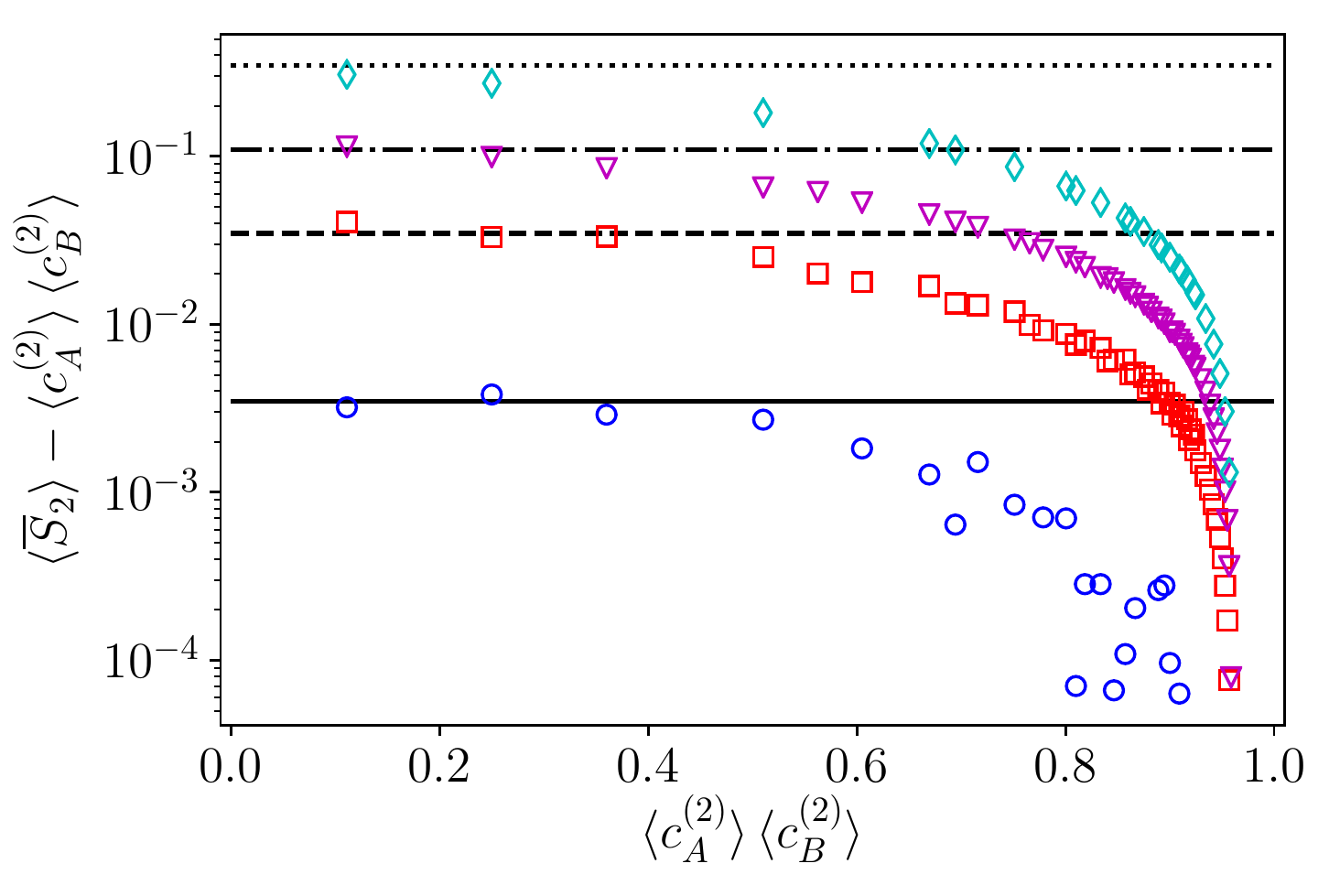}
    \caption{\label{fig:diff_avg_S2_inf} 
    The difference of $\langle \overline{S}_2 \rangle$ and initial state
    coherence $\langle c_A^{(2)} \rangle \langle c_B^{(2)}\rangle$ as a 
    function of the initial state coherence for various interaction strengths
    $\Lambda = 10^{-6}$ ($\color{blue}{\Circle}$), $10^{-4}$
    ($\color{red}{\square}$), $10^{-3}$ ($\color{magenta}{\bigtriangledown}$), 
    $10^{-2}$ ($\color{cyan}{\diamondsuit}$) are shown where
    dimensionality $N_A = N_B = 50$. The horizontal black lines show
    Eq.~\eqref{eq:S2_jk_inf} for various $\Lambda$ where the solid line 
    corresponds $\Lambda = 10^{-6}$ and dotted line is for $\Lambda = 10^{-2}$.}
\end{figure}

Numerical data reveal that the effect of eigenstate rotation is 
quite evident during the initial entanglement production phase. For any 
$\Lambda > 0^+$, a linear-in-time growth is found, which happens to be the same
as that of \EE initial states given in Eq.~\eqref{eq:S2_jk_te_short} for 
$t \approx 0$. The theory derived for $\Lambda \rightarrow 0^+$ in 
Eq.~\eqref{eq:S2inf_TP_0p} fails for this initial entanglement growth phase 
and after including the correction due to the rotation of eigenstates, the 
entanglement production for short times is given by
\begin{equation}
 \langle S_2(t;\Lambda) \rangle \approx  4 \pi t \sqrt{\Lambda} + 2 \langle
   c_A^{(2)} \rangle \langle c_B^{(2)} \rangle t^2 \ , 
   \label{eq:S2_weak_combined}
\end{equation}
which quantifies a competition between quadratic entanglement growth 
and the generally expected linear behavior. Depending on the circumstances, 
the quadratic or linear term may dominate, and the ratio of the terms generates 
a \emph{crossover} time scale given by
\begin{equation}
    t^* = \frac{2 \pi \sqrt{\Lambda}}{\langle c_A^{(2)} \rangle \langle
    c_B^{(2)} \rangle}
\end{equation}
at which the linear and quadratic growth terms contribute equally.  
For $t \ll t^*$ a prominent linear growth is seen and for $t > t^*$ it 
becomes increasingly quadratic.  An interesting aspect of the expression for
the crossover time scale is that for small enough $\Lambda$ and large enough quantum 
coherence in the initial state, $t^* \rightarrow 0$, the linear regime collapses, 
and short time entanglement production effectively displays purely quadratic 
behavior.  On the other hand, for values of $\Lambda$ and quantum coherence 
leading to a $t^*$ which is an appreciable fraction of unity (greater than 
the short scaled time regime), this quadratic regime ceases to exist, and 
only the linear regime behavior results. Note that quadratic growth has been
found to exist in strongly coupled holographic systems \cite{Liu_2014a,
Liu_2014b}, where quadratic behavior transitions to a linear growth, and in
random local Gaussian circuits \cite{Zhuang_2019}.

The crossover between linear and quadratic entanglement growth during the 
initial phase of the time evolution is shown in 
Fig.~\ref{fig:avg_S2_TP_06_short_t}. In Fig.~\ref{fig:avg_S2_TP_06_short_t} (a) 
the average entanglement growth for fixed $\Lambda = 10^{-6}$ and initial states
of type \RR with various $K$ values are shown (\CC type initial states show 
similar behavior and are not displayed here), where the crossover time
scales $t^* = 0.056, 0.017, 0.0068$, for $K = 2, 4, 50$, respectively.
For small $K_A$ and $K_B$ values, the initial states are more like \EE type and 
they show a prominent linear-in-time behavior for a
long period during the initial phase as opposed to the initial states whose 
$K_A \sim N_A$ and $K_B \sim N_B$ where the linear-in-time behavior transitions
to a quadratic behavior rather quickly.
As the interaction strength increases, a dominant linear-in-time entanglement
growth is found stretching to longer and longer times predicted by the time 
scale $t^*$. This is displayed in Fig.~\ref{fig:avg_S2_TP_06_short_t} (b) where 
$K = 50$ and $\Lambda = 10^{-6}, 10^{-5}, 10^{-4}$ whose crossover time scales
are $t^* = 0.0068, 0.021, 0.068$, respectively. Here $\langle
S_2(t;\Lambda)\rangle$ is scaled by $4 \pi \sqrt{\Lambda}$ so that all the
curves for various $\Lambda$ have the same slope for easier visual 
comparison.

The influence of eigenstate rotation can be seen in 
the difference of $\langle \overline{S}_2 \rangle$ and 
$\langle c_A^{(2)} \rangle \langle c_B^{(2)} \rangle$ for various 
interaction strengths  as shown in Fig.~\ref{fig:diff_avg_S2_inf}. This 
difference is expected to be of the order $\mathcal{O}(\sqrt{\Lambda})$ based on 
what is known for the \EE case in Eq.~\eqref{eq:S2_jk_inf}, illustrated in the 
plot as horizontal black lines. As the initial state coherence increases, 
there are increasing deviations from this expectation due to the corrections 
that strongly depend on the initial state coherence.  For near-maximal 
initial state coherence, the eigenstate rotation has negligible affect on 
the saturation value $\langle \overline{S}_2\rangle$, and 
the difference vanishes.
\begin{figure}[h!]
\centering
    \includegraphics[width=\columnwidth]
    {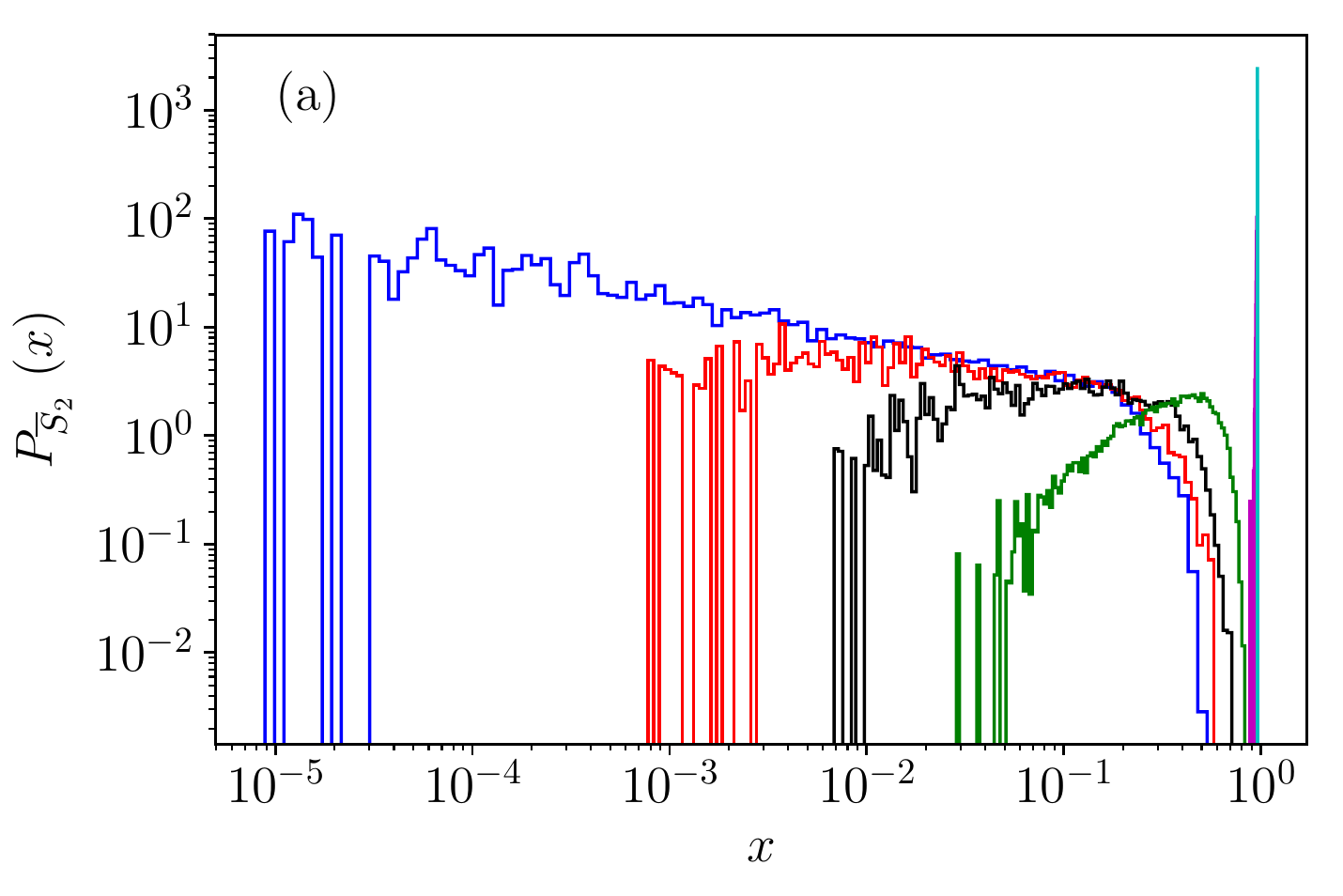}
    \includegraphics[width=8cm]
    {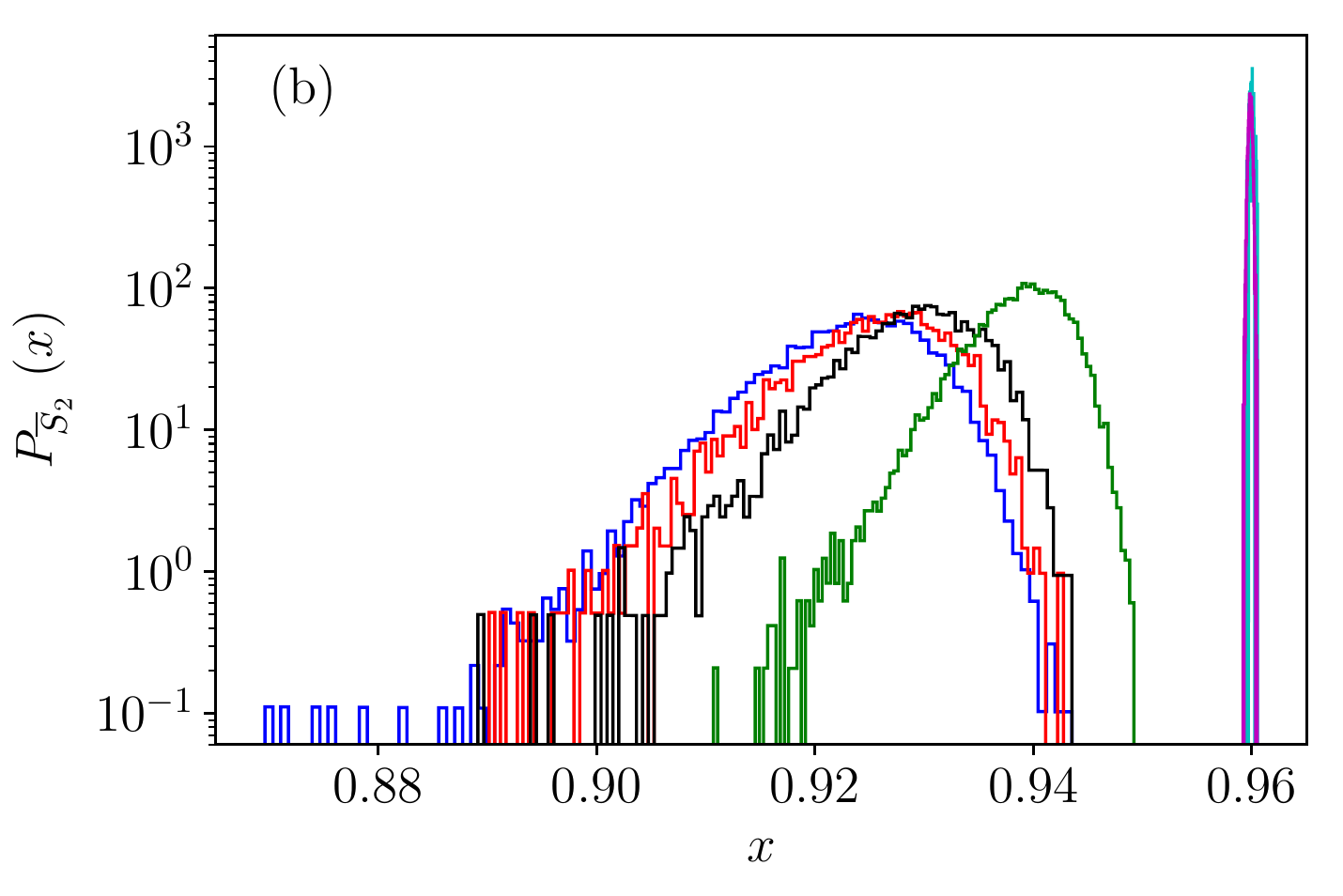}
    \caption{\label{fig:S2inf_K2_dist} Probability density of $\overline{S}_2$
    values for \RR type initial states for (a) $K = 2$, (b) $K = 50$. 
    $\Lambda = 10^{-6}$ (blue), $10^{-4}$ (red), $10^{-3}$ (black), $10^{-2}$ 
    (green), $1$ (magenta), $10$ (cyan).}
\end{figure}
\begin{figure}[h!]
\centering
\includegraphics[width=\columnwidth]
    {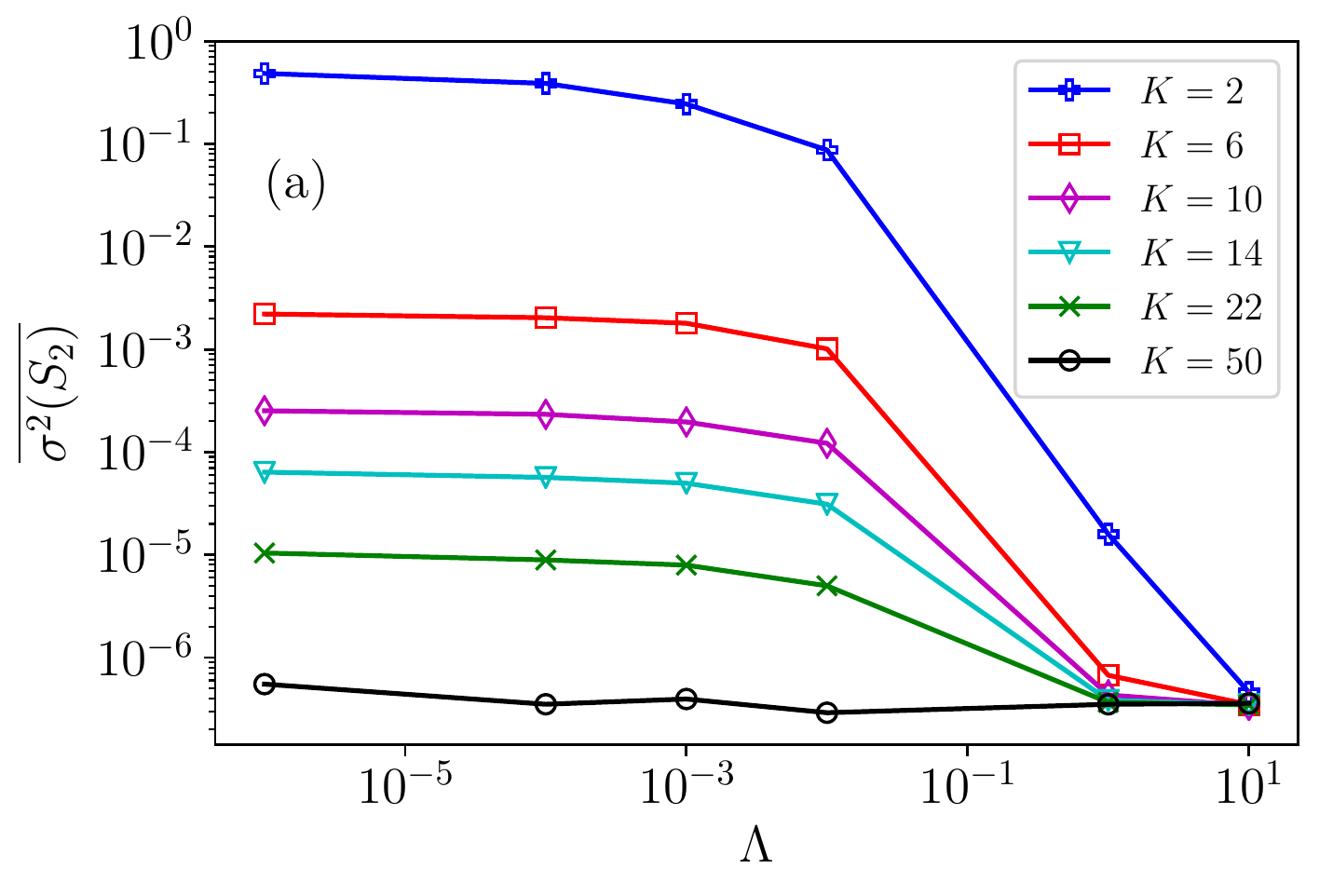}
\includegraphics[width=\columnwidth]
    {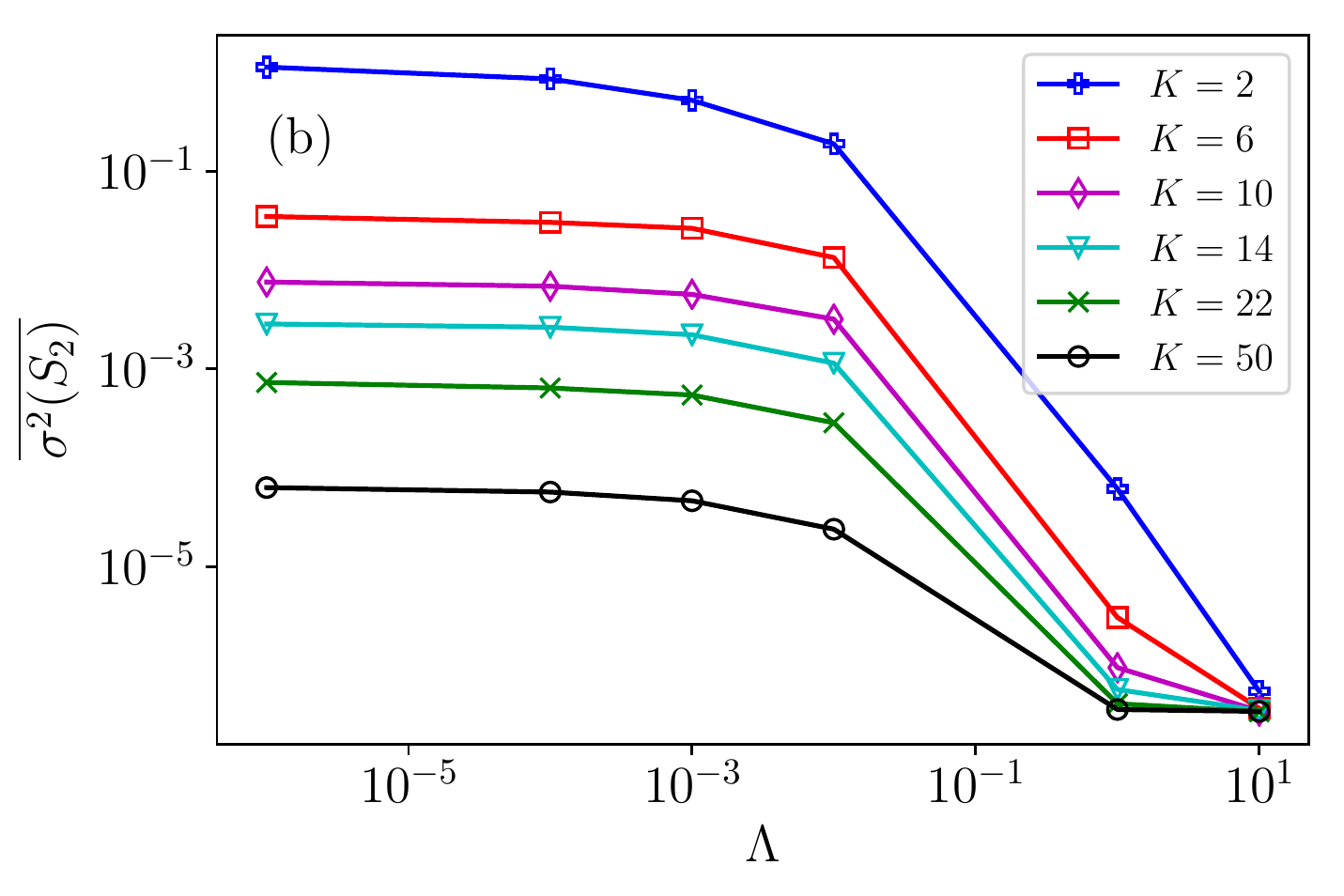}  
    \caption{\label{fig:rel_S2_TP_all} The relaxation measure
    $\overline{\sigma^2(S_2)}$ as a function of $\Lambda$ for (a) \CC and
    (b) \RR type initial states. Various $K$ values with $N_A = N_B = 50$.}
\end{figure}

Furthermore, the probability density of $\overline{S}_2$ becomes increasingly 
localized as the interaction strength increases, which is not surprising.  The
change in shape is more pronounced for $K_A \ll N_A$, $K_B \ll N_B$, where it 
transitions from the broad heavy-tail like behavior to a localized 
Gaussian-like probability density around $\Lambda = 10^{-2}$ as shown in
Fig.~\ref{fig:S2inf_K2_dist} (a).  On the other hand, for $K_A \sim N_A$, 
$K_B \sim N_B$ the density is a narrow, distorted Gaussian-like shape, which
becomes even narrower as the interaction strength is increased; see 
Fig.~\ref{fig:S2inf_K2_dist} (b). This also is not surprising because for 
small $K_A,\,K_B$ and $\Lambda \rightarrow 0^+$ the probability density is 
broad whereas for large $K_A,\,K_B$ it is quite localized as seen in 
Figs.~\ref{fig:S2inf_dist_TP0p} (a),~\ref{fig:S2_TP_0p_fluct_CC_RR} (a). 
So as the interaction strength increases, for initial states with small 
$K_A, \, K_B$ more and more unperturbed eigenstates are perturbatively added 
leading to a more localized Gaussian-like distribution by the virtue of 
central limit theorem along with self-canceling of terms involving oscillating 
eigencomponents with random frequencies that appear in the $\overline{S}_2$
expression.  This also means that as the interaction strength increases, more 
and more initial states relax to the equilibrium as observed in 
Figs.~\ref{fig:rel_S2_TP_all} (a) and (b) where both \CC and \RR type initial 
states are shown, respectively. As the interaction strength increases, a 
significant change in $\overline{\sigma^2(S_2)}$ occurs only for $K=2$.  
Although, for most $K$, a drop occurs across the intermediate strength 
perturbation regime. Interestingly, for maximally coherent \CC type initial 
states ($K=50$), the relaxation measure is nearly constant and much less than 
unity, even extending to the intermediate and strong regimes, implying that 
the system attains thermal equilibrium eventually.  Note that the time 
taken to reach saturation gets shorter and shorter as $\Lambda$ increases.
\emph{In other words, the maximal coherence in the initial product states is an 
alternate path to thermalization of the system under time evolution with 
$\Lambda$ determining how rapidly it reaches thermal equilibrium 
regardless of whether the eigenstates of the system are thermal or not.}	

\subsection{Intermediate perturbation regime}
\begin{figure}[h!]
\centering
\includegraphics[width=\columnwidth]
    {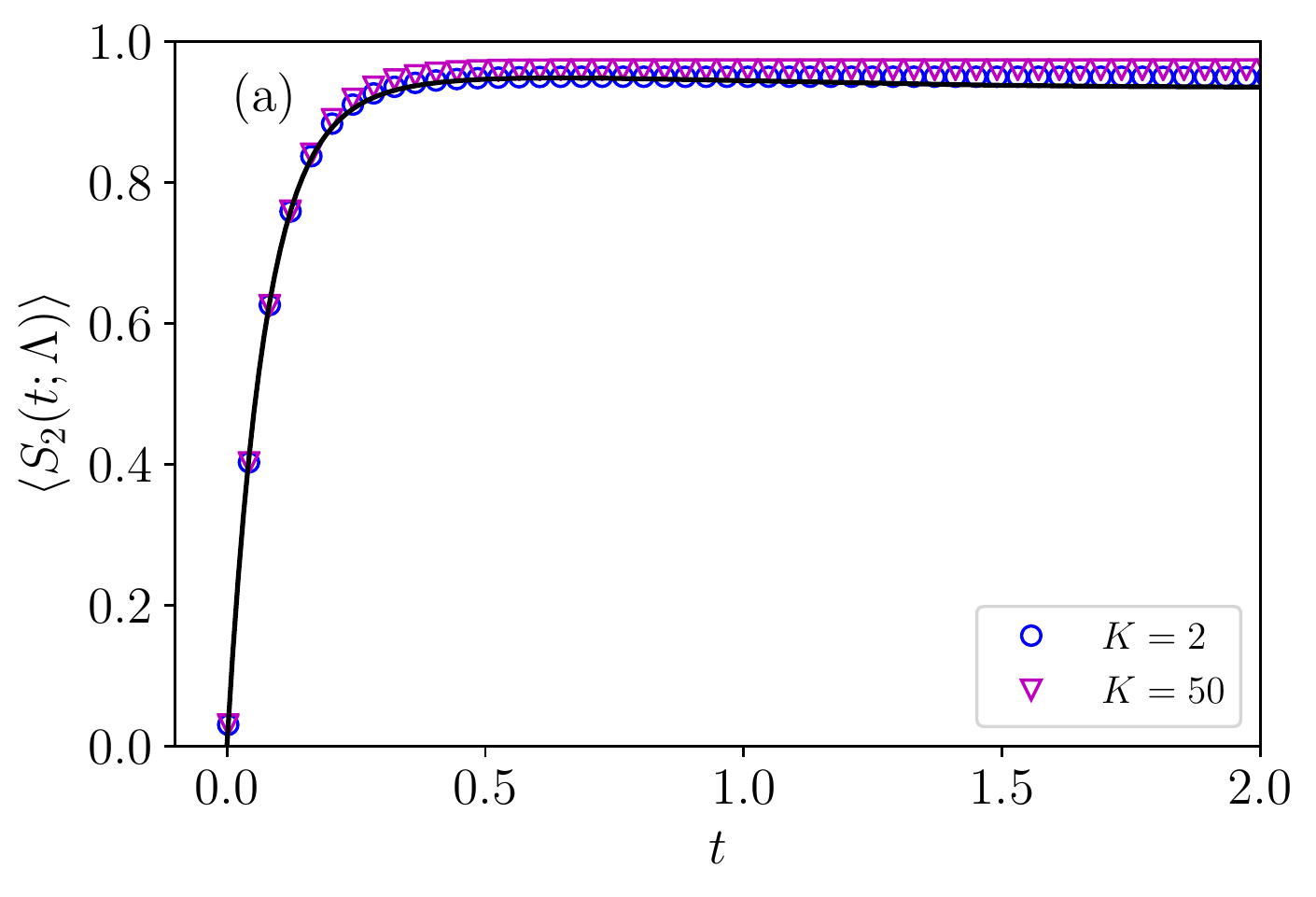}
\includegraphics[width=\columnwidth]
    {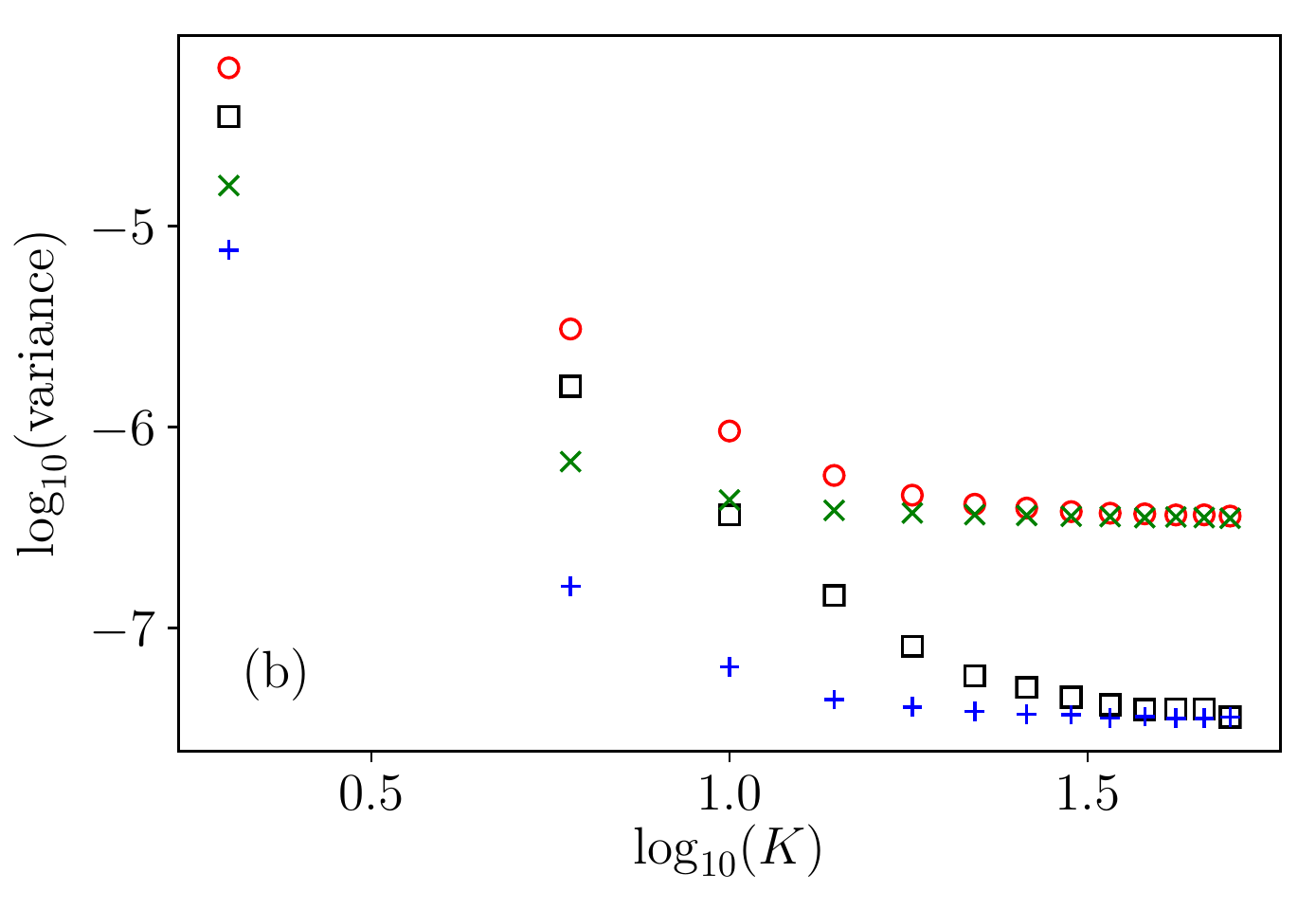}
    \caption{\label{fig:S2_avg_var_intermediate} (a) Ensemble averaged linear 
    entropy $\langle S_2(t;\Lambda) \rangle$ as a function of 
    rescaled time $t$ is shown for $K = 2,\,50$ \RR type initial states. 
    The black solid curve shows the \EE type result 
    in Eq.~\eqref{eq:avg_S2_te_intermediate} for the non-perturbative regime.
    (b) The two fluctuation measures, $\sigma^2(\overline{S}_2)$ 
    and $\overline{\sigma^2(S_2)}$ for \CC ($\color{blue}{+}$ and 
    $\color{Green}{\times}$, respectively) and \RR type 
    ($\color{black}{\square}$ and $\color{red}{\Circle}$, respectively). For
    both (a) and (b) $N_A = N_B = 50$ and $\Lambda  = 1$.}
\end{figure}
This regime can be viewed as an interpolation or transition from the 
weak to strong interaction regime, where a weak dependence on the initial 
state coherence is observed which vanishes as the interaction strength 
increases.  From this observation, the expression for 
$\langle S_2(t;\Lambda) \rangle$ derived in the study \cite{Jethin_PRE2020} for
\EE initial states in the non-perturbative regime
\begin{equation}
    \langle S_2(t;\Lambda) \rangle \approx \langle \overline{S}_2^\infty
    \rangle \Bigg[1 - \exp(- \frac{C(2;t) \sqrt{\Lambda}}{\langle
    \overline{S}_2^\infty \rangle})\Bigg], \label{eq:avg_S2_te_intermediate}
\end{equation}
can be used as a decent approximation to the entanglement produced for any 
initial product states. This is illustrated in 
Fig.~\ref{fig:S2_avg_var_intermediate} (a) for $\Lambda = 1$. The 
saturation values for $K = 2,\,50$ can be seen to be slightly higher than 
that the theory curve in Eq.~\eqref{eq:avg_S2_te_intermediate} and $K=50$ curve 
saturates above that of $K=2$. Note that in the strong interaction regime 
discussed previously in Sect.~\ref{subsec:strong} the above equation was 
used to show that the average entanglement generated is largely independent 
of the initial (product) state.

The probability density of $\overline{S}_2$ for $\Lambda = 1$ is highly 
localized in contrast to the ones in the weak perturbative regime; see 
Fig.~\ref{fig:S2inf_K2_dist}. The initial state coherence appears to play a role
in the width of the density.  The $K=50$ case is almost identical to that of
$\Lambda = 10$ in the strong interaction regime.
Furthermore, the fluctuation measures obtained numerically for \CC and \RR type 
initial state ensembles show that a majority of the initial states eventually 
attain the equilibrium value $\langle\overline{S}_2\rangle$ and remain there 
as shown in Fig.~\ref{fig:S2_avg_var_intermediate} (b).

\subsection{Discussion}

It is worthwhile to collect all the results discussed so far to
get an overall picture of how relaxation and equilibration depend on the quantum 
coherence in an initial product state and the interaction strength between the 
subsystems. Together, the coherence and transition 
parameter govern the entanglement growth and its saturation to an 
equilibrium (if it exists) and the nature of the equilibration.  

In the ultra-weak interaction regime, where the rotation of eigenstates are 
neglected, an existence of equilibrium can be established based on the 
equilibrium measure where the equilibrium becomes sharper as the average 
coherence in the initial state ensemble increases. 
The coherence of the initial states also 
affects its ability to equilibrate. A weak to strong equilibration is found to 
occur as the average coherence of the initial state ensemble increases. In 
particular, for maximally coherent initial product states the system thermalizes 
after sufficiently long time and remains thermalized for almost all the initial
states as indicated by negligibly small fluctuations about the equilibrium.
Furthermore, a quadratic initial entanglement growth is seen in this regime
characterized by the average coherence of the ensemble under consideration.

A similar story can be said for weak interaction regime, except for the new
effects that arise due to the rotation of eigenstates. Firstly, for initial
states whose coherence is vanishing an equilibrium is questionable and hence no
relaxation can be seen. The initial entanglement growth is linear-in-time and is
purely due to the rotation of eigenstates and the rate is dictated by the
transition parameter. The average entanglement saturation value is perturbative
and is also dictated strongly by the transition parameter. Secondly, for initial
state ensembles with non-vanishing average coherence, a competition between
linear and quadratic growths occurs in the initial phase. While the
linear-in-time behavior is same for any kind of initial state ensemble, the
quadratic growth depends strongly on the average coherence. Furthermore, the
entanglement saturation shows strong dependence on the transition parameter for
initial state ensembles with low average coherence while a weak dependence is
seen for ensembles with high coherence. 

For intermediate and strong interaction regimes, little to negligibly small
dependence on the initial states is observed, including the ones with vanishing
coherence. In these regimes, a strong equilibration is seen. A noteworthy point
here is that the initial state ensemble with maximal coherence thermalizes
regardless of the strength of the interaction but how rapidly it thermalizes
is controlled by the transition parameter. Larger the transition parameter,
shorter it takes to thermalize. Table~\ref{tab:table} summarizes the key
results.
\begin{longtable*}{c|c|c|c|c}
    \caption{\label{tab:table} Various initial state ensembles and nature of the
    equilibrium and initial entanglement growth in various interaction regimes.}
    \\
    \hline \hline
    \multirow{2}{*}{Initial state ensemble} & \multicolumn{4}{|c}{Interaction
    regime}  \\
    \cline{2-5}
      & Ultra-weak & Weak & Intermediate & Strong \\

    \hline \hline 
    E $\otimes$ E, E $\otimes$ C, E $\otimes$ R  & 
    no equilibration, & 
    no equilibration, & 
    strong equilibration, &
    strong thermalization, \\
    (vanishing coherence) & 
    initial linear growth &
    initial linear growth &
    initial linear growth &
    initial linear growth \\
    \hline
    C $\otimes$ C, \RR & 
    weak equilibration, &
    weak equilibration, &
    strong equilibration, &
    strong thermalization, \\
    (low coherence) &
    initial quadratic growth &
    late linear to quadratic crossover & 
    initial linear growth &
    initial linear growth \\

    \hline
    C $\otimes$ C, \RR &
    strong equilibration, &
     strong equilibration, & 
     strong equilibration, & 
     strong thermalization, \\
    (high coherence) &
    initial quadratic growth &
    early linear to quadratic crossover & 
    initial linear growth &
    initial linear growth \\
    \hline
    \CC &
    strong thermalization &
    strong thermalization &
    strong thermalization &
    strong thermalization \\
    (maximal coherence) &
    initial quadratic growth &
    early linear to quadratic crossover  &
    initial linear growth &
    initial linear growth \\
    \hline
\end{longtable*}

\section{Summary and outlook}

In this paper, some of the direct consequences of quantum coherence towards the 
entanglement production, equilibration and thermalization are presented. 
In an earlier study \cite{Jethin_PRE2020}, the entanglement production of 
special \EE type initial states was found to 
depend only on a universal transition parameter $\Lambda$, and the rescaled 
time $t$, which as shown in this study, generalizes the scope of these universal 
quantities to generic initial states. As guided by perturbation theory,
various perturbation regimes were identified.

In the ultra-weak regime $\Lambda \rightarrow 0^+$ where the diagonal elements 
of the perturbation matrix are relevant and the effect of rotation of 
eigenstates is neglected, it was analytically shown that the entanglement 
production is a Gaussian as a function of time and 
saturates to the product of coherence 
measures (a coherence measure based on the $l_2$-norm) of the subsystems in
their respective \emph{preferred} eigenstate basis.  This implies an unusual 
initial quadratic time dependent increase in the entanglement.  Whether 
quadratic or linear behavior is to be expected is contained in 
Eq.~(\ref{eq:S2_weak_combined}).  It was found that for
initial (product) states whose subsystem coherence measures are near maximal 
after long time the system saturates to the maximal entanglement.  Furthermore,
such states show thermalization as evident from the distribution of infinite 
time average of linear entropy and the relaxation measure, which is a remarkable 
result noting that the interaction between the subsystems is ultra-weak, and 
the full system eigenstates are barely different from the non-interacting case. 
This shows that quantum coherence acts as a resource for equilibration and 
thermalization.

In the weak regime, where the eigenstate rotation is relevant in the
perturbation theory, it was found that the probability density of the 
infinite time linear entropy average of initial states whose product of 
subsystem coherence measures is vanishing has a broad heavy-tail like 
behavior and shows no relaxation. 
Entanglement production is of the order $\mathcal{O}(\sqrt{\Lambda})$ on 
average and shows a slow convergence to the average due to the heavy-tailed
nature of the distribution. In the initial entanglement growth phase, it can be
shown that the entanglement production is linear in time and the rate is
proportional to $\sqrt{\Lambda}$. This initial linear-in-time entanglement
growth is seen regardless of the initial state coherence and is universal, even
for large $\Lambda$. The effect of eigenstate rotation is also evident in
the entanglement saturation values, especially for initial states whose
coherence is close to minimal. On the other hand, initial states with near-maximal
coherence are not dependent on this. Lastly,
in the intermediate regime and beyond, the entanglement production appears to
have little to no dependence on the initial product states and shows an
exponential behavior.

In the present study, an RMT transition ensemble was used to mimic a bipartite
system whose subsystems are fully quantum chaotic. For a real dynamical system 
either single-particle or many-body, features like eigenstate scarring
\cite{Heller84}, $k$-body interactions in case of many-body systems (see
the review \cite{Brody81}) and other dynamical 
features may give rise to some system specific deviations to the universal 
features presented in this paper. It would be interesting to know how the universal
features presented here would change if the subsystems are not fully chaotic,
where the notion of the transition parameter may not exist due to selection
rules and existence of local integrals of motion (see the review 
\cite{Abanin_RMP2019}).
As seen in this study, the initial product state coherence plays a crucial role
in determining the fate of the interacting bipartite system at long times.
This study may shed light on understanding the phase transition towards
thermalization not just with the interaction strength, but also due to coherence
in the initial state.
Moreover, it would be interesting to understand the entanglement
production in a semiclassical sense similar to the earlier fidelity studies 
that related the transition parameter to the classical action diffusion 
coefficient and the phase space volume in the ultra-weak perturbation
regime~\cite{Cerruti02, Cerruti2003}.

\acknowledgments

We are grateful for Washington State University's Kamiak High Performance 
Computer at the Center for Institutional Research Computing, which was used 
extensively for the numerical calculations of the present study. This research 
was partially funded by the Deutsche Forschungsgemeinschaft (DFG, German 
Research Foundation) --  497038782.

\appendix
\section{Derivation of variances of $\mathcal{V}$ matrix elements 
\label{App:HaarUnitary}}

Starting from Eq.~\eqref{eq:Vmatjk_def}, the expression for the variance is
\begin{align}
    \langle |{\cal V}_{jk,j'k'}|^2 \rangle & = \Big \langle \sum_{ab,a'b'} 
    u_{ja}^{A*}\, u_{kb}^{B*} \, u_{j'a}^A \, u_{k'b}^B \,(2 \pi \xi_{ab}) 
    \nonumber \\
    & \qquad \times u_{ja'}^{A} \, u_{kb}^{B} \,
    u_{j'a}^{A*} \, u_{k'b}^{B*} \, (2 \pi \xi_{a'b'}) \Big \rangle.
\end{align}
Given the probability density of $\xi_{ab}$, Eq.~\eqref{eq:V_def_ab}, and 
independence
\begin{equation}
    \langle \xi_{ab} \, \xi_{a'b'} \rangle = \delta_{ab,a'b'} \frac{1}{12},
    \label{eq:xi_ab_var}
\end{equation}
the variance reduces to the expression
\begin{equation}
    \langle | \mathcal{V}_{jk,j'k'}|^2 \rangle = \frac{\pi^2}{3}
    \Big\langle \sum_{ab}
    |u_{ja}^A|^2\,|u_{j'a}^A|^2\,|u_{kb}^B|^2\,|u_{k'b}^B|^2 \Big\rangle,
    \label{eq:Vmatjk_var_deriv}
\end{equation}
where Haar averaging over unitary groups remains to be performed.
Using the known results~\cite{Puchala11} gives
\begin{equation}
    \langle |u_{ja}^A|^2\,|u_{j'a}^A|^2 \rangle = \frac{1+\delta_{jj'}}{N_A
    (N_A+1)},
\end{equation}
and similarly for subsystem $B$ in Eq.~\eqref{eq:Vmatjk_var_deriv}. 
Summing over the variables $a,b$ leads to the result 
in Eq.~\eqref{eq:Vmatjk_var}.

The covariance, $\langle x_{jk}\, x_{j'k'} \rangle$, in
Eq.~\eqref{eq:x_jkDiagCov} is given by
\begin{equation}
    \langle x_{jk} x_{j'k'} \rangle = \frac{\langle \mathcal{V}_{jk,jk} \,
    \mathcal{V}_{j'k',j'k'} \rangle}{\sqrt{\langle |\mathcal{V}_{jk,jk}|^2
    \rangle} \sqrt{\langle |\mathcal{V}_{j'k',j'k'}|^2 \rangle} },
\end{equation}
where making use of $\sqrt{ \langle |\mathcal{V}_{jk,jk}|^2 \rangle} = 
\sqrt{ \langle |\mathcal{V}_{j'k',j'k'}|^2 \rangle}$ and 
Eq.~\eqref{eq:Vmatjk_var} gives
\begin{equation}
\sqrt{\langle |\mathcal{V}_{jk,jk}|^2 \rangle} 
    \sqrt{\langle |\mathcal{V}_{j'k',j'k'}|^2 \rangle} = \frac{4
    \pi^2}{3(N_A+1)(N_B+1)} \,.
\end{equation}
That leaves the $\langle \mathcal{V}_{jk,jk}\,\mathcal{V}_{j'k',j'k'}\rangle$
computation to be done.  Starting from Eq.~\eqref{eq:Vmatjk_def} and using
Eq.~\eqref{eq:xi_ab_var} gives
\begin{equation}
    \langle \mathcal{V}_{jk,jk} \,\mathcal{V}_{j'k',j'k'} \rangle =
    \frac{\pi^2}{3} \Big \langle \sum_{ab}
    |u_{ja}^A|^2\,|u_{j'a}^A|^2\,|u_{kb}^B|^2\,|u_{k'b}^B|^2 \Big\rangle
\end{equation}
and performing Haar averaging on the subsystem unitary groups,
as in Eq.~\eqref{eq:Vmatjk_var_deriv}, establishes
Eq.~\eqref{eq:x_jkDiagCov}.

\section{Details of numerical calculations \label{App:numerics}}

All the calculations presented in this article are based on realizations of 
the random matrix transition ensemble defined in 
Eq.~\eqref{eq:GenericFloquetRMT}  using subsystem dimensionality 
$N_A = N_B = 50$.
The sample size details of time evolution raw data for various \CC and 
\RR type initial state ensembles and all $\Lambda$ values are shown in
Table.~\ref{tab:numerics} as initial state sampled per realization $\times$ total
number of realizations.
\begin{table}[H]
    \caption{\label{tab:numerics} Various $\Lambda$ values and corresponding
    initial state samples per realization $\times$ number of realizations.}
    \begin{ruledtabular}
        \begin{tabular}{llllll}
            $\Lambda$ & \CC & \RR & \EE & E $\otimes$ C & E $\otimes$ R \\
            \hline 
            $10^{-6}$ & 1250 $\times$ 5 & 2500 $\times$ 5 & 2500 $\times$ 20 &
            2500 $\times$ 20 & 2500 $\times$ 5 \\
            $10^{-4}$ & 1250 $\times$ 5 & 750 $\times$ 5 & N/A & N/A & N/A \\
            $10^{-3}$ & 750 $\times$ 5  & 750 $\times$ 5 & N/A & N/A & N/A\\
            $10^{-2}$ & 2500 $\times$ 5 & 2500 $\times$ 5 & N/A & N/A & N/A \\
            $1$       & 2500 $\times$ 2 & 2500 $\times$ 2 & N/A & N/A & N/A \\
            $10$      & 50 $\times$ 5 & 50 $\times$ 5 & N/A & N/A & N/A \\
        \end{tabular}
    \end{ruledtabular}
\end{table}
For the short time behavior depicted in Fig.~\ref{fig:avg_S2_TP_06_short_t} and for
various $\Lambda$ values 1000 $\times$ 5 initial states were used instead. In
addition, for Fig.~\ref{fig:S2inf_EE_dist} data generated for the $\overline{S}_2$
probability density based on the infinite time average expression in 
Eq.~\eqref{eq:ITA_purity} used 2500 $\times$ 10 initial states.

\section{Eigenvector statistics of unitary ensemble
\label{App:EigenvectorStatistics}}

Consider an eigenvector $\ket{\alpha}$ of an $N$-dimensional CUE matrix of 
unitary ensemble. Represented in some fixed basis $\ket{i}$, it is given by
$\ket{\alpha} = \sum_{i=1}^N z_i \ket{i}$, where $z_i$ are complex coefficients.
The only constraint is the normalization, $\sum_i
|z_i|^2 = 1$. The probability density of the eigenvector components are
given by~\cite{Brody81, Haake, Lakshminarayan2008}
\begin{equation}
    P(z_1,z_2,\ldots,z_N) = \frac{(N-1)!}{\pi^N}
    \delta\Big(\sum_{i=1}^N|z_i|^2-1\Big),   
\end{equation}
for which reduced probability density can be found by integrating out $N-l$
variables resulting in
\begin{equation}
    P_l(z_1,z_2,\ldots,z_l) = \frac{\Gamma(N)}{\pi^l \Gamma(N-l)}\Big(1 -
    \sum_{i=1}^l |z_i|^2\Big)^{N-l-1}.
\end{equation}
Using the above reduced probability density various moments of $z_i$ can be
computed analytically. Below a list of useful moments relevant to the main text
are given
\begin{equation}
    \langle |z_i|^4 \rangle = \frac{2}{N(N+1)},
\end{equation}
\begin{equation}
    \langle |z_i|^2 |z_j|^2\rangle  = \frac{1}{N(N+1)} \quad \text{for}\, i 
    \neq j,
\end{equation}
\begin{equation}
    \langle |z_i|^8 \rangle  = \frac{24}{N(N+1)(N+2)(N+3)},
\end{equation}
and
\begin{equation}
    \langle |z_i|^4 |z_j|^4\rangle  = \frac{4}{N(N+1)(N+2)(N+3)} \quad 
    \text{for} \, i\neq j.
\end{equation}

\section{$C(2;t)$ and $C_2(2;t)$ functions \label{App:Cfunctions}}

The $C(2;t)$ and $C_2(2;t)$ functions are defined in a previous 
work~\cite{Jethin_PRE2020} for the 
context of the ensemble averaged unperturbed eigenstate time evolution, 
Eq.~\eqref{eq:TimeEvoljk}. The function $C_2(2;t)$ is given by
\begin{align}
    \Big \langle \sum_{l > 1} \lambda_{l,jk}^2(t;\Lambda) \Big \rangle &= 
    C_2(2;t) \sqrt{\Lambda},
\end{align}
where
\begin{align}
    C_2(2;t) & = \int_0^\infty \dd{w} \int_{-\infty}^\infty \dd{z} \,
    \Bigg(\frac{4 z}{z^2+4w} \Bigg)^2 \exp(-w)  \nonumber \\
    & \quad \times \sin^4\Big(\frac{t}{2}\sqrt{z^2+4w}\Big).
\end{align}
The function $C(2;t)$ defined in Eq.~\eqref{eq:S2_unpert_eig_evol} is given by
\begin{align}
& C(2;t) = \pi t \Big( 3 \ue^{-t^2} - \frac{1}{2}\ue^{-4 t^2} \Big) + 
    \pi^{3/2} \text{erf}(t) \Big( \frac{1}{2} + 3 t^2 \Big) \nonumber \\
& \qquad \qquad + \pi^{3/2} \text{erf}(2 t) \Big( \frac{1}{8} - 3 t^2 \Big),
    \label{eq:C2_func}
\end{align}
where $\text{erf}(x) = (2/\sqrt{\pi}) \int_0^x \ue^{-t^2} \dd{t}$ is the 
error function.


\bibliographystyle{cpg_unsrt_title_for_phys_rev.bst}

\bibliography{abbrevs,extracted,further,new_additions,rmtmodify}

\end{document}